%% file: main.tex
\documentclass[%
reprint,
superscriptaddress,
preprintnumbers,
nofootinbib,
amsmath,amssymb,
aps,
longbibliography,
prx,
]{revtex4-1}

\DeclareMathOperator{\Tr}{Tr}
\usepackage{graphicx}% Include figure files

\usepackage{dcolumn}% Align table columns on decimal point
\usepackage{bm}% bold math
\usepackage[colorlinks, linkcolor=blue, citecolor=blue, urlcolor=blue]{hyperref}

\usepackage{tikz}
\usetikzlibrary{matrix,arrows,decorations.pathmorphing}
\usepackage{tikz-cd}


\AtBeginDocument{%
\newwrite\bibnotes
\def\bibnotesext{Notes.bib}
\immediate\openout\bibnotes=\jobname\bibnotesext
\immediate\write\bibnotes{@CONTROL{REVTEX41Control}}
\immediate\write\bibnotes{@CONTROL{%
    apsrev41Control,author="08",editor="1",pages="1",title="0",year="1"}}
\if@filesw
\immediate\write\@auxout{\string\citation{apsrev41Control}}%
\fi
}%

\begin{document}

\title{
	Universal entanglement signatures of interface conformal field theories 
}
\preprint{YITP-23-100}
\preprint{RIKEN-iTHEMS-Report-23}

\author{Qicheng Tang}
\thanks{These two authors contributed equally.}
%\email{tangqicheng@westlake.edu.cn}
\affiliation{Department of Physics, School of Science, Westlake University, Hangzhou 310030, China}
\affiliation{Institute of Natural Sciences, Westlake Institute for Advanced Study, Hangzhou 310024, China}

\author{Zixia Wei}
\thanks{These two authors contributed equally.}
%\email{zixiawei@fas.harvard.edu}
\affiliation{
	Yukawa Institute for Theoretical Physics, Kyoto University, Kyoto 606-8502, Japan}
\affiliation{Interdisciplinary Theoretical and Mathematical Sciences Program (iTHEMS), RIKEN, Wako 351-0198, Japan}

\author{Yin Tang}
%\email{tangyin@westlake.edu.cn}
\affiliation{Department of Physics, School of Science, Westlake University, Hangzhou 310030, China}
\affiliation{Institute of Natural Sciences, Westlake Institute for Advanced Study, Hangzhou 310024, China}

\author{Xueda Wen}
%\email{wenxueda@gatech.edu}
\affiliation{Department of Physics, Harvard University, Cambridge, MA 02138, USA}
\affiliation{Department of Physics, University of Colorado, Boulder, CO 80309, USA}

\author{W. Zhu}
\email{zhuwei@westlake.edu.cn}
\affiliation{Department of Physics, School of Science, Westlake University, Hangzhou 310030, China}%
\affiliation{Institute of Natural Sciences, Westlake Institute for Advanced Study, Hangzhou 310024, China}

%\date{\today}% It is always \today, today,
%  but any date may be explicitly specified

\begin{abstract} 
	An interface connecting two distinct conformal field theories hosts rich critical behaviors. 
	In this work, we investigate the entanglement properties of such critical interface theories for probing the underlying universality. 
	As inspired by holographic perspectives, we demonstrate vital features of various entanglement measures regarding such interfaces based on several paradigmatic lattice models. 
	Crucially, for two subsystems adjacent at the interface, 
	the mutual information and the reflected entropy exhibit identical leading logarithmic scaling, 
	giving an effective interface central charge that takes the same value as the smaller central charge of the two conformal field theories. 
	Our work demonstrates that the entanglement measure offers a powerful tool to explore the rich physics in critical interface theories. 
\end{abstract}

\maketitle

%\tableofcontents

Entanglement offers an exotic path to characterize the universal information about conformal symmetry at quantum critical points~\cite{HOLZHEY1994443, vidal2003entanglement, Calabrese_2004, fradkin2006EE2p1Critical, fradkin2009EE2p1, william2015cornerEE}. 
Especially, when conformal symmetry is partially broken by boundaries and defects into a subset, 
entanglement is sensitive to their presence and can capture their intrinsic features~\cite{Laflorencie2006_boundaryEE, Schollwock2006_boundaryEE, Sorensen_2007_impurityEE, Sorensen_2007_impurityEE2, Affleck_2009review_boundaryEE, igloi_2009_EE_interface, Vasseur2014_EE_QuantumWire, Herzog2016_boundaryEE}. 
%Generally speaking, both boundaries and defects can be viewed as interfaces gluing two theories, i.e. the former glues a critical theory with a trivial gapped theory, and the latter glues two identical critical theories. 
In this letter, we explore an interface gluing two distinct conformal field theories (CFTs) with different values of the central charge: $c^{\rm (I)}$ for CFT$^{\rm (I)}$ and $c^{\rm (II)}$ for another CFT$^{\rm (II)}$, and focus on possible universal entanglement signatures about the interface. 
Such kinds of interfaces 
can naturally appear in various scenarios, 
like the junction of two quantum wires~\cite{Affleck2014_QuantumWire, Oshikawa2021_junction_inequivalent}, 
renormalization group (RG) interfaces between QFTs~\cite{Brunner_2008_RG_DW, Gaiotto2012_RG_DW, Konechny_2014_RG_defect, Brunner_2016_transmission, Cardy2017_boundary_bulkRG, Konechny2021_RGinterface}, evaporation of black holes~\cite{Myers2020_defect_island_1, Myers2020_defect_island_2, Sonner2022_Island}, just to name a few.

When two CFTs are glued in a scale-invariant way~\cite{Bachas_2002_permeable}, the theory is called an \textit{interface CFT} (ICFT). 
Existing attempts on ICFT are mainly based on a simple folding picture~\cite{Affleck1994folding, oshikawa1997_Ising_defect, Bachas_2002_permeable} % Affleck_1992_BCFT_Kondo, 
which converts the interface to a boundary condition of the folded theory. 
While this tool is powerful for investigating two-point functions and transmission properties, the entanglement properties are in general not under analytical control and more difficult to access~\cite{Quella_2007_transmission, Kazuhiro_Sakai_2008, Peschel_2012_EE_defect, Calabrese_2012_EE_junction, Eisler_2012_defect, brehm_2015_EE_interface_ising}, especially for our interested case of $c^{\rm (I)} \neq c^{\rm (II)}$. 
This problem is particularly challenging in the context of CFT, therefore motivates us to consider a holographic estimation and lattice simulations.

\begin{figure}\centering
	\includegraphics[width=\columnwidth]{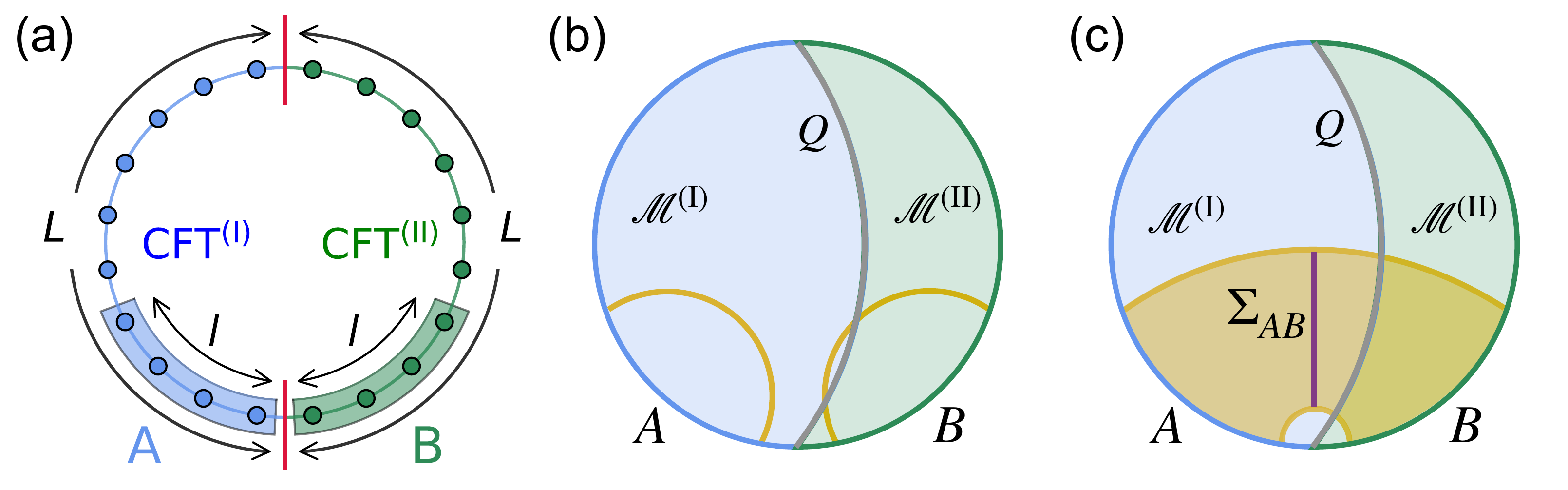}
	\caption{
		\label{fig:schematics}
		%A schematic of calculating entanglement properties in AdS/ICFT.
		A schematic of ICFT and the corresponding holographic model. 
		(a) Two distinct CFTs with length $L$ are glued into an ICFT on a circle with length $2L$. Two subsystems $A$ (blue shade) and $B$ (green shade) are located on two sides of the interface (red). 
		(b-c) The thin-brane model for realizing a holographic ICFT$_2$, contains two AdS$_3$ manifolds $\mathcal{M}^{\rm (I)}$ and $\mathcal{M}^{\rm (II)}$ with different AdS radii $\alpha^{\rm (I)}$ and $\alpha^{\rm (II)}$ joined by a 2D thin brane $Q$ in gray. For convenience, we let $c^{\rm (I)} < c^{\rm (II)}$, and hence $\alpha^{\rm (I)} < \alpha^{\rm (II)}$. Yellow lines in (b) represent the RT surface of $A/B$ for calculating the holographic EE, and purple line in (c) is the entanglement-wedge cross-section $\Sigma_{AB}$ of $A$ and $B$. 
	}
\end{figure}

Here, we consider two distinct CFTs with the same length $L$ glued into a circle with length $2L$ through an interface. 
To access the entanglement structure of its ground state, 
we start by investigating a holographic thin-brane model~\cite{Erdmenger2015_ICFT, vanRaamsdonk2020_Holoween, Bachas2020_Interface, Bachas2021_Hol_ICFT, Karch2021_interfaceEE, yuya2022_reflected, Sonner2022_Island, karch_2022_universal_EE_ICFT} for realizing ICFT$_2$. 
While such a construction is extremely special, it might be the simplest example of ICFTs with nontrivial interfaces whose entanglement properties are analytically tractable. 
Based on the insights from the holographic ICFT, we numerically study two paradigmatic lattice models. 
As shown in Fig.~\ref{fig:schematics}(a), a symmetric entanglement-cut configuration allows us to extract universal information about the interface. 
In particular, we uncover a selection rule of an effective interface central charge $c_{\rm eff} = {\text{min}} \{ c^{\rm (I)}, c^{\rm (II)} \}$ from the reflected entropy (RE), which offers a peek into the underlying physics of interface. 

\textit{Insights from AdS/ICFT.---} The gravity dual of a holographic ICFT$_2$ can be constructed in a bottom-up fashion using the thin brane model~\cite{Azeyanagi_2008, yuya2022_reflected, Sonner2022_Island}. 
As shown in Fig.~\ref{fig:schematics}(b)\&(c), two 3D anti-de Sitter (AdS$_3$) spacetime $\mathcal{M}^{\rm (I)}$ and $\mathcal{M}^{\rm (II)}$ with different AdS radii $\alpha^{\rm (I)}$ and $\alpha^{\rm (II)}$ are joined on a tensile brane $Q$, to mimic an ICFT$_2$ of gluing two distinct CFTs. The AdS radii on the gravity side and the central charges on the ICFT$_2$ side are related by~\cite{BH_1986_ads3cft2}
\begin{align}
	c^{\rm (I,II)} = {3\alpha^{\rm (I,II)}}/{2G_N}, 
\end{align}
where $G_N$ is the Newton constant, and we let $c^{\rm (I)} < c^{\rm (II)}$ in the following. 
Meanwhile, the location of the brane $Q$ is determined by solving a junction condition between $\mathcal{M}^{\rm (I)}$ and $\mathcal{M}^{\rm (II)}$, which reflects nontrivial interaction between CFT$^{\rm (I)}$ and CFT$^{\rm (II)}$. 
For a discussion on the standard AdS/CFT correspondence and the thin brane model for realizing ICFT$_2$, see Supple. Mat.~\cite{SM} and also Ref.~\cite{Maldacena1999AdSCFT, GKP_1998_adscft, Witten_1998_adscft, Sonner2022_Island, Coleman1980}.

The holographic ICFT$_2$ can be considered to be living on the asymptotic boundary of the current AdS$_3$ setup. In AdS/ICFT, the EE for a subsystem $A$ in the ICFT can be computed from the length of the geodesic $\gamma_A$ which connects the endpoints of $A$ as~\cite{RTformula2006,RT2006Aspects}
\begin{align}
	S_A = \frac{{\rm Length}(\gamma_A)}{4 G_N}. 
\end{align}
Here, $\gamma_A$ is called the Ryu-Takayanagi (RT) surface of $A$.
Fig.~\ref{fig:schematics}(b) shows how the RT surfaces look like for a single interval $A$ in $\mathcal{M}^{\rm (I)}$ and a single interval $B$ in $\mathcal{M}^{\rm (II)}$. Note that it is possible for an RT surface to penetrate the brane, which results in a diverse behavior of the EE. However, for an interval $A$ living in CFT$^{\rm (I)}$, $\gamma_A$ always lies inside $\mathcal{M}^{\rm (I)}$ and we find
\begin{align}\label{eq:single_interval_I}
	S_{A \in {\rm CFT^{(I)}}} &= \frac{c^{\rm (I)}}{3} \ln \left( \frac{2L}{\pi\epsilon} \sin\left(\frac{\pi l}{2L}\right) \right) \nonumber\\
	&= \frac{c^{\rm (I)}}{3} \ln {\frac{l}{\epsilon}} + \mathcal{O}\left( \left(\frac{l}{L}\right)^2\right), 
\end{align}
where $l$ is the length of $A$ and $\epsilon$ is a UV cutoff corresponding to the lattice distance. We can also get a clean result when $A$ is an interval with length $2l$ and is symmetric with respect to the interface. Let us call the EE in this case the {\it symmetric EE}, and it turns out to be 
\begin{align}\label{eq:symmetic_EE}
	S_{\rm symm} = \frac{c^{\rm (I)} + c^{\rm (II)}}{6} \ln \left( \frac{2L}{\pi\epsilon} \sin\left(\frac{\pi l}{L}\right) \right) + {\rm const.}.
\end{align}
This relation is not only accessible via a holographic calculation, but also can be derived by using the folding trick and the Cardy-Tonni approach~\cite{2016CardyTonni} in the context of CFT, see details in Supple. Mat.~\cite{SM}.

Another useful correlation measure that reflects entanglement structures to study is the RE. Initially proposed in the context of AdS/CFT~\cite{Faulkner2021_reflected}, the RE has attracted considerable attention~\cite{ryu2021_MI_LN_RE_quench, zou2021_MarkovGap, yuya2020_RE_local_quench, rath2020_EWCS, ryu2020_EWCS_quench, Bueno2020_reflected_fermion, Bueno2020_reflected_scalar, venkatesa2020_island_RE, chenbin_2021_topo_RE_CS, zhou2020_RE_evaporateBH,Wei_22_microstates}. For a (generally mixed) state $\rho_{AB}$ on the subsystem $A\cup B$, we can diagonalize it as $\rho_{AB} = \sum_{i} p_i |\varphi_i\rangle\langle\varphi_i|$. The canonical purification of $\rho_{AB}$ is accordingly defined as 
$|\sqrt{\rho_{AB}}\rangle = \sum_{i} \sqrt{p_i} |\varphi_i\rangle|\varphi_i^*\rangle$,
where for each $|\varphi_i\rangle \in \mathcal{H}_A \otimes \mathcal{H}_B$, $|\varphi_i^*\rangle \in \mathcal{H}_{A^*} \otimes \mathcal{H}_{B^*}$ is its CPT conjugate. The RE between $A$ and $B$ is defined as the EE of the canonical purification, 
\begin{align}
	S^R_{A:B} = S_{A A^*} (| \sqrt{\rho_{AB}}) \rangle.
\end{align}
Notably, as shown in Fig.~\ref{fig:schematics}(c), in holographic theories, RE can also be computed geometrically as~\cite{Faulkner2021_reflected}
\begin{align}
	S^R_{A:B} = \frac{2{\rm Length}(\Sigma_{AB})}{4G_N}, 
\end{align}
where $\Sigma_{AB}$ is the minimal surface crossing the region surrounded by the entanglement wedge~\cite{Czech_2012_EntanglementWedge, Wall_2012_EntanglementWedge, Headrick_2014_EntanglementWedge} $\gamma_{AB} \cup A \cup B$ of subsystem $A \cup B$, so called the entanglement-wedge cross-section~\cite{UT_18_EoP, Swingle_17_EoP, KR_18_negativity, Tamaoka_18_odd, rath2020_EWCS, Faulkner2021_reflected}. 
For two adjacent subsystems $A$ and $B$ with size $l = l_A = l_B$ that touch at the interface, $\Sigma_{AB}$ always locates inside $\mathcal{M}^{\rm (I)}$ and one finds~\cite{SM, yuya2022_reflected}
\begin{align}\label{eq:log_RE_interface_main}
	S^R_{A:B} &= \frac{c^{\rm (I)}}{3} \ln \left(\frac{2L}{\pi\epsilon} \tan\left(\frac{\pi l}{2L}\right) \right) \nonumber\\
	&= \frac{\min\{c^{\rm (I)}, c^{\rm (II)} \}}{3} \ln {\frac{l}{\epsilon}} + \mathcal{O}\left( \left(\frac{l}{L}\right)^2\right). 
\end{align}
which depends only on the smaller central charge. 
Note that, compared to previous results~\cite{Karch2021_interfaceEE, yuya2022_reflected, Sonner2022_Island, karch_2022_universal_EE_ICFT} where the setups were on an infinite line, we present the very first analysis of holographic entanglement entropy in holographic ICFT defined on a compact space constructed by the thin brane model. Although we have just presented analytic formulas for some special choices of the subsystem, results for generic subsystems can be found in Supple. Mat.~\cite{SM}. In the analysis for generic subsystems, taking into account the nontrivial saddle points, where the RT surface crosses the thin brane twice~\cite{Sonner2022_Island}, turns out to be very important. 

Below, we will introduce two paradigmatic lattice models and numerically test if the behaviors observed above also hold in them. Before proceeding, we would like to note that, while it is natural to expect that \eqref{eq:symmetic_EE} holds generically~\cite{Karch2021_interfaceEE}, it would be very surprising to find \eqref{eq:single_interval_I} and \eqref{eq:log_RE_interface_main} hold in generic cases. To see this, we may consider a ``trivial" ICFT with no interaction between CFT$^{\rm (I)}$ and CFT$^{\rm (II)}$. In this case, for an interval $A$ lying in CFT$^{\rm (I)}$ and ending at the interface, the leading order of $S_A$ would be $(c^{\rm (I)}/6) \ln l$, which is roughly a half of \eqref{eq:single_interval_I}. 
As for the RE, since $\rho_{AB} = \rho_{A} \otimes \rho_{B}$ in this case, $S^{R}_{A:B}$ would be zero which differs a lot from \eqref{eq:log_RE_interface_main}. On the other hand, \eqref{eq:symmetic_EE} still holds. Therefore, up to this point, it is natural to expect that \eqref{eq:single_interval_I} and \eqref{eq:log_RE_interface_main} reflect the uniqueness of the interface interaction exhibiting in AdS/ICFT.
However, surprisingly, we will see that all of \eqref{eq:single_interval_I}, \eqref{eq:symmetic_EE} and \eqref{eq:log_RE_interface_main} hold in the lattice models studied below, which suggests that they may generically hold in nontrivial ICFTs.

\begin{figure*}\centering
	\includegraphics[width=\textwidth]{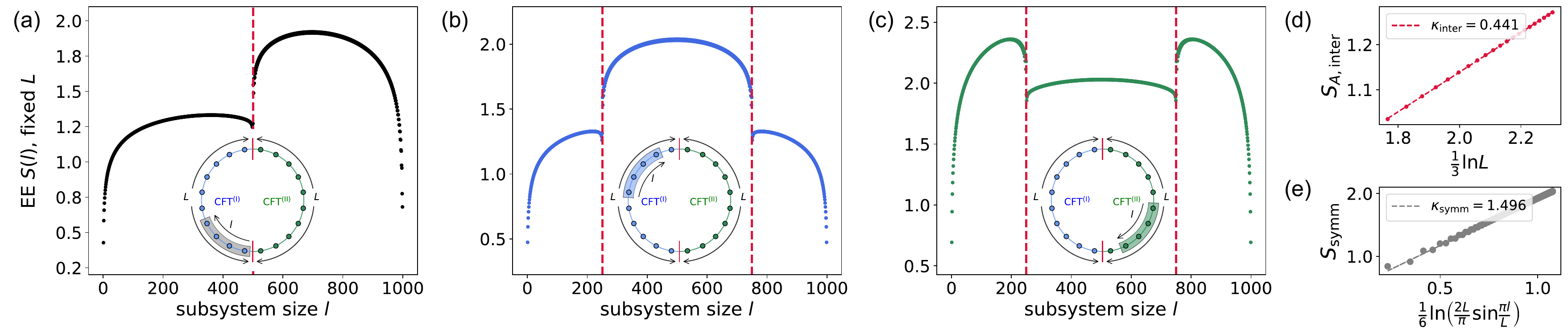}
	\caption{
		\label{fig:EE_sub_size_main}
		The bipartite EE in a fermionic model of gluing the real fermion CFT ($c^{\rm (I)} = \frac{1}{2}$) at left and the complex fermion CFT ($c^{\rm (II)} = 1$) at right, under a periodic boundary condition. 
		(a-c) The dependence of EE on the subsystem size $l$, with fixed total system size $2L = 1000$. 
		The insets show corresponding entanglement-cut configurations: fix one end (a) at the interface, (b) in the middle of CFT$^{\rm (I)}$, and (c) in the middle of CFT$^{\rm (II)}$. 
		The red dash lines represent the phase boundaries that one end of the subsystem touches the interface, which hosts a jump on the bulk degrees of freedom. 
		(d) The dependence of EE on the total system size $2L$, where both ends of the subsystem $A$ lie on the interface. 
		% The subsystem size is always $L$ and the corresponding EE displays a logarithmic scaling as $S_{A, {\rm inter}} \propto \ln L$. 
		% 
		(e) The dependence of symmetric EE on the subsystem size $l$, with fixed total system size $2L = 1000$. 
		% By considering the finite-size scaling form of \eqref{eq:symmetic_EE}, we find a universal perfactor $\kappa_{\rm symm} = c^{\rm (I)} + c^{\rm (II)} \approx 1.496$. 
		% 
		These numerical results are consistent with holographic calculations (see Supple. Mat.~\cite{SM}).  }
\end{figure*}

\textit{Lattice models and numerical method.---} 
In what follows, we consider two lattice models for realizing ICFT$_2$. 
The first one is the O’Brien-Fendley (OF) model~\cite{fendley_2018_triIsing} with an inhomogeneous coupling constant
\begin{align}\label{eq:interface_Ising}
	H_1 = H_{\rm TFI} 
	+ g_L \sum_{n \le -1} H_{\rm int} (n)
	+ g_R \sum_{n \ge 0} H_{\rm int} (n) , 
\end{align}
where $H_{\rm{TFI}} = \sum_n \sigma^{x}_{n} \sigma^{x}_{n+1} -  \sigma^z_n$, $H_{\rm{int}} = \sigma^{x}_{n-1} \sigma^{x}_{n} \sigma^{z}_{n+1} + \sigma^z_{n-1} \sigma^{x}_{n} \sigma^{x}_{n+1}$, and the site index $n$ run over $[-L, L-1]$. 
The anisotropy between $g_L$ and $g_R$ creates an interface at the bond connecting spins at site $-1$ and site $0$. 
Since we are considering a periodic chain, there is another symmetric interface bond between site $-L$ and site $L-1$. 
In the homogeneous case of $g = g_L = g_R$, the OF model realizes a tricritical Ising fixed point at $g = g_c$ that separates a phase with Ising universality class for $g < g_c$ and a gapped phase for $g > g_c$.\footnote{Theoretically, one can confirm $g_c < 0.5$, but the exact value of $g_c$ can be only numerically obtained and would be modified by finite size or the interface setting. In the homogeneous case, we find the previously reported critical value $g_c \approx 0.428$ in Ref.~\cite{fendley_2018_triIsing} is faithful, but it is modified in the interface case as $g_c  \sim 0.41$ for our considered total system size.}
In the context of CFT, tuning the coupling constant $g$ away from $g_c$ can be understood as adding an $\Phi_{1,3}$ operator that triggers an RG flow from tricritical Ising CFT to Ising CFT or massive IR, depending on the sign of $\Phi_{1,3}$~\cite{huse1984_rg_A_series}.
As setting $g_L = 0$ and $g_R = g_c$, the lattice Hamiltonian in \eqref{eq:interface_Ising} offers an appropriate playground for an interface of gluing the Ising CFT at the left part (with $c^{\rm (I)}_1 = \frac{1}{2}$) and the tricritical Ising CFT at the right part ($c^{\rm (II)}_1 = \frac{7}{10}$).

The second one is a non-interacting fermionic model with inhomogeneous pairing
\begin{align}\label{eq:interface_Fermion}
	H_2 = \sum_{n \le -1} H_{\rm RF}(n) 
	+ \sum_{n \ge 0} H_{\rm CF}(n) , 
\end{align}
where $H_{\rm RF}(n) = H_{\rm CF}(n) + (f_n f_{n+1} + h.c.) - 2 f_n^\dagger f_n$ and $H_{\rm CF}(n) = - f_n^\dagger f_{n+1} + h.c.$. 
Here, the left half chain with pairing terms realizes a real (Majorana) fermion CFT with $c^{\rm (I)}_2 = \frac{1}{2}$, but the right half chain realizes a complex (Dirac) fermion CFT with $c^{\rm (II)}_2 = 1$. 
Again, an interface of gluing two distinct CFTs is created between site $-1$ and site $0$. 
Upto a Jordan-Winger transformation, this model is dual to a spin model of gluing an Ising chain and an XX chain.

For accessing entanglement properties of these lattice models, we perform a numerical simulation based on matrix product states (MPS) techniques~\cite{Schollwock2011_age_mps}.\footnote{Here we note that, the second fermionic model is non-interacting and Gaussian, which allows an exact solution of the EE and MI from the correlation matrix techniques~\cite{Peschel_2003calculation, Peschel_2009reduced}. } % However, a similar approach cannot be directly applied for calculating the RE for pairing free fermions.
First, the ground state of the model is solved by the density matrix renormalization group algorithm~\cite{white1992dmrg} with a bond dimension $\chi$. 
At this step, one can easily obtain the bipartite EE. 
Second, for calculating mutual information (MI) and RE, 
we need to evaluate reduced density matrices for a continuous region (the subsystems $A$, $B$ and their complement $A \cup B$), 
for which the computational complexity grows exponentially. 
An efficient simulation requires further compressing the dimension of local Hilbert space of the cutting subsystem (the dimension of reduced density matrix $\rho_A$) to  $\widetilde{d}_A$ by applying a standard MPS coarse-graining procedure to the physical leg of the subsystem's local wavefunction (see Supple. Mat.~\cite{SM}).  
Through this approach, we are able to calculate the multipartite entanglement measures -- MI and RE with high accuracy and affordable computational complexity: $\chi = 100$ and $\widetilde{d}_A = 100$ for the Hamiltonian in \eqref{eq:interface_Ising} and~\eqref{eq:interface_Fermion} with total system size $2L$ up to $300$ under a periodic boundary condition.

\begin{figure*}\centering
	\includegraphics[width=0.95\textwidth]{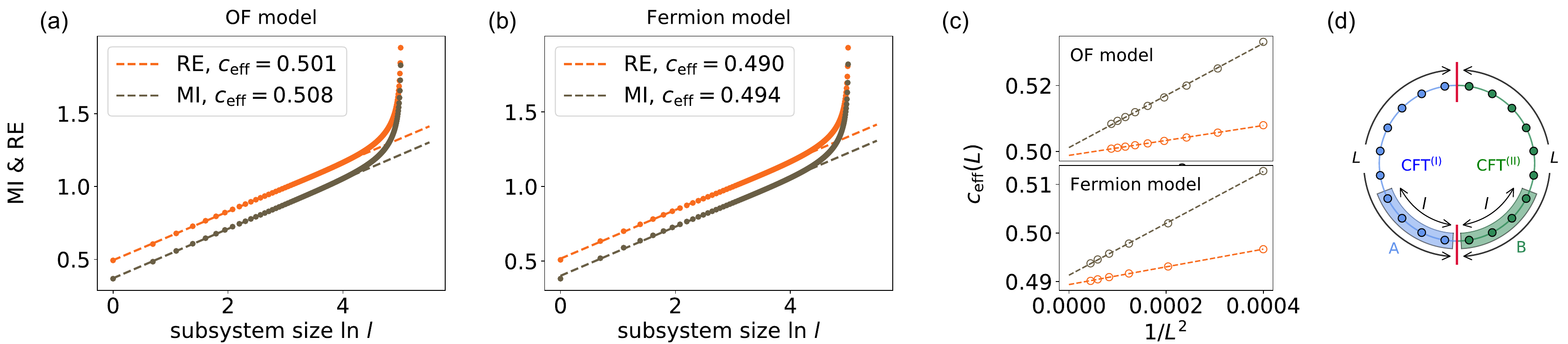}
	\caption{
		\label{fig:MI_RE_pbc}
		The scaling behavior of MI $I_{A:B}$ and RE $S^R_{A:B}$ for two adjacent subsystems $A$ and $B$, of which the touching point is located at the interface, for (a) the inhomogeneous OF model of gluing the Ising CFT ($c^{\rm (I)} = \frac{1}{2}$) and the tricritical Ising CFT ($c^{\rm (II)} = \frac{7}{10}$), and 
		(b) a fermionic model of of gluing the real fermion CFT ($c^{\rm (I)} = \frac{1}{2}$) and the complex fermion CFT ($c^{\rm (II)} = 1$), with total system size $2L = 300$. 
		The dash lines represent linear fits in the form of $S^{R}_{A:B} \sim I_{A:B} = \frac{c_{\rm eff}}{3} \ln l + b$ under $l \ll L$. 
		(c) A finite-size scaling of the extracted $c_{\rm eff}(L)$ from RE and MI, under various total system size $2L \in [100, 300]$. 
		The dash lines represent linear fits in the form of $c_{\rm eff}(L) = k/L^2 + c_{\rm eff}(L \to \infty)$, giving $c_{\rm eff, 1}(L \to \infty, {\rm MI}) \approx 0.501$, $c_{\rm eff, 1}(L \to \infty, {\rm RE}) \approx 0.499$ for the OF model and $c_{\rm eff, 2}(L \to \infty, {\rm MI}) \approx 0.491$, $c_{\rm eff, 2}(L \to \infty, {\rm RE}) \approx 0.489$ for the fermionic model. 
		(d) A schematic of the entanglement-cut configuration, where two adjacent subsystems $A$ and $B$ with the same length $l$ touch at the interface. 
	}
\end{figure*}

\textit{Entanglement entropy.---} Let us begin with inspecting the dependence of EE on the subsystem size. 
In Fig.~\ref{fig:EE_sub_size_main}, we present the result on the fermionic Hamiltonian of \eqref{eq:interface_Fermion}, as its non-interacting nature allows an exact solution of the EE (see inhomogeneous OF model in Supple. Mat.~\cite{SM}). 
Remarkably, we find a good agreement between these lattice results and a holographic calculation on the thin-brane model (see Supple. Mat.~\cite{SM}) for various entanglement-cut configurations. 
The subsystem-size dependence of EE shows a clear change in bulk degrees of freedom across the interface, corresponding to the two distinct bulk central charges on each side of the interface. 
Moreover, when both ends of the subsystem lie on the interface (subsystem $A = \{-L,-L+1,\cdots,-1\}$), we find that the corresponding EE exhibits a logarithmic scaling $S_{A, {\rm inter}} \propto \ln L$, as shown in Fig.~\ref{fig:EE_sub_size_main}(d). 
This provides strong evidence that a massive RG flow is not triggered in our lattice model, while the prefactor of logarithmic EE of cutting along the interface is generally not of universal meaning in the case of ${\rm CFT^{(I)} \neq CFT^{(II)}}$~\cite{uhlemann2023splitting}.

We now try to extract possible universal information about the interface from the finite-size scaling forms obtained from holographic derivation. 
In the case of cutting a subsystem $A$ in CFT$^{\rm (I)}$, holographic calculation on the thin-brane model gives the result shown in \eqref{eq:single_interval_I}, as a pure CFT$^{\rm (I)}$. 
In lattice simulations, we find that this scaling form is valid at $l \ll L$, even when $A = [-l, -a], a \to 0$ is very close to the interface. % On the other hand, the numerical results differ from \eqref{eq:single_interval_I} when $l$ becomes large and both two ends are close to the interface. 
Another solvable case is the symmetric EE of cutting a subsystem $A = [-l, l]$ that is symmetrically around the interface at $x = 0$. 
Holographic calculation suggests the scaling form in a finite system is given by \eqref{eq:symmetic_EE}. 
This scaling form also appears in numerical simulation on lattice models with high accuracy (see Fig.~\ref{fig:EE_sub_size_main}(e)). 
For characterizing the interface, one may consider extracting the interface entropy (See Supple. Mat.~\cite{SM} for a definition) from lattice simulations. 
However, different from the case of gluing two identical CFTs~\cite{Peschel_2012_EE_defect, Calabrese_2012_EE_junction, brehm_2015_EE_interface_ising, Roy2022_EE_topo_defect, pollmann_2022_EE_Ising_defect}, here we do not have a simple way to separate the interface entropy from the non-universal correction in the sub-leading term of EE. 
Moreover, it is worth noting that the discussion in this part focuses on the logarithmic dependence of EE on the subsystem size $l$. 
This is in general different from considering the logarithmic dependence on the UV cutoff $\epsilon$, for which a universal relation of the prefactor is expected~\cite{Karch2021_interfaceEE, karch_2022_universal_EE_ICFT, karch_to_appear}.

\textit{Reflected entropy and mutual information.---} 
Let us then move on to study the RE and MI.
In pure CFTs, a symmetric entanglement-cut configuration of separating two adjacent subsystems $A$ and $B$ with the same length $l$ leads to $S^R_{A:B} \sim I_{A:B} \sim \frac{c}{3} \ln l$. 
By putting the touching point of $A$ and $B$ onto the interface (see a schematic in Fig.~\ref{fig:MI_RE_pbc}), holographic calculation suggests that the RE remains the same logarithmic scaling in ICFTs, as shown in \eqref{eq:log_RE_interface_main}. 
The only difference appears in the prefactor with replacing the central charge $c$ to an effective value $c_{\rm eff} = \text{min} \{ c^{\rm (I)}, c^{\rm (II)} \}$. 
While this behavior was observed in a simple thin-brane model, we will see that, surprisingly, it also precisely holds in both of the two lattice models considered here. 

As shown in Fig.~\ref{fig:MI_RE_pbc}(a)\&(b), for a given finite total system size $2L$, the RE $S^R_{A:B}$ exhibits a logarithmic dependence on $l$. 
A further finite-size scaling (see Fig.~\ref{fig:MI_RE_pbc}(c)) on the prefactor of logarithmic RE suggests $c_{\rm{eff}, 1} \approx 0.499$ , approaching $\text{min} \{ \frac{1}{2}, \frac{7}{10} \}$, and $c_{\rm{eff}, 2} \approx 0.489$, approaching $\text{min} \{ \frac{1}{2}, 1 \}$, in the thermodynamic limit $L \to \infty$.  
Moreover, holographic calculation implies that the MI $I_{A:B} = S_A + S_B - S_{AB}$ (and consequently the Markov gap $S^{R}_{A:B} - I_{A:B}$~\cite{Hayden2021_MarkovGap}) in ICFTs has a convoluted dependence on the subsystem size $l$, since $S_B$ involves a non-trivial phase of EE scaling (see details in Supple. Mat.~\cite{SM}). 
Numerically, we also find that the RE and MI exhibit distinct scaling behaviors on the subsystem size $l$ when $l$ becomes comparable with the total system size $2L$. 
Nevertheless, we behold a logarithmic MI $I_{A:B} \sim \frac{c'_{\rm eff}}{3} \ln l$ under $l \ll L$, sharing the same selection rule of $c_{\rm eff} = c'_{\rm eff} = \text{min} \{ c^{\rm (I)}, c^{\rm (II)} \}$ (a finite-size scaling gives $c'_{\rm{eff}, 1}(L \to \infty) \approx 0.501$ on the OF model and $c'_{\rm{eff}, 2} \approx 0.491$ on the fermion model). 
To summarize, we conclude that there is a universal scaling of tripartite entanglement measure in critical interface theories as $S^{R}_{A:B} \sim I_{A:B} \sim \frac{c_{\rm eff}}{3} \ln l$, with a single effective central charge satisfying the universal selection rule of $c_{\rm eff} = \text{min} \{ c^{\rm (I)}, c^{\rm (II)} \}$.

\textit{Discussions \& Outlooks.---} We have explored possible universal entanglement signatures in ICFTs through numerical simulations on the representative lattice models. 
Some initiations were provided from a holographic perspective, by considering a simple brane construction in AdS$_3$ to mimic the ICFT$_2$. 
Surprisingly, numerical results obtained from lattice models resembles a lot of observations in AdS/ICFT. 
One of the most important features is that the effective central charge appearing in the reflected entropy is given by $c_{\rm eff} = \text{min} \{ c^{\rm (I)}, c^{\rm (II)} \}$. 

These common features between lattice models and AdS/ICFT are surprising because they do not hold in general ICFTs, and one can easily construct a counterexample, e.g. by considering an ICFT without interaction between the two sides. 
In general, the value of $\frac{1}{3} \text{min} \{ c^{\rm (I)}, c^{\rm (II)} \}$ is expected to be the upper bound for the prefactor of reflected entropy, and the condition of saturating it is not clear. 
Intuitively, saturating the upper bound requires (almost) perfect transmission associate with the interface. 
On lattice models, this means that we should let the (dominate) bond coupling at the interface take the same value as in the bulk of the connected two half-spaces. Otherwise, the transmission rate of the interface would be strongly reduced.
Both of the two considered lattice models are constructed based on this consideration. 
Moreover, it is tested that local perturbations on the interface do not lead to a qualitative change of the universal logarithmic scaling of mutual information and reflected entropy, which indicates an RG stability of our critical theories. 
It motivates us to conjecture that the observed features are universal for a class of critical interface theories with an RG stability, which needs further study to demonstrate.

Moreover, we would like to point out that, the inhomogeneous OF model realizes a specific case of \textit{RG interfaces} between nearby minimal models (tricritical Ising CFT at UV and Ising CFT at IR)~\cite{Gaiotto2012_RG_DW, Konechny2021_RGinterface}. 
Universal information about the RG flow is expected to be traceable through two-point correlations~\cite{Brunner_2008_RG_DW, Gaiotto2012_RG_DW, Konechny_2014_RG_defect, Konechny2021_RGinterface}, which was investigated by a recent work~\cite{cogburn2023cft} with introducing a different lattice model.
Our results are potentially helpful for extracting this information from the entanglement structure. 
Another free fermionic interface model provides a particular approach to studying symmetry breaking in ICFTs, where the $U(1)$ symmetry of Dirac fermion is broken to $Z_2$ of Majorana fermion on half space. 
Besides, it would also be interesting to explore other interface models with more complicated structures, e.g. an interface separating a unitary CFT from a non-unitary CFT (e.g. Ising to Lee-Yang fixed point~\cite{Quella_2007_transmission, Konechny_2017_Ising}). 
We leave these to future investigations.

%%%%%%

%\begin{acknowledgments}
	\textit{Acknowledgments.---}
	We would like to thank Yuya Kusuki, Masahiro Nozaki, and Shan-Ming Ruan for discussion. 
	We would also like to thank Hao Geng, Andreas Karch, Zhu-Xi Luo, Christoph Uhlemann, and Mianqi Wang for valuable comments on a draft of this paper. 
	WZ was supported by the Key R\&D Program of Zhejiang Province under 2022SDXHDX0005, 2021C01002, the National Key R\&D Program under 2022YFA1402200, and NSFC under No. 92165102. ZW was supported by Grant-in-Aid for JSPS Fellows No. 20J23116. XW is supported by the Simons Collaboration on Ultra-Quantum Matter (UQM), which is funded by grants from the Simons Foundation (651440, 618615). 
%\end{acknowledgments}

%\bibliographystyle{apsrev4-1}
\bibliography{interface}

\clearpage
\include{interface_sup}

\end{document}

%% file: interface_sup.tex
%\documentclass[reprint, superscriptaddress, nofootinbib, amsmath,amssymb, aps, longbibliography, prx]{revtex4-1}

%\DeclareMathOperator{\Tr}{Tr}
%\usepackage{graphicx}% Include figure files
%\usepackage{multirow}

%\makeatletter

%\makeatother
%\usepackage{dcolumn}% Align table columns on decimal point
%\usepackage{bm}% bold math
%\usepackage[colorlinks, linkcolor=blue, citecolor=blue, urlcolor=blue]{hyperref}
%\usepackage{slashed}

%\begin{document}

\newcommand{\beginsupplement}{%
	\setcounter{table}{0}
	\renewcommand{\thetable}{S\arabic{table}}%
	\setcounter{figure}{0}
	\renewcommand{\thefigure}{S\arabic{figure}}%
	\setcounter{section}{0}
	\renewcommand{\thesection}{\Roman{section}}%
	\setcounter{equation}{0}
	\renewcommand{\theequation}{S\arabic{equation}}%
}
\clearpage

\onecolumngrid

\beginsupplement

\begin{center}
	\textbf{\large Supplementary Materials for `` Universal entanglement signatures of interface conformal field theories  ''}
\end{center}
\vspace{2mm}

\tableofcontents

\section{Holographic ICFT}

In this appendix, we present analytical calculations of entanglement entropy, mutual information and reflected entropy in a holographic ICFT$_2$. Here,  ``holographic ICFT'' means an ICFT with a semiclassical gravity dual. 

In the following, we will firstly review the standard AdS/CFT correspondence and how to compute the entanglement entropy and reflected entropy in the CFT from the AdS dual. Then we will present our AdS/ICFT model, and perform the computation of these quantities in it. 

\subsection{AdS/CFT, holographic entanglement entropy, and holographic reflected entropy}

\subsubsection{AdS/CFT correspondence}
The standard AdS/CFT \cite{Maldacena1999AdSCFT} correspondence states that a quantum gravity defined in an asymptotically AdS spacetime $\mathcal{M}$ is equivalent to a CFT defined on its asymptotic boundary $\Sigma$. The correspondence is often phrased as the equivalence between the partition functions on the two sides, the so-called GKP-Witten relation \cite{GKP_1998_adscft,Witten_1998_adscft}:
\begin{align}
    \int Dg_{\mu\nu} e^{-I_{\rm grav.}[{\mathcal{M}}]} = Z_{\rm CFT}[\Sigma].
\end{align}
While the left-hand side is a gravitational path integral and hence one should sum over all the possible geometries in principle, by taking the semiclassical limit $G_N \rightarrow 0$, one can approximate the left-hand side with the most dominated saddle point and the analysis gets greatly simplified. Such a CFT is often called a holographic CFT. In the following, we would like to focus on the AdS$_3$/CFT$_2$ case. The most simple example of AdS$_3$/CFT$_2$ is the case where the gravity side is the standard Einstein gravity,
\begin{align}
    I_{\rm grav.} = -\frac{1}{16\pi G_N} \int_{\mathcal{M}} \sqrt{g} \left(R + \frac{2}{\alpha^2}\right) -\frac{1}{8\pi G_N} \int_{\Sigma} \sqrt{\gamma} B .
\end{align}
Here, the first and the second terms are the Einstein-Hilbert term and the Gibbons-Hawking term, respectively. Besides, $g_{\mu\nu}$ is the metric in the bulk AdS $\mathcal{M}$, $R$ is its Ricci scalar, $\gamma_{ij}$ is the induced metric on $\Sigma$, and $B$ is its extrinsic curvature. 

The Newton constant $G_N$ on the AdS$_3$ side and the central charge $c$ on the CFT$_2$ side are related by the Brown-Henneaux relation \cite{BH_1986_ads3cft2}, 
\begin{align}\label{eq:BHrelation}
    c = \frac{3}{2} \frac{\alpha}{G_N}.
\end{align}
From this, we can see that the semiclassical limit $G_N \rightarrow 0$ on the AdS$_3$ side corresponds to the large $c$ limit on the CFT$_2$ side. We will only consider such cases in this appendix. 

\begin{figure*}[h]
        \centering
	\includegraphics[width=12cm]{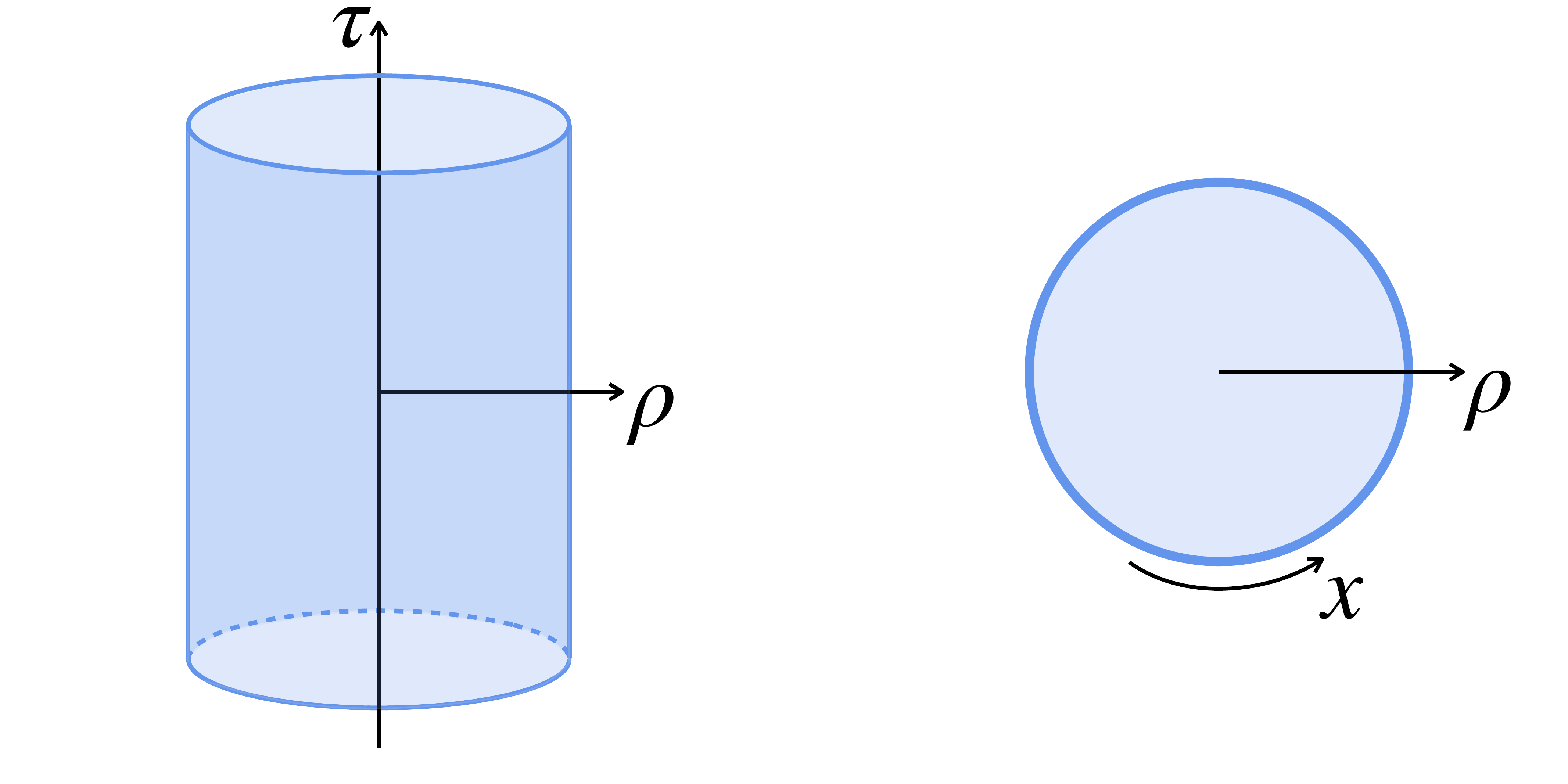}
	\caption{
		\label{fig:globalAdS}	
  A sketch of the global AdS$_3$ geometry (left) and its time slice $\tau = {\rm const.}$. The asymptotic boundary $\rho\rightarrow\infty$ is $\Sigma = S\times \mathbb{R}$, which is the manifold where the corresponding CFT$_2$ is defined. Since the current setup possesses a time translation symmetry, we can just consider its time slice. 
  }
\end{figure*}

As the most simple example, let us consider a CFT$_2$ defined on $\Sigma = S \times \mathbb{R}$ where $S$ is the spatial direction parameterized by $x\in [-L,L)$ with $x\sim x+2L$ and $\mathbb{R}$ is the Euclidean time direction parameterized by $\tau \in (-\infty,\infty)$. This path integral realizes the ground state of the CFT on the time slice $\tau = 0$. The gravity dual is given by the global AdS$_3$ with AdS radius $\alpha$. The metric of $\mathcal{M}$ can be written as 
\begin{align}\label{eq:gloabl_AdS}
    ds^2 = \alpha^2 \left( d\rho^2 + \cosh^2 \rho \left(\frac{\pi}{L}\right)^2 d\tau^2 + \sinh^2\rho \left(\frac{\pi}{L}\right)^2 dx^2  \right)
\end{align}
See Fig.~\ref{fig:globalAdS} for a sketch. Just as shown in the figure, since the current setup has a time translation symmetry, it is sufficient to just consider its time slice to recover the information of the corresponding CFT ground state. Therefore, we may just focus on the time slice in the following analysis. 

\subsubsection{Entanglement entropy and the Ryu-Takayanagi formula}

\begin{figure*}[h]
        \centering
	\includegraphics[width=8cm]           {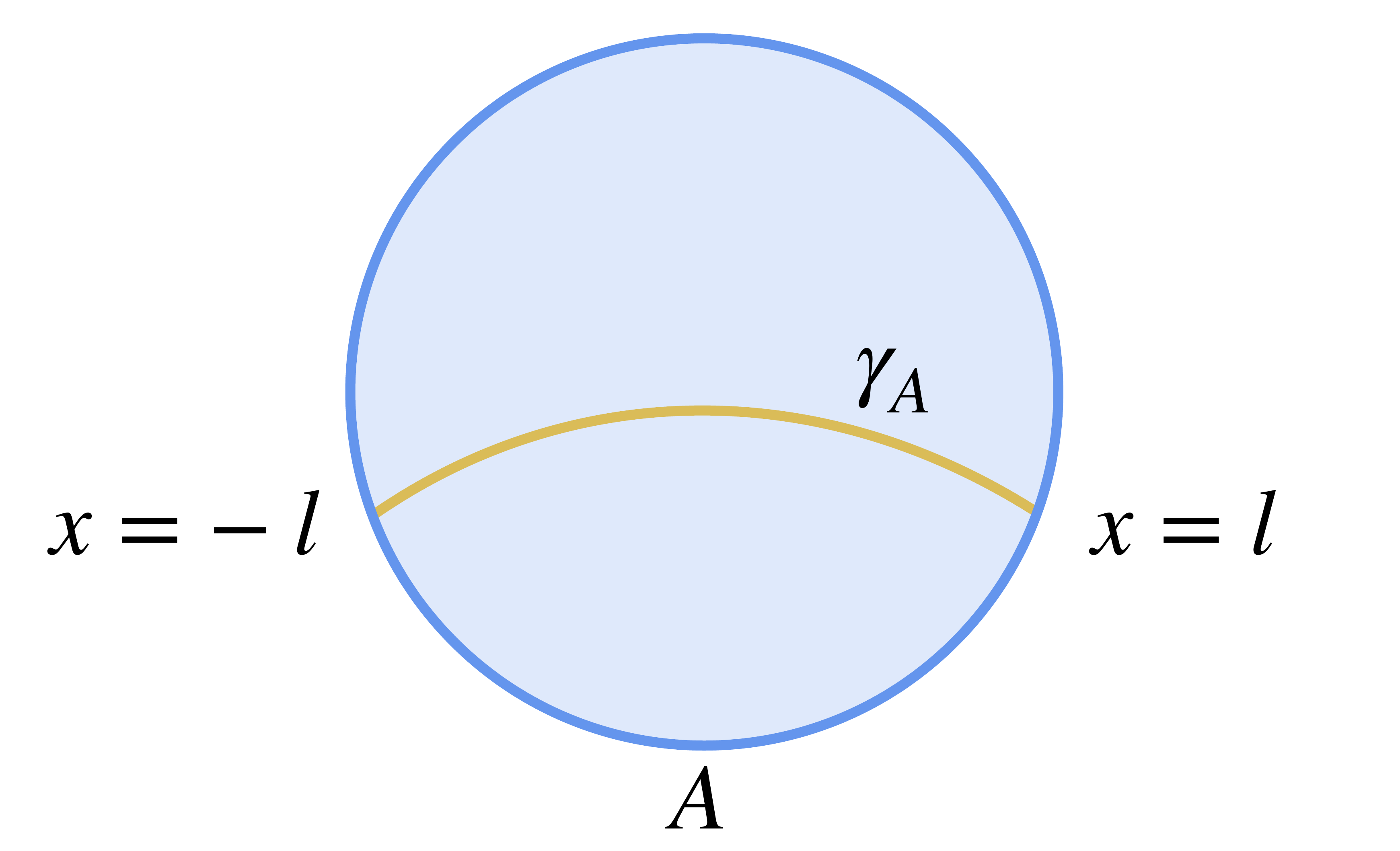}
        \includegraphics[width=8cm]{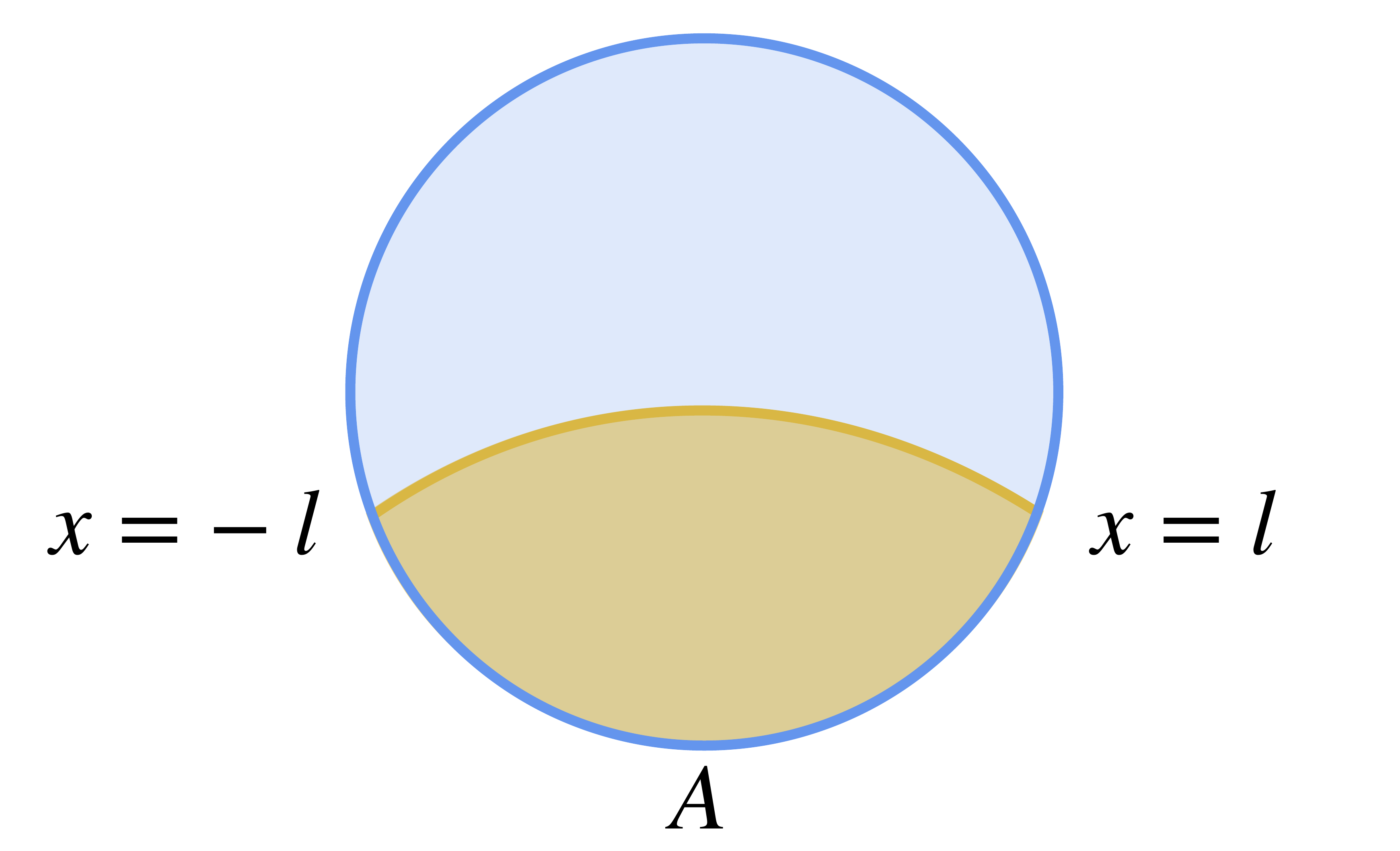}
	\caption{
		\label{fig:RTsurface}	
  A time slice of the global AdS$_3$ which corresponds to the ground state of the corresponding CFT$_2$. The whole CFT system is divided into $A = [-l,l]$ and its complement. The entanglement entropy between $A$ and its complement can be computed from the area of the corresponding Ryu-Takayanagi surface $\gamma_A$ shown by the orange line in the left figure. $\gamma_A$ is the minimum surface which shares end points with $A$ and is homologous to it. The region surrounded by $\gamma_A \cup A$ is shaded orange in the right figure and called the entanglement wedge of $A$
  }
\end{figure*}

Let us then explain how to use the gravity dual to compute the entanglement entropy \cite{RTformula2006,RT2006Aspects} for the corresponding CFT state. Dividing the whole spatial region of the CFT into $A$ and its complement, if the current setup is static or time-translational symmetric, then the entanglement entropy between $A$ and its complement can be computed as 
\begin{align}
    S_A = \frac{{\rm Area}(\gamma_A)}{4G_N} 
\end{align}
where $\gamma_A$ is a codimension-2 surface in the AdS spacetime which satisfies the following conditions: 
\begin{itemize}
    \item $\gamma_A$ shares the boundary with the subsystem $A$, i.e. $\partial \gamma_A = \partial A$. 
    \item $\gamma_A$ is homologous to $A$. 
    \item The area of $\gamma_A$ takes the minimal value. 
\end{itemize}
This formula is known as the Ryu-Takayanagi formula \cite{RTformula2006,RT2006Aspects}, and $\gamma_A$ is often called the Ryu-Takayanagi surface. See the left part of Fig.~\ref{fig:RTsurface} for a sketch. 
In this paper, we will only focus on static cases. One can find generalizations to Lorentzian time-dependent cases\cite{HRT_07_HEE}, generalizations to Euclidean time-dependent case\cite{20_PseudoEntropy}, and subleading quantum corrections of the RT formula\cite{Faulkner_13_quantumHEE,EW_14_QES} in the literature listed above. 

As one of the most simple examples, let us consider the ground state of a CFT$_2$ defined on a circle with perimeter $2L$. The corresponding AdS geometry, as we have already seen above, is given by the global AdS \eqref{eq:gloabl_AdS}. Let $A$ be a single interval given by $-l\leq x \leq l$. Then the RT surface is a geodesic which connects the following two points: 
\begin{align}\label{eq:partialA}
    (\tau,\rho,x) = (0, \rho_\infty, -l), (0, \rho_\infty, l)
\end{align}
Here, $\rho_\infty$ is an IR cutoff on the AdS side, which corresponds to the UV cutoff on the CFT side. In order for the UV cutoff on the CFT side to be $\epsilon$, it should be related to $\rho_\infty$ as 
\begin{align}\label{eq:UVandIRcutoff}
    \sinh \rho_\infty = \frac{L}{\pi\epsilon}.
\end{align}
We will use this relation later to compute the holographic entanglement entropy. The corresponding setup is shown in the left part of Fig.~\ref{fig:RTsurface}. 

Since we are considering global AdS, it is more convenient to embed the geometry into a higher dimensional spacetime. It is well-known that a Euclidean global AdS$_3$ (or equivalently a 3D hyperbolic spacetime $H^3$) with AdS radius $\alpha$ can be given by 
\begin{align}
    -X_0^2 + X_1^2 + X_2^2 + X_3^2 = -\alpha^2,
\end{align}
where $(X_0,X_1,X_2,X_3)$ parameterize a $\mathbb{R}^{1,3}$. The distance between two points $(X_0,X_1,X_2,X_3)$ and $(Y_0,Y_1,Y_2,Y_3)$ on the AdS$_3$ is given by 
\begin{align}
    d(X,Y) = \alpha \cosh^{-1} \left(\frac{X_0Y_0 - X_1Y_1 - X_2Y_2 - X_3Y_3}{\alpha^2}\right). 
\end{align}
The coordinates used in \eqref{eq:gloabl_AdS} are related to $(X_0,X_1,X_2,X_3)$ as 
\begin{align}
    &X_0 = \alpha \cosh \rho \cosh \left(\frac{\pi \tau}{L}\right), \\
    &X_1 = \alpha \sinh \rho \cos \left(\frac{\pi x}{L}\right), \\
    &X_2 = \alpha \sinh \rho \sin \left(\frac{\pi x}{L}\right), \\
    &X_3 = \alpha \cosh \rho \sinh \left(\frac{\pi \tau}{L}\right).
\end{align}
By applying this relation, the two edges of the RT surface are given by 
\begin{align}
    (X_0,X_1,X_2,X_3) = &\left(\alpha \cosh\rho_\infty, \alpha \sinh\rho_\infty \cos\left(\frac{\pi l}{L}\right), -\alpha \sinh\rho_\infty \sin\left(\frac{\pi l}{L}\right),0\right), \\
    &\left(\alpha \cosh\rho_\infty, \alpha \sinh\rho_\infty \cos\left(\frac{\pi l}{L}\right), \alpha \sinh\rho_\infty \sin\left(\frac{\pi l}{L}\right),0\right), 
\end{align}
respectively. Accordingly, the area of the RT surface is
\begin{align}
    {\rm Area}(\gamma_A) = \alpha \cosh^{-1} \left(
    \cosh^2 \rho_\infty - \sinh^2 \rho_\infty \cos\left(\frac{2\pi l}{L}\right)
    \right) \simeq 2\alpha \ln \left( \frac{2L}{\pi\epsilon} \sin\left(\frac{\pi l}{L}\right) \right),
\end{align}
where we have used the relation \eqref{eq:UVandIRcutoff} and the fact that $\rho_\infty$ is very large. It is then straightforward to get the holographic entanglement entropy as follows 
\begin{align}
    S_A = \frac{\alpha}{2G_N} \ln \left( \frac{2L}{\pi\epsilon} \sin\left(\frac{\pi l}{L}\right) \right) = \frac{c}{3} \ln \left( \frac{2L}{\pi\epsilon} \sin\left(\frac{\pi l}{L}\right) \right), 
\end{align}
where we have used the Brown-Henneaux relation \eqref{eq:BHrelation} in the last step.

\subsubsection{Reflected entropy and entanglement wedge cross section}

We would like to review how to holographically compute the reflected entropy \cite{Faulkner2021_reflected} in the following. To begin with, let us introduce some geometric objects on the AdS side, which will lead to the notion of the entanglement wedge cross section. 

Let us consider a time-reflection symmetric time slice of an AdS/CFT setup. Let $A$ be a subregion on the CFT side, then there exists a corresponding RT surface $\gamma_A$ on the AdS side. The region surrounded by $\gamma_A \cup A$ on the time slice is called the entanglement wedge \cite{Czech_2012_EntanglementWedge,Wall_2012_EntanglementWedge,Headrick_2014_EntanglementWedge} of $A$. 

Here, the region we call the entanglement wedge is a codimension-1 region from the AdS point of view. See the right part of Fig.~\ref{fig:RTsurface} for a sketch. 
In general, the terminology ``entanglement wedge" is also used to call a codimension-0 region which includes the current codimension-1 entanglement wedge as a time-reflection symmetric surface. 
The entanglement wedge of $A$ is considered to be the gravity dual of the CFT density matrix $\rho_A$ on $A$ \cite{Czech_2012_EntanglementWedge,Wall_2012_EntanglementWedge,Headrick_2014_EntanglementWedge}. 

\begin{figure*}[h]
        \centering
	\includegraphics[width=8cm]           {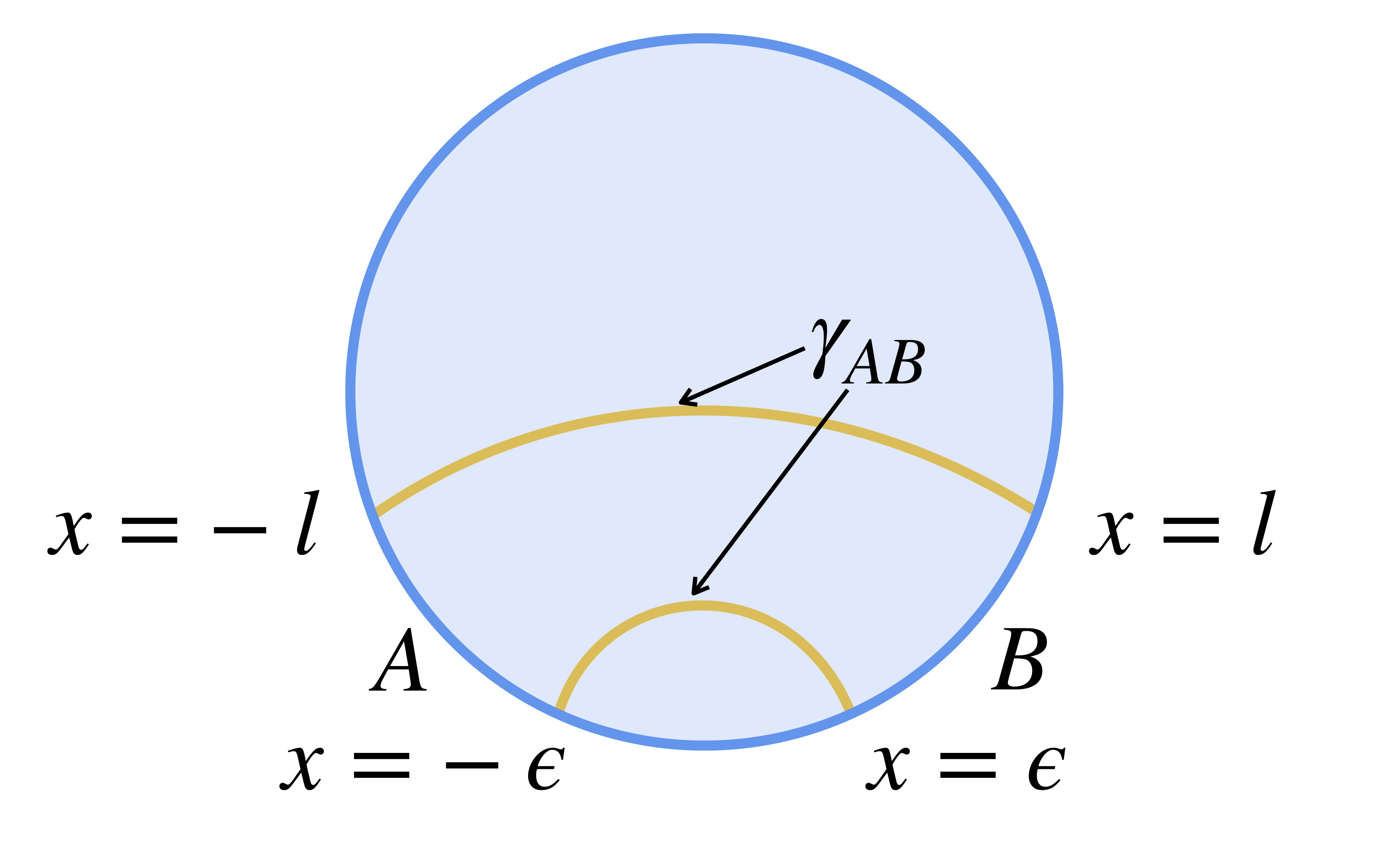}
        \includegraphics[width=8cm]{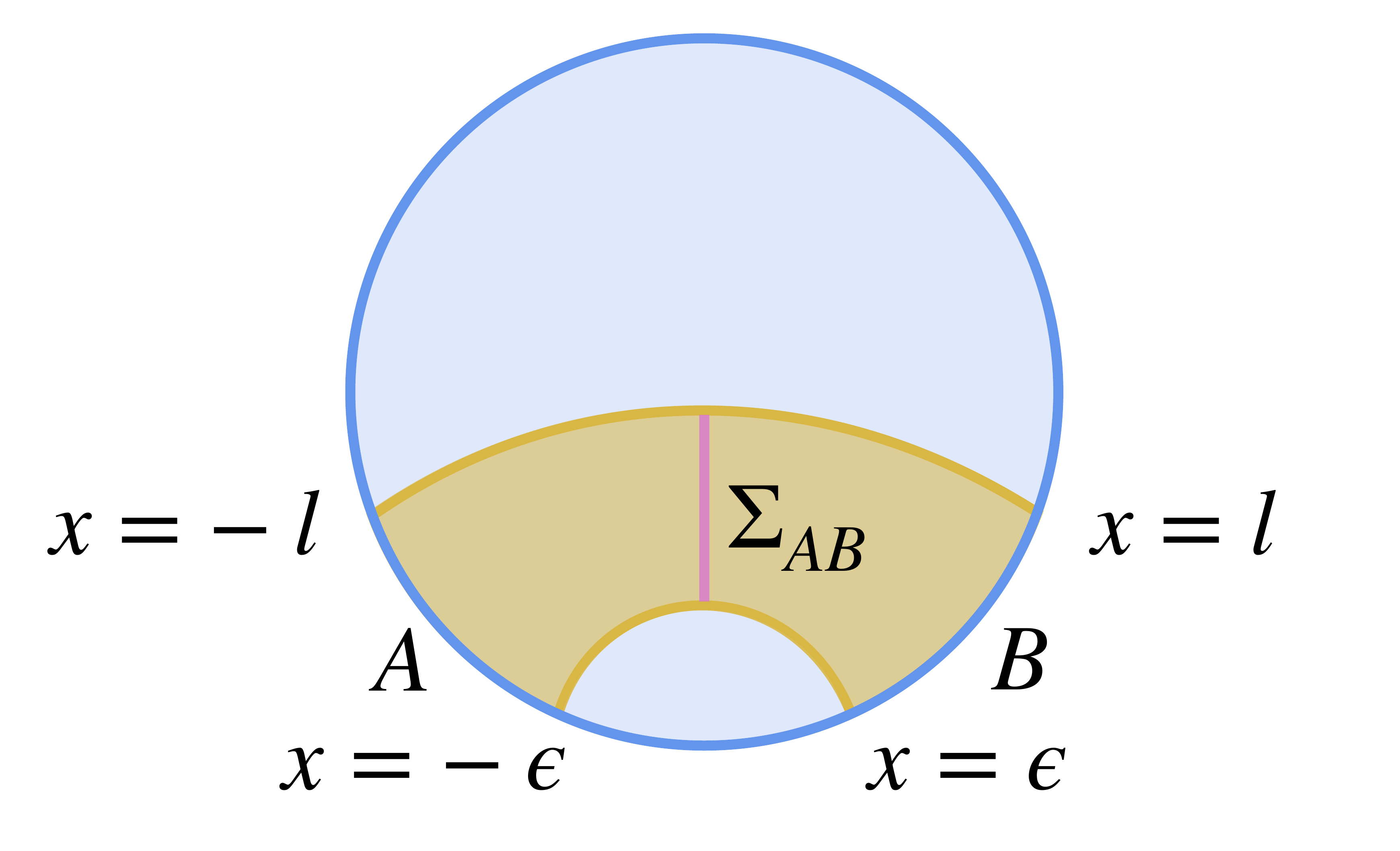}
	\caption{
		\label{fig:EWCS}	
 A time slice of the global AdS$_3$ which corresponds to the ground state of the corresponding CFT$_2$. The whole CFT system is divided into $A = [-l,\epsilon]$, $B = [\epsilon, l]$ and their complement. The Ryu-Takayanagi surface $\gamma_{AB}$ associated with subsystem $A\cup B$ is shown by the solid orange line. The shaded orange region surrounded by $\gamma_{AB}\cup A \cup B$ shows the entanglement wedge of $A\cup B$. The pink solid line shows the minimal surface $\Sigma_{AB}$ which divides the entanglement wedge into two with $A$ and $B$ on each side. The entanglement wedge cross section is evaluated from the area of $\Sigma_{AB}$. Note that we will consider $\epsilon/L \ll 1$. In this figure, we sketch in a way as if $\epsilon/L$ was large just for convenience. 
  }
\end{figure*}

Let us then consider a subregion $A \cup B$, which is formally divided into two parts $A$ and $B$. Accordingly, we have the RT surface $\gamma_{A\cup B}$ and the entanglement wedge corresponding to it. In this setup, the entanglement wedge cross section between $A$ and $B$ is defined as 
\begin{align}
    E^W_{A:B} = \frac{{\rm Area} (\Sigma_{AB})}{4G_N}
\end{align}
where $\Sigma_{AB}$ is a codimension-2 surface in the AdS spacetime which satisfies the following conditions: 
\begin{itemize}
    \item $\Sigma_{AB}$ ends on $\gamma_{AB} \cup A \cup B$. 
    \item $A$ and $B$ are on the different two sides of $\Sigma_{AB}$. 
    \item The area of $\Sigma_{AB}$ takes the minimal value. 
\end{itemize}
See Fig.~\ref{fig:EWCS} for a sketch. The entanglement wedge cross section was firstly proposed as the holographic dual of the entanglement of purification \cite{UT_18_EoP,Swingle_17_EoP}. Its relation with entanglement negativity \cite{KR_18_negativity}, odd entropy \cite{Tamaoka_18_odd}, and reflected entropy \cite{Faulkner2021_reflected} was later discussed. These colorful correspondences related to the entanglement wedge cross section enable us to apply quantum information to characterize quantum gravity. For example, one can argue that constructing classical spacetime requires tripartite entanglement \cite{rath2020_EWCS}, and lower bound the number of orthogonal atypical black hole microstates \cite{Wei_22_microstates} based on these results. In this paper, we will only focus on the reflected entropy, and we will not go into the discussion about quantum gravity.

\begin{figure*}[h]
        \centering
	\includegraphics[width=12cm]           {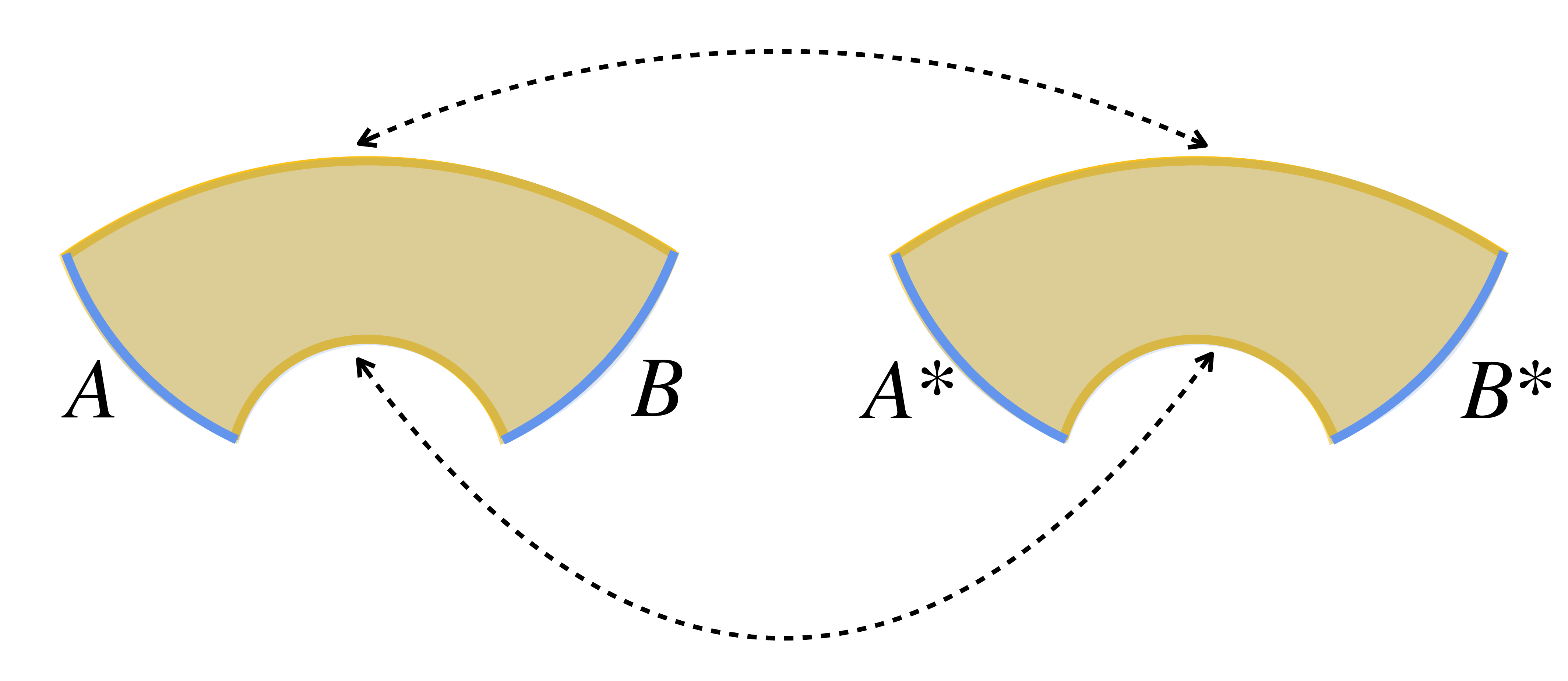}
	\caption{
		\label{fig:canonical_purification}	
 The gravity dual of $|{\sqrt{\rho_{AB}}}\rangle$, the canonical purification of $\rho_{AB}$, is obtained by pasting the entanglement wedge of $A\cup B$ (left) and a copy of it (right) along the Ryu-Takayanagi surface. $A^*$ and $B^*$ are the copies of $A$ and $B$, respectively. By appling the Ryu-Takayanagi formula to the subsystem $A\cup A^*$, it is straightforward to find the reflected entropy and the entanglement wedge cross section are related to each other as \eqref{eq:hol_RE}.
  }
\end{figure*}

In short, it is shown that \cite{Faulkner2021_reflected} the reflected entropy $S^R_{A:B}$ between $A$ and $B$ for a CFT state $\rho_{AB}$ is given by 
\begin{align}\label{eq:hol_RE}
    S^R_{A:B} = 2 E^W_{A:B}. 
\end{align}
To roughly understand why this is the case, let us remind ourselves how the reflected entropy is defined. The first step to define the reflected entropy between $A$ and $B$ on a (generally) mixed state $\rho_{AB}$ is to consider its canonical purification. Diagonalizing $\rho_{AB} \in \mathcal{D}(\mathcal{H}_A \otimes \mathcal{H}_B)$ into 
\begin{align}
    \rho_{AB} = \sum_{i} p_i |\varphi_i\rangle\langle\varphi_i|, 
\end{align}
its canonical purification is defined as 
\begin{align}
    |\sqrt{\rho_{AB}}\rangle = \sum_{i} \sqrt{p_i} |\varphi_i\rangle|\varphi_i^*\rangle,
\end{align}
where for each $|\varphi_i\rangle \in \mathcal{H}_A \otimes \mathcal{H}_B$, $|\varphi_i^*\rangle \in \mathcal{H}_{A^*} \otimes \mathcal{H}_{B^*}$ is its CPT conjugate. Accordingly, $A^*$ ($B^*$) serves as a copy of $A$ ($B$). While the gravity dual of $\rho_{AB}$ is given by the corresponding entanglement wedge, the gravity dual of $|\sqrt{\rho_{AB}}\rangle$ can be obtained by considering two copies of the entanglement wedge and paste them together along the RT surfaces. See Fig.~\ref{fig:canonical_purification} for a sketch. Based on these, the reflected entropy $S^R_{A:B}$ is defined as the entanglement entropy between $A\cup A^*$ and $B\cup B^*$ for the state $|\sqrt{\rho_{AB}}\rangle$. It is then natural to expect \eqref{eq:hol_RE} by directly applying the RT formula to the gravity dual of $|\sqrt{\rho_{AB}}\rangle$ shown in Fig.~\ref{fig:canonical_purification}.  

Before ending this subsection, let us show an example of holographically computing the reflected entropy. Again, let the whole system to be a circle with length $2L$ parameterized by $x\in[-L,L)$ and consider the ground state on it. For simplicity, we consider adjacent subregions $A=[-l,-\epsilon]$ and $B=[\epsilon,l]$, where $\epsilon$ is a UV cutoff to avoid divergence. The corresponding setup is sketched in Fig.~\ref{fig:EWCS}. 

To compute the entanglement wedge cross section $E^{W}_{A:B}$ in this setup, it is convenient to introduce another coordinate as follows 
\begin{align}
    &X_0 = \frac{\alpha^2}{2z_P} \left(1 + \frac{\tau_P^2 + x_P^2 + z_P^2}{\alpha^2}\right), \\
    &X_1 = \frac{\alpha^2}{2z_P} \left(1 - \frac{\tau^2_P + x^2_P + z^2_P}{\alpha^2}\right), \\
    &X_2 = \alpha \frac{x_P}{z_P}, \\
    &X_3 = \alpha \frac{\tau_P}{z_P}.
\end{align}
This coordinate covers the Poincar\'e patch of the global AdS, and hence is often called the Poincar\'e AdS. The metric turns out to be 
\begin{align}
    ds^2 = \alpha^2 \frac{dz_P^2 + d\tau_P^2 + dx_P^2}{z_P^2},
\end{align}
in the Poincar\'e coordinate. Note that this metric is invariant under rescaling. For example, we can introduce a dimensionless version of Poincar\'e coordinate as 
\begin{align}
    \left(\hat{z}_P, \hat{\tau}_P, \hat{x}_P\right) = \left(\frac{z_P}{\alpha}, \frac{\tau_P}{\alpha}, \frac{x_P}{\alpha}\right), 
\end{align}
with which the metric is still 
\begin{align}\label{eq:Poincare}
    ds^2 = \alpha^2 \frac{d\hat{z}_P^2 + d\hat{\tau}_P^2 + d\hat{x}_P^2}{\hat{z}_P^2}. 
\end{align}
While whether using $\left(\hat{z}_P, \hat{\tau}_P, \hat{x}_P\right)$ or $(z_P, \tau_P, x_P)$ will not make a big difference in this subsection, we will find the dimensionless coordinate $\left(\hat{z}_P, \hat{\tau}_P, \hat{x}_P\right)$ more convenient when discussing holographic ICFT later. 

\begin{figure*}[h]
        \centering
        \includegraphics[width=8cm]{fig_appendix/Holography/EWCS2.pdf}
        \includegraphics[width=8cm]{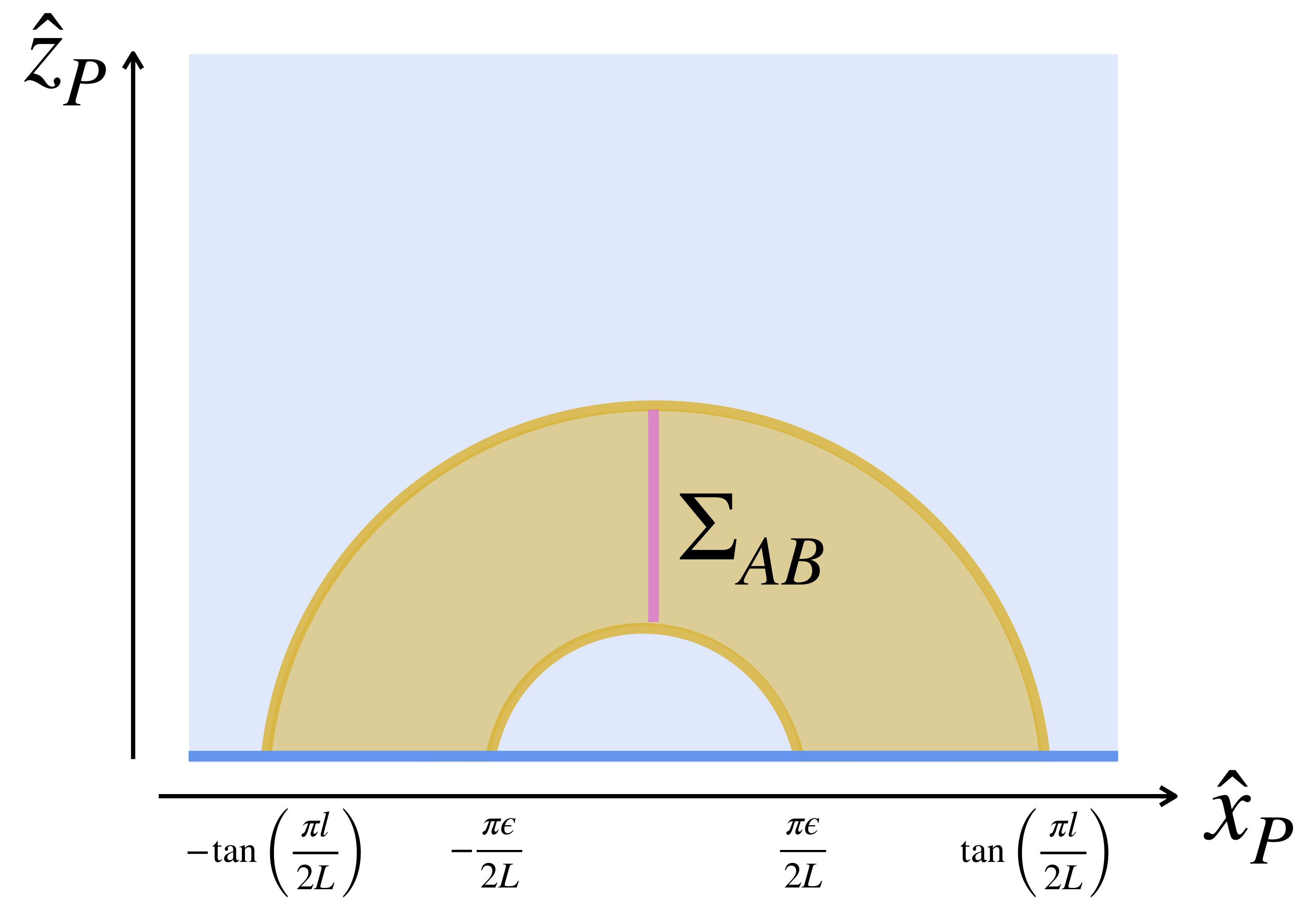}
	\caption{
		\label{fig:EWCS_Poincare}	
The entanglement wedge of $A\cup B$ in the global coordinate (left) and the Poincar\'e coordinate (right). From the right figure, we can see that the two end points of $\Sigma_{AB}$ are given by $\left(\hat{z}_P, \hat{x}_P\right) = \left(\frac{\pi \epsilon}{2 L}, 0\right), \left(\tan\left(\frac{\pi l}{2L}\right), 0\right)$. }
\end{figure*}

One feature of the Poincar\'e coordinate is the geodesic connecting two points 
\begin{align}
    \left(\hat{z}_P, \hat{\tau}_P, \hat{x}_P\right) = (0, 0, a), (0,0,b),
\end{align}
is given by 
\begin{align}
    \left(\hat{x}_P-\frac{(a+b)}{2}\right)^2 + \hat{z}_P^2 = \left(\frac{b-a}{2}\right)^2. 
\end{align}
This fact greatly simplifies our computation of the reflected entropy. The two end points of $A$
\begin{align}
    (\rho, \tau, x) = (\rho_\infty, 0, -l), (\rho_\infty, 0, -\epsilon),
\end{align}
are mapped to 
\begin{align}
    \left(\hat{z}_P, \hat{\tau}_P, \hat{x}_P\right) = \left(\frac{\pi \epsilon}{2L} \frac{1}{\cos^2\left(\frac{\pi l}{2L}\right)},0,-\tan\left(\frac{\pi l}{2L}\right)\right), \left(\frac{\pi\epsilon}{2L}, 0, -\frac{\pi\epsilon}{2L}\right),
\end{align}
in the Poincar\'e coordinate. Note that we have used the relation \eqref{eq:UVandIRcutoff} and the fact that $\epsilon/L \ll 1$. Similarly, the two end points of $B$
\begin{align}
    (\rho, \tau, x) = (\rho_\infty, 0, \epsilon), (\rho_\infty, 0, l),
\end{align}
are mapped to 
\begin{align}
    \left(\hat{z}_P, \hat{\tau}_P, \hat{x}_P\right) = \left(\frac{\pi\epsilon}{2L}, 0, \frac{\pi\epsilon}{2L}\right), \left(\frac{\pi \epsilon}{2L} \frac{1}{\cos^2\left(\frac{\pi l}{2L}\right)},0, \tan\left(\frac{\pi l}{2L}\right)\right).
\end{align}
Referring to the right part of Fig.~\ref{fig:EWCS_Poincare}, the entanglement wedge cross section can be computed as 
\begin{align}
    E^{W}_{A:B} = \frac{\alpha}{4G_N} \int_{\frac{\pi\epsilon}{2L}}^{\tan\left(\frac{\pi l}{2L}\right)}\frac{d \hat{z}_P}{\hat{z}_P} = \frac{\alpha}{4G_N} \ln \left(\frac{2L}{\pi\epsilon} \tan\left(\frac{\pi l}{2L}\right) \right). 
\end{align}
Accordingly, the reflected entropy is given by 
\begin{align}
    S^R_{A:B} = 2 E^W_{A:B} = \frac{c}{3} \ln \left(\frac{2L}{\pi\epsilon} \tan\left(\frac{\pi l}{2L}\right) \right). 
\end{align}
Let us also present the mutual information for reference: 
\begin{align}
    I_{A:B} = S_A + S_B - S_{AB} = 2\times \frac{c}{3} \ln \left( \frac{2L}{\pi\epsilon} \sin\left(\frac{\pi l}{2L}\right) \right) - \frac{c}{3} \ln \left( \frac{2L}{\pi\epsilon} \sin\left(\frac{\pi l}{L}\right) \right) = \frac{c}{3} \ln\left(\frac{L}{\pi\epsilon} \tan\left(\frac{\pi l}{2L}\right)\right). 
\end{align}
The Markov gap \cite{Hayden2021_MarkovGap} is 
\begin{align}
    S^R_{A:B} - I_{A:B} = \frac{c}{3} \ln 2.
\end{align}
These results may be used for later reference. 

\subsection{AdS/ICFT}
In this subsection, we will first introduce a simple gravitational setup to realize a holographic dual of ICFT. After that, we will compute the entanglement entropy, mutual information, and reflected entropy for some relevant subregions. We will end up commenting on how to understand some phenomena we observed in the main text using intuitions from holography.

\subsubsection{A thin brane model of AdS/ICFT}
\begin{figure*}[h]
        \centering
        \includegraphics[width=8cm]{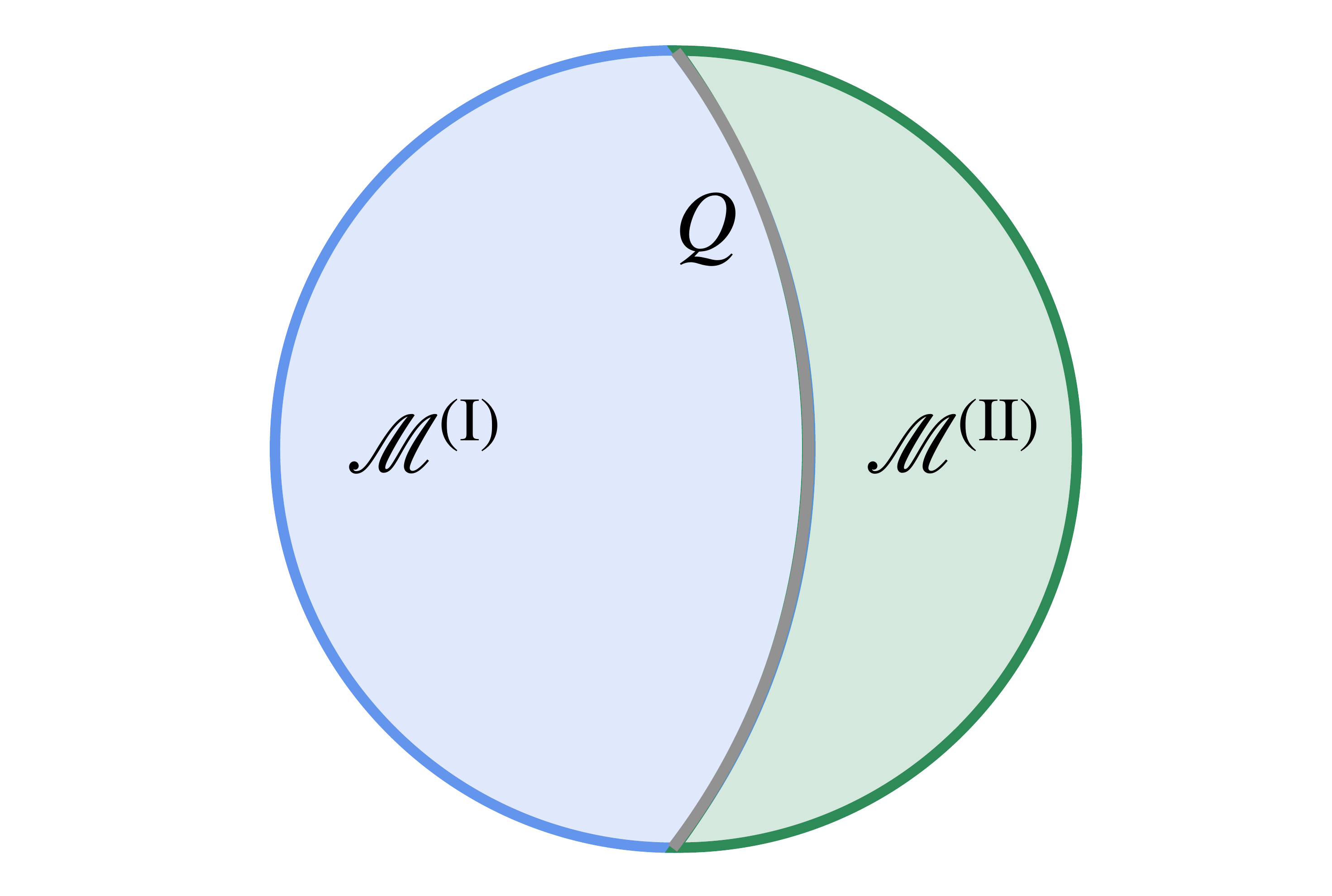}
        \includegraphics[width=8cm]{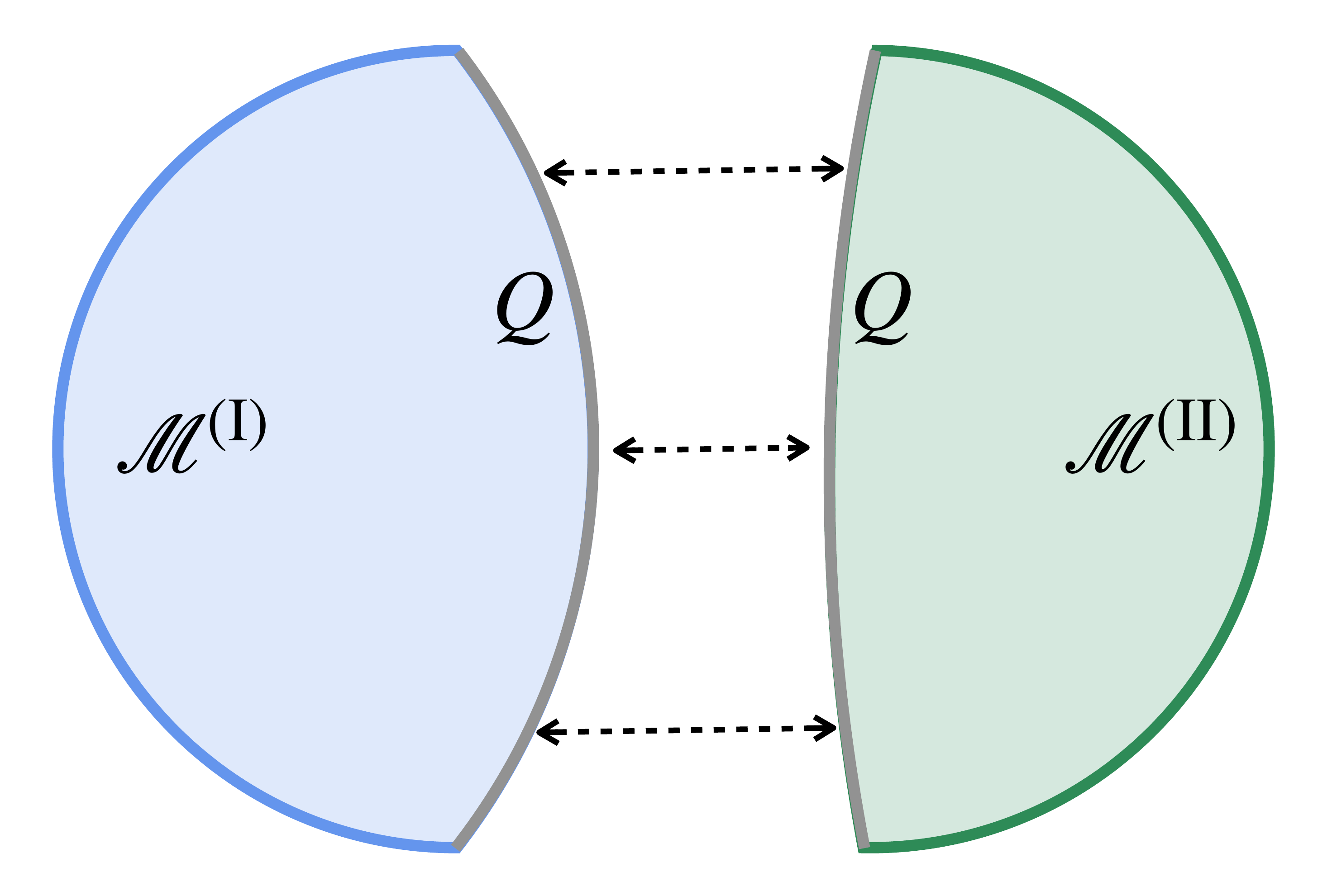}
	\caption{A time slice of the thin brane model for holographic ICFT. In the left figure, a codimension-1 object, the thin brane $Q$, is introduced in the bulk gravity. The two sides are AdS gravity with different cosmological constants. The location of the brane $Q$ shown in the left figure is not accurate. In general, as shown in the right figure, one needs to solve the equation of motion associated with $Q$ on each side and then paste them together in practice. For simplicity, however, we will draw a holographic ICFT as in the left figure in the following part of this particle. 
		\label{fig:Hol_ICFT}	
 }
\end{figure*}

An ICFT is realized by connecting two (in general) different CFT through an interface with junction condition maximally preserving the conformal symmetries. Corresponding to this structure, the gravity dual of an ICFT is also expected to have some object to play the role of an interface. In the minimal model of a holographic ICFT, this is realized by introducing a codimension-1 object called a thin brane, and let the two sides of the thin brane to have different cosmological constants \cite{Azeyanagi_2008,Bachas2020_Interface,vanRaamsdonk2020_Holoween,Bachas2021_Hol_ICFT,Karch2021_interfaceEE,Sonner2022_Island}.  See Fig.~\ref{fig:Hol_ICFT} for a skectch. The minimal action is given by 
\begin{align}
    I_{\rm grav.} = &-\frac{1}{16\pi G_N} \int_{\mathcal{M}^{{\rm (I)}}} \sqrt{g^{{\rm (I)}}} \left(R^{{\rm (I)}} + \frac{2}{\left(\alpha^{{\rm (I)}}\right)^2}\right) 
    -\frac{1}{16\pi G_N} \int_{\mathcal{M}^{{\rm (II)}}} \sqrt{g^{{\rm (II)}}} \left(R^{{\rm (II)}} + \frac{2}{\left(\alpha^{{\rm (II)}}\right)^2}\right) \nonumber\\
    &-\frac{1}{8\pi G_N} \int_{Q} \sqrt{h} \left(K^{{\rm (I)}} - K^{{\rm (II)}} - T\right) ,
\end{align}
where the first two terms are the Einstein-Hilbert terms on the two sides of the thin brane $Q$, and the third term is the thin brane term. Here, $K^{{\rm (I)}}$ and $K^{{\rm (II)}}$ are the extrinsic curvatures of $Q$ computed from the ${\rm (I)}$ side and the ${\rm (II)}$ side, respectively. For both cases, we choose the normal vector such that it points from ${\rm (I)}$ to ${\rm (II)}$. Besides, $T$ is the tension of the brane $Q$, and $h_{ab}$ is the induced metric on it. Note that we have omitted the counterterms at the asymptotic boundary of the spacetime. By considering the variation at the vicinity of the brane, one can get the following junction condition 
\begin{align}
    \left(K^{{\rm (I)}}_{ab} - K^{{\rm (II)}}_{ab}\right) - h_{ab} \left(K^{{\rm (I)}} - K^{{\rm (II)}}\right) = -T h_{ab}. 
\end{align}
The central charges of the two CFT are given by 
\begin{align}
    c^{{\rm (I)}} = \frac{3}{2} \frac{\alpha^{{\rm (I)}}}{G_N}, ~~ c^{{\rm (II)}} = \frac{3}{2} \frac{\alpha^{{\rm (II)}}}{G_N},
\end{align}
respectively. For simplicity, we assume $c^{\rm (I)} \leq c^{\rm (II)}$ throughout this appendix.

In the following, we would like to consider the ground state of an ICFT defined on a circle parameterized by $x\in[-L,L)$ on which CFT$^{{\rm (I)}}$ lives on the $x\in(-L,0)$ part and CFT$^{{\rm (II)}}$ lives on the $x\in(0,L)$ part. The corresponding gravity dual is realized by sewing two portions of global AdS geometries with AdS radius $\alpha^{\rm (I)}$ and $\alpha^{\rm (II)}$ respectively, in a way that satisfies the junction condition. To this end, let us first introduce some coordinates in global AdS$_3$ which are useful to describe the gravity dual of the ground state of a holographic ICFT. 

\subsubsection{Useful coordinates in global AdS}
\begin{figure*}[h]
        \centering
        \includegraphics[width=8cm]{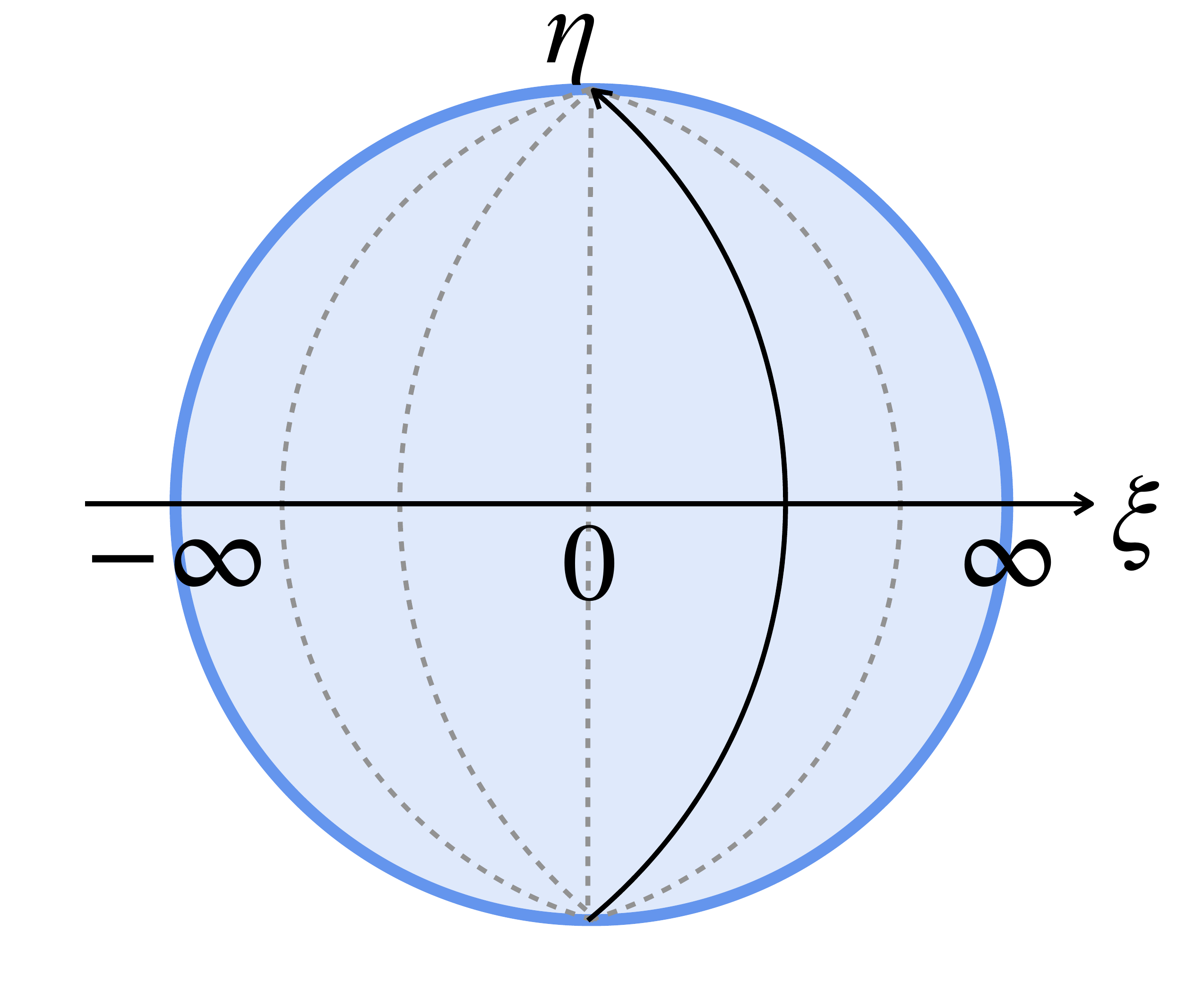}
        \includegraphics[width=8cm]{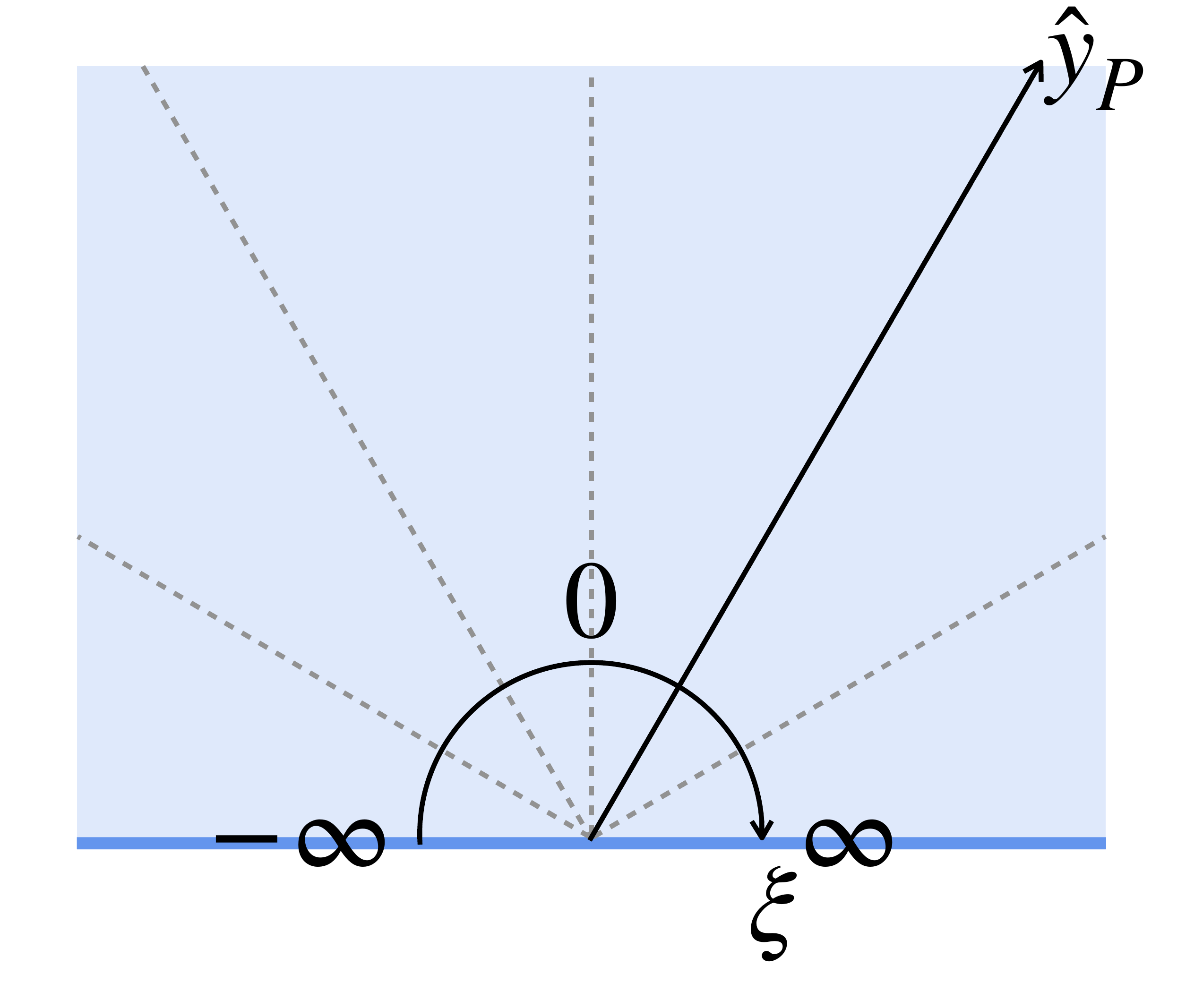}
	\caption{Coordinates slicing global AdS$_3$ into global AdS$_2$. The left figure shows a time slice of a global AdS$_3$. The right figure shows a time slice of a Poincar\'e AdS$_3$.
		\label{fig:slicing_coordinate}	
 }
\end{figure*}

When describing a 3D sphere, it is sometimes useful to slice it into 2D spheres with different radii. Similarly, we can also consider slicing a global AdS$_3$ into global AdS$_2$ with different AdS radii. By properly choosing the coordinates, the metric of a global AdS$_3$ can be written as
\begin{align}
    ds^2 = \alpha^2 \left(d\xi^2 + \cosh^2 \xi \left(\cosh^2\eta~ d\sigma^2 + d\eta^2\right)\right).
\end{align} 
It is straightforward to see that each $\xi = {\rm const.}$ slice gives a global AdS$_2$ spacetime, where the AdS radius is given by $\alpha \cosh \xi$. One can also introduce a similar coordinate system which only covers the Poincar\'e patch, with which the metric is 
\begin{align}
    ds^2 = \alpha^2 \left(d\xi^2 + \cosh^2 \xi \left(\frac{d\hat{y}_P^2 + d\hat{\tau}_P}{\hat{y}_P^2}\right)\right).
\end{align}
In this case, each $\xi = {\rm const.}$ slice gives a Poincar\'e AdS$_2$ with radius $\alpha \cosh \xi$. 
Also note that this coordinate system is related to the standard Poincar\'e coordinate \eqref{eq:Poincare} via 
\begin{align}
    \hat{z}_P = \frac{\hat{y}_P}{\cosh \xi}, ~\hat{x}_P = \hat{y}_P \tanh \xi.
\end{align}
See Fig.~\ref{fig:slicing_coordinate} for a sketch of these coordinate systems. 

\subsubsection{The ground state configuration}

We will then move on to present the gravity dual of the ground state of the corresponding holographic ICFT. Let us use the coordinates introduced above and use (I) and (II) to label each side. By solving the equation of motion associated with the brane $Q$, one can find $Q$ locates at $\xi^{\rm (I)} = \xi^{\rm (I)}_*$ ($\xi^{\rm (II)} = \xi^{\rm (II)}_*$) when seen from the $\mathcal{M}^{\rm (I)}$ ($\mathcal{M}^{\rm (II)}$) side. Here, $\xi^{\rm (I)}_*$ and $\xi^{\rm (II)}_*$ are constants given by 
\begin{align}\label{eq:brane_location_1}
    \tanh \left(\xi^{\rm (I)}_*\right) &= \frac{\alpha^{\rm (I)}}{2T} \left(T^2 + \frac{1}{\left(\alpha^{\rm (I)}\right)^2} - \frac{1}{\left(\alpha^{\rm (II)}\right)^2}\right), \\
    \tanh \left(\xi^{\rm (II)}_*\right) &= \frac{\alpha^{\rm (II)}}{2T} \left(T^2 + \frac{1}{\left(\alpha^{\rm (II)}\right)^2} - \frac{1}{\left(\alpha^{\rm (I)}\right)^2}\right). \label{eq:brane_location_2}
\end{align}
Therefore, the holographic dual looks like the right part of Fig.~\ref{fig:Hol_ICFT}. Referring to the left part of Fig.~\ref{fig:slicing_coordinate}, we can see that when $\xi_*$ is positive, the brane looks convex, and when $\xi_*$ is negative, the brane looks concave. The right part of Fig.~\ref{fig:Hol_ICFT} shows a case where both $\xi^{\rm (I)}_*$ and $\xi^{\rm (II)}_*$ are positive. The left part of Fig.~\ref{fig:Hol_ICFT} shows a case where $\xi^{\rm (I)}_*>0$ and $\xi^{\rm (I)}_* < 0$. There is another junction condition which tells us how the branes are identified with each other on the two sides: on the brane $Q$, the $y$-coordinates on the two sides are related with each other as 
\begin{align}
    \hat{y}_{P}^{\rm (I)} = \hat{y}_{P}^{\rm (II)}.
\end{align}
Refer to \cite{Sonner2022_Island} for details of derivation. 

We can see that if one takes 
\begin{align}\label{eq:Trange}
    \frac{1}{\alpha^{\rm (I)}} - \frac{1}{\alpha^{\rm (II)}} < T < \frac{1}{\alpha^{\rm (I)}} + \frac{1}{\alpha^{\rm (II)}},
\end{align}
then the right-hand sides of \eqref{eq:brane_location_1} and \eqref{eq:brane_location_2} are consistent with the range of the $\tanh$ function on the left-hand sides. In fact, the $1/\alpha^{\rm (I)} - 1/\alpha^{\rm (II)}$ is known as the  Coleman-De Lucia bound below which the gravitational configuration becomes unstable \cite{Coleman1980}. On the other hand, $1/{\alpha^{\rm (I)}} + 1/{\alpha^{\rm (II)}}$ is the bound for the brane $Q$ to be AdS. 

One key feature of the brane when taking $T$ from \eqref{eq:Trange} is that, while $\xi_*^{\rm (II)}$ can be both positive and negative, $\xi_*^{\rm (I)}$ is always positive. This feature plays a crucial role when discussing holographic entanglement entropy and holographic reflected entropy below. 

\subsubsection{Holographic entanglement entropy and reflected entropy in AdS/ICFT}

\begin{figure*}[h]
        \centering
        \includegraphics[width=7.5cm]{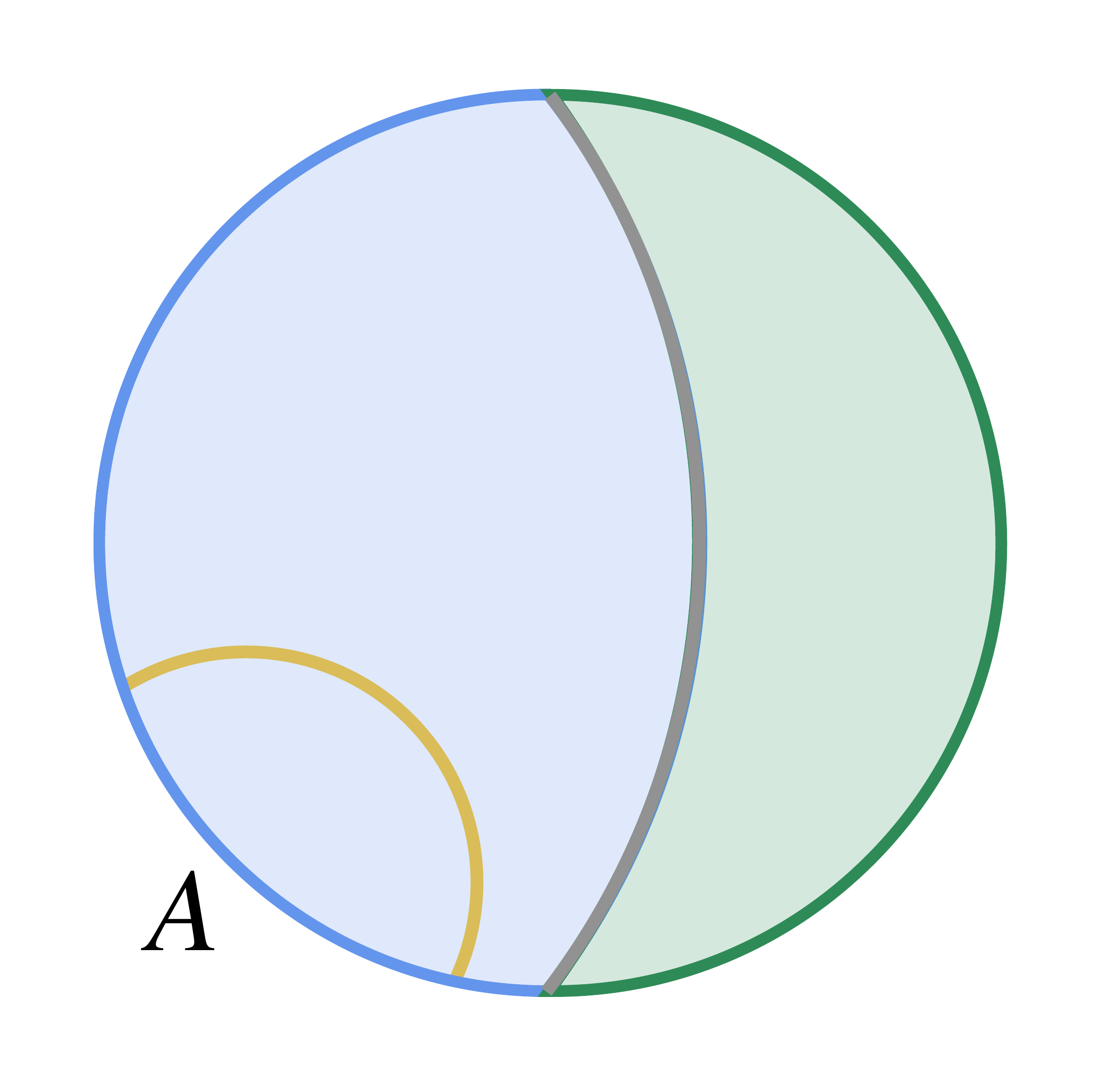}
        \includegraphics[width=7.5cm]{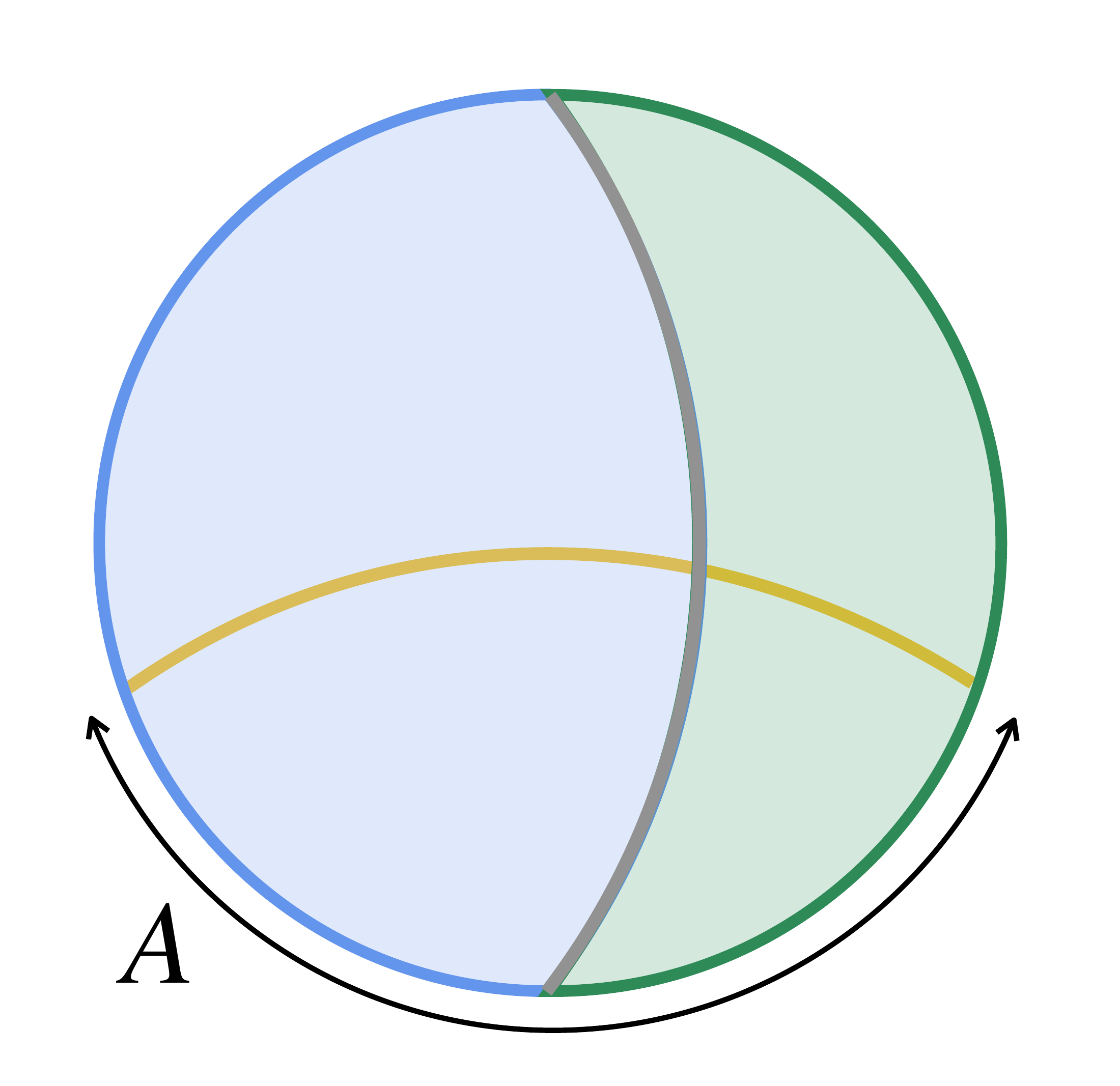}	\caption{(Left) The RT surface of a single interval $A$ lying in (I) is shown in orange. The RT surface lies in $\mathcal{M}^{\rm (I)}$. (Right) The RT surface of a single interval $A$ going across (I) and (II) is shown in orange. The RT surface lies in both $\mathcal{M}^{\rm (I)}$ and $\mathcal{M}^{\rm (II)}$ and intersects the brane $Q$ once. 
        \label{fig:interval_I_across}	
 }
\end{figure*}

\paragraph{Single intervals in (I).} As the very first example, let us consider a single interval $A$ in (I). Thanks to the properties that the geometry $\mathcal{M}^{\rm (I)}$ is just a portion of AdS$_3$ with radius $\alpha^{(I)}$, $\alpha^{(I)} < \alpha^{(II)}$, and $\xi_*^{\rm (I)} > 0$, the RT surface of $A$ sits in $\mathcal{M}^{\rm (I)}$. See the left part of Fig.~\ref{fig:interval_I_across} for a sketch. When the length of $A$ is $l$, the holographic entanglement entropy of $A$ is 
\begin{align}
    S_A = \frac{\alpha^{\rm (I)}}{2G_N} \ln \left( \frac{2L}{\pi\epsilon} \sin\left(\frac{\pi l}{2L}\right) \right) = \frac{c^{\rm (I)}}{3} \ln \left( \frac{2L}{\pi\epsilon} \sin\left(\frac{\pi l}{2L}\right) \right). 
\end{align}
The situation is exactly the same as the ground state of a standard holographic CFT. 

\paragraph{Single intervals across (I) and (II).} The next simple example is a single interval across (I) and (II). The RT surface, in the case, lives on both sides and intersects $Q$ once. See the right part of Fig.~\ref{fig:interval_I_across} for a sketch. If we take the subsystem $A$ as $A=\{x|-l\leq x \leq l\}$, which is symmetric with respect to the interface, the holographic entanglement entropy turns out to be 
\begin{align}\label{eq:symmetic_EE_SM}
    S_A &= \frac{\alpha^{\rm (I)}}{4G_N} \ln \left( \frac{2L}{\pi\epsilon} \sin\left(\frac{\pi l}{L}\right) \right) + \frac{\alpha^{\rm (II)}}{4G_N} \ln \left( \frac{2L}{\pi\epsilon} \sin\left(\frac{\pi l}{L}\right) \right) + \frac{\alpha^{\rm (I)}}{4G_N} \xi_*^{\rm (I)}  + \frac{\alpha^{\rm (II)}}{4G_N} \xi_*^{\rm (II)}  \nonumber \\
    &= \frac{c^{\rm (I)}}{6} \ln \left( \frac{2L}{\pi\epsilon} \sin\left(\frac{\pi l}{L}\right) \right) + \frac{c^{\rm (II)}}{6} \ln \left( \frac{2L}{\pi\epsilon} \sin\left(\frac{\pi l}{L}\right) \right) + \frac{c^{\rm (I)}}{6} \xi_*^{\rm (I)}  + \frac{c^{\rm (II)}}{6} \xi_*^{\rm (II)}. 
\end{align}
where $\xi_*^{\rm (I)}$ and $\xi_*^{\rm (II)}$ are determined by \eqref{eq:brane_location_1} and \eqref{eq:brane_location_2}. 
This result can be easily obtained by noticing that the RT surface of subsystem $A=\{x|-l\leq x \leq l\}$ is given by $\hat{y}_P = \tan\left(\frac{\pi l}{2L}\right)$ in the $(\xi, \hat{y}_P, \hat{\tau}_P)$ coordinate in a standard AdS/CFT setup, and the junction conditions in AdS/ICFT is $\hat{y}_P^{\rm (I)} = \hat{y}_P^{\rm (II)}$. We will not write down the analytic expression of the holographic entanglement entropy for a general interval since it is complicated and not so useful for our discussion, but one can find a closed form in section 3.4 of \cite{Sonner2022_Island}. 

\begin{figure*}[h]
        \centering
        \includegraphics[width=7.5cm]{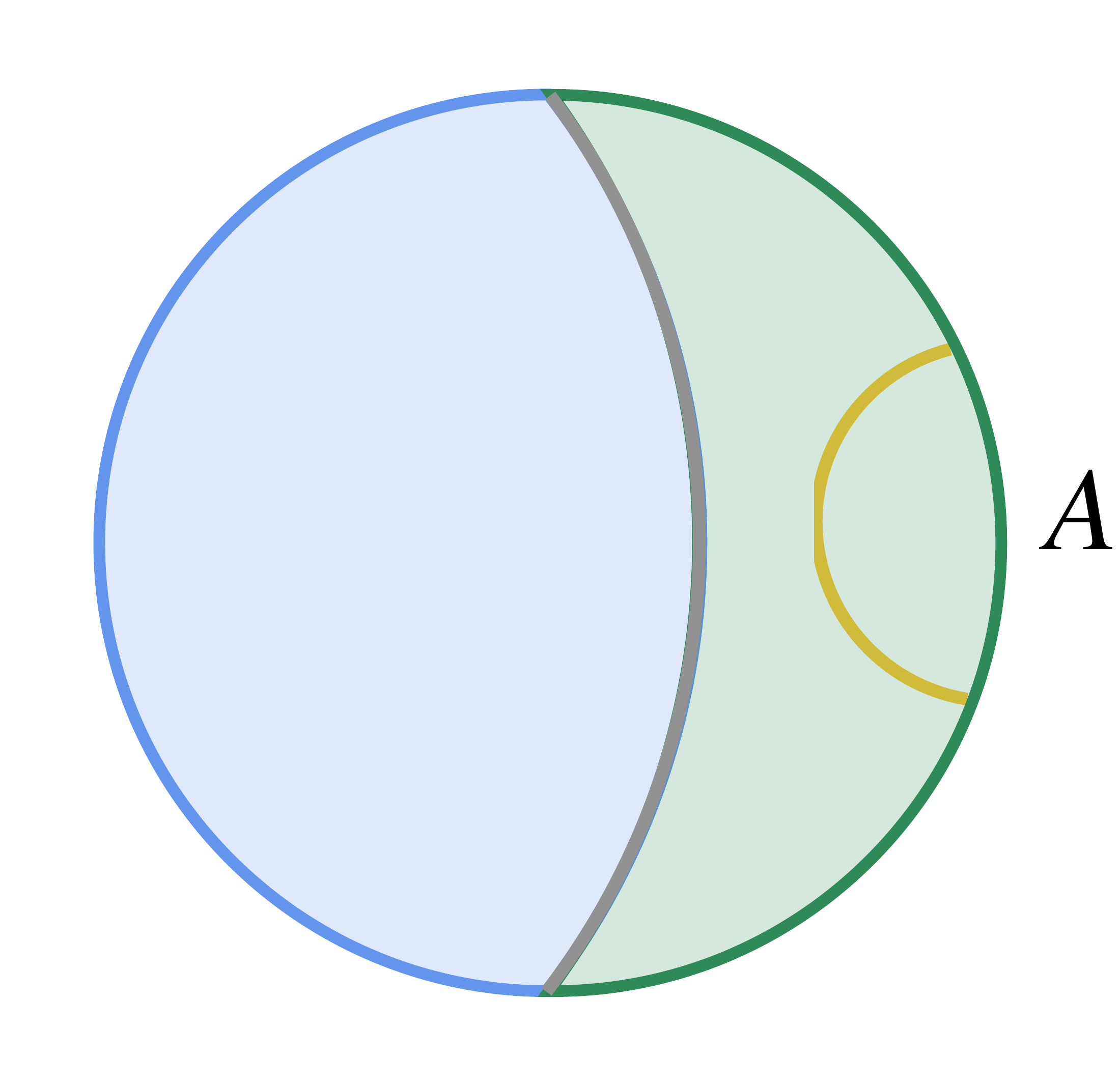}
        \includegraphics[width=7.5cm]{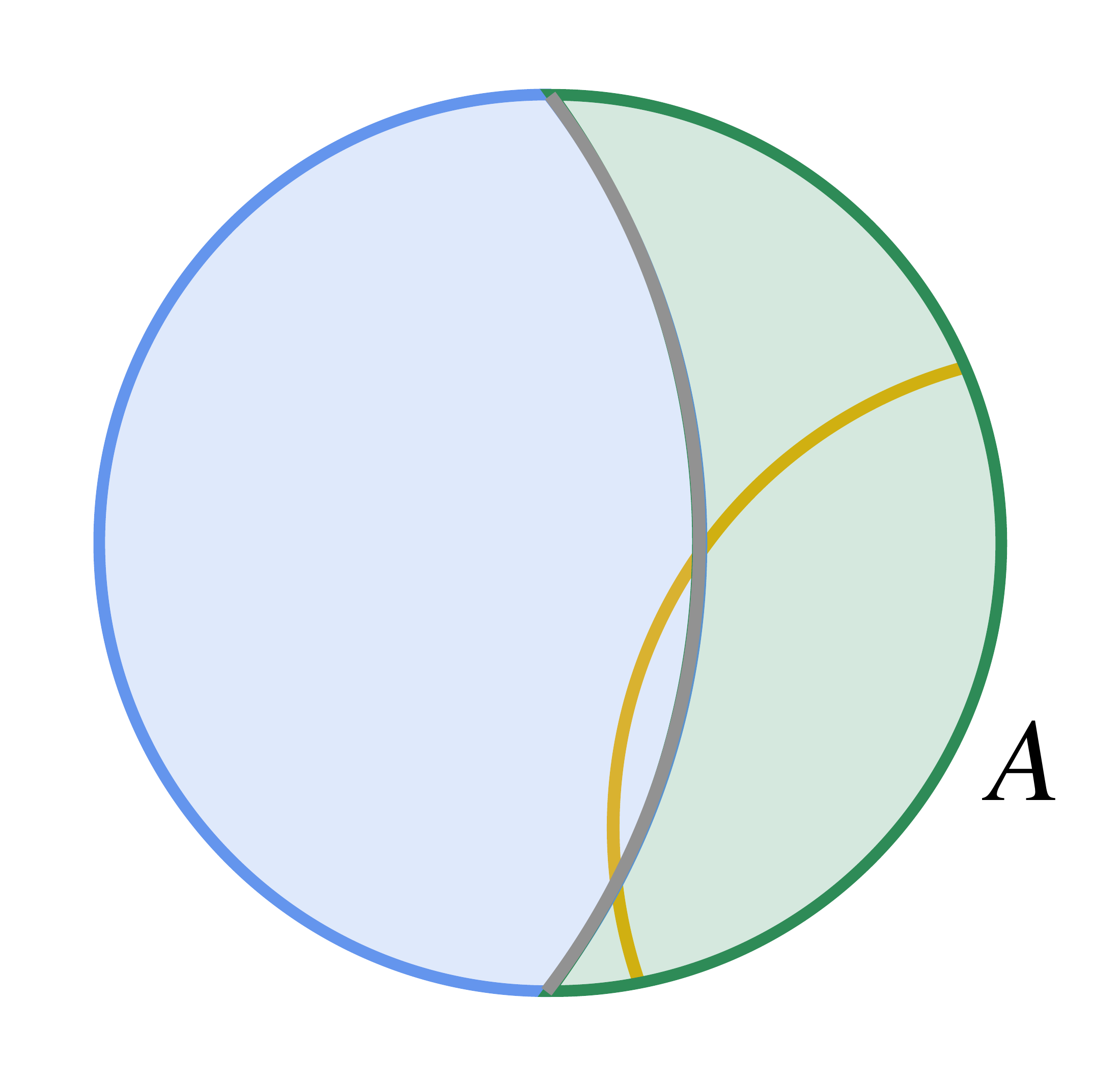}	\caption{Possibilities of RT surfaces for an interval $A$ lying in (II). The RT surface may live in $\mathcal{M}^{\rm (II)}$ as shown in the left figure, or go across the brane $Q$ as shown in the right figure. Note that the shape of the RT surface shown in the right figure is not accurate.  
        \label{fig:interval_II}	
 }
\end{figure*}

\paragraph{Single intervals in (II).} For single intervals, the most nontrivial phenomena occur when considering a single interval $A$ in (II). While the two end points of the corresponding RT surface sit on the asymptotic boundary of $\mathcal{M}^{\rm (II)}$, since $\alpha^{\rm (I)} < \alpha^{\rm (II)}$, there is a chance that the RT surface can go beyond the brane $Q$ to $\mathcal{M}^{\rm (I)}$ to become shorter. This can occur no matter $\xi^{\rm (II)}_*$ is negative or positive. 
As a result, there are two possibilities for the RT surface. The first possibility is that the whole RT surface lie in $\mathcal{M}^{\rm (II)}$, and the second possibility is that the RT surface goes across $Q$ to $\mathcal{M}^{\rm (I)}$ and then comes back to $\mathcal{M}^{\rm (II)}$. See Fig.~\ref{fig:interval_II} for a sketch. We will not show the analytic form of the length of the second RT surface since it is rather complicated. One can find an analytic form in section 3.4 of \cite{Sonner2022_Island}. 

There are, however, some subtle issues when computing the holographic entanglement entropy using the RT surface which crosses the thin brane twice, as shown in the right half of Fig.~\ref{fig:interval_II}. The length of such an RT surface is written as an analytic form in section 3.4 of \cite{Sonner2022_Island}, which includes an undetermined parameter. This undetermined parameter is expected to be determined by solving an equation, which has multiple solutions. In \cite{Sonner2022_Island}, the authors pointed out that only some of the solutions are physical, in the sense that they have straightforward geometric interpretation, and one should only use physical solutions to compute the holographic entanglement entropy. However, we numerically found that an ``unphysical" solution seems to give more sensible results to the holographic entanglement entropy. For example, we found that this ``unphysical" solution correctly recovers the behavior in a BCFT at $c^{\rm (II)}/c^{\rm (I)} \gg 1$. Therefore, in the following, our numerical computation involving the RT surface crossing the brane twice are performed with this ``unphysical" solution. We leave the geometric interpretation of this ``unphysical" solution and whether our results give the correct holographic entanglement entropy as future questions.

\paragraph{Some plots for single intervals.} 

\begin{figure*}[h]
        \centering
        \includegraphics[width=\textwidth]{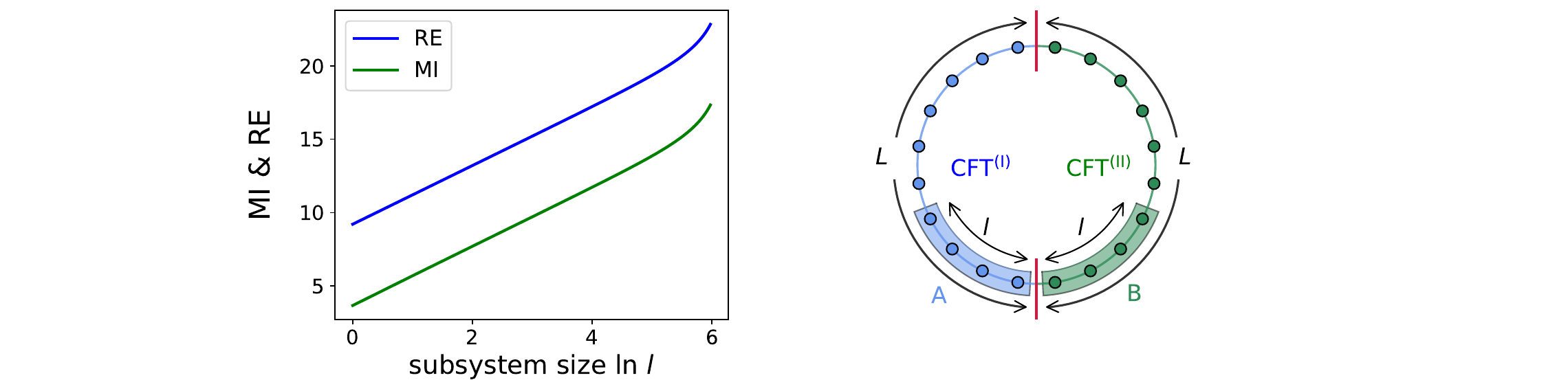}
        
	\caption{Reflected entropy $S_{A:B}^R$ and mutual information $I_{A:B}$ for adjacent intervals $A = (-l,-\epsilon)$ and $B = (\epsilon, l)$ on the two sides of the interface. Here, we set $L = 500$, $T=1$, $\alpha^{\rm (I)}=1$, $\alpha^{\rm (II)}=2$ and $\epsilon = 0.01$. The vertical axis shows $6S_{A:B}^R/c^{\rm (I)}$ (orange) or $6I_{A:B}/c^{\rm (I)}$ and the horizontal axis shows $\ln l$. 
 }
\label{fig:RE_and_MI}
\end{figure*}

While we just show the analytic form of the holographic entanglement entropy for some rather simple cases, it is useful to show some numerical results for more complicated cases to grasp their behaviors. Let us consider an ICFT defined on a circle with length $2L$. Let the $-L/2 < x < 0$ be CFT$^{\rm I}$ with central central charge $c^{\rm (I)}$ and let the $0 < x < L/2$ be CFT$^{\rm II}$ with central central charge $c^{\rm (II)}$, respectively. We consider a single interval whose left edge is fixed, change the position of the right edge and compute the holographic entanglement entropy. The plots for several cases are shown below. By comparing the following plots with %Fig.~{\blue 2} in the main text %
Fig.~\ref{fig:EE_sub_size_main} in the main text, 
one can see that the qualitative behavior of the bipartite entanglement entropy is very similar in holographic ICFT and free fermion ICFT. We also compute reflected entropy $S_{A:B}^R$ and mutual information $I_{A:B}$ for adjacent intervals $A = (-l,-\epsilon)$ and $B = (\epsilon, l)$ on the two sides of the interface, shown in Fig.~\ref{fig:RE_and_MI}.

\begin{figure*}[h]
        \centering
        \includegraphics[width=\textwidth]{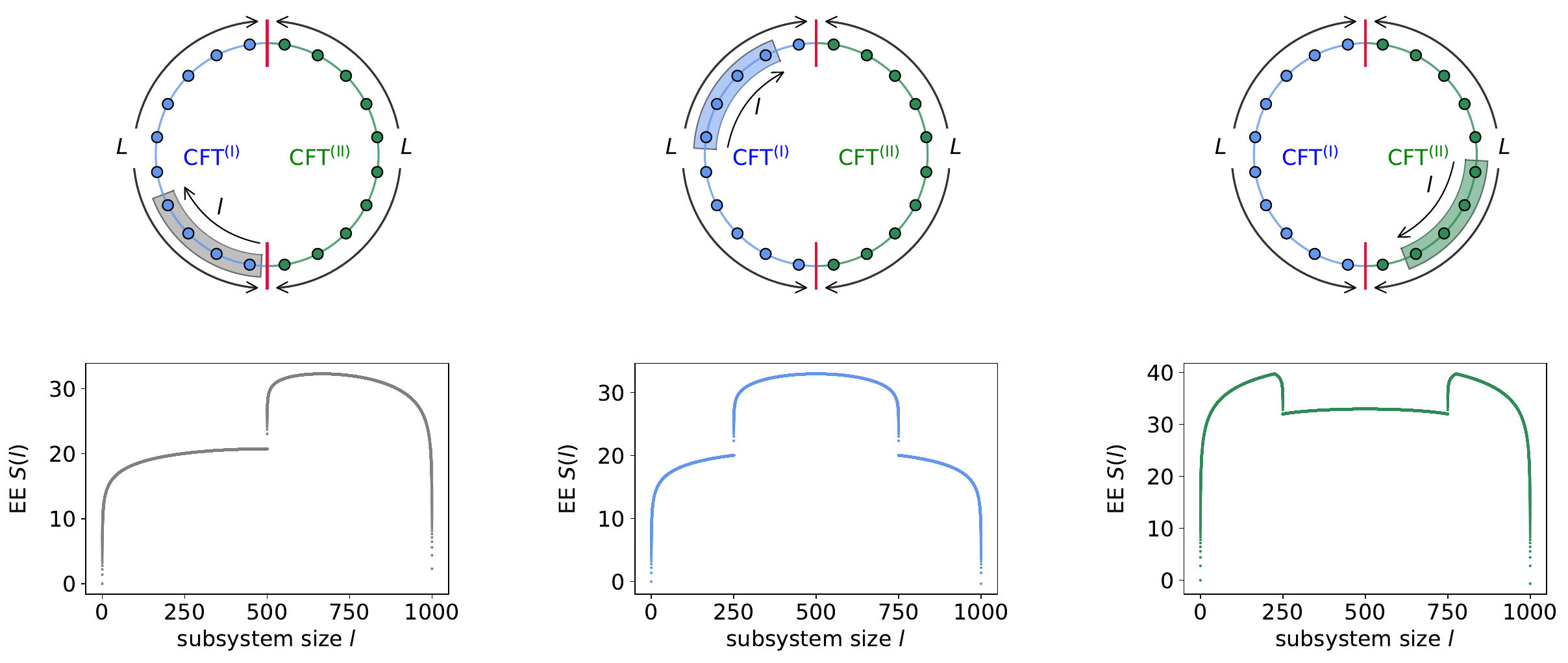}
	\caption{ The bipartite EE in the thin brane model of holographic ICFT defined on a circle with length $2L$. Here, we set $T=1$, $\alpha^{\rm (I)} = 1$ and $\alpha^{\rm (II)} = 2$. The ratio of the corresponding central charges then turns out to be $c^{\rm (II)}/c^{\rm (I)} = \alpha^{\rm (II)}/\alpha^{\rm (I)} = 2$. 
    Similar to the setup shown in %Fig.~{\blue 2} in the main text %
    Fig.~\ref{fig:EE_sub_size_main} in the main text, 
    we set $L=500$, fix one endpoint of the subsystem and change the length of the subsystem. The 
    plots show $6S_A/c^{\rm (I)}$, and the horizontal axis represents the length of the subsystem. We have set the UV cutoff to $\epsilon=0.01$. 
 }
\end{figure*}

\begin{figure*}[h]
        \centering
        \includegraphics[width=7.5cm]{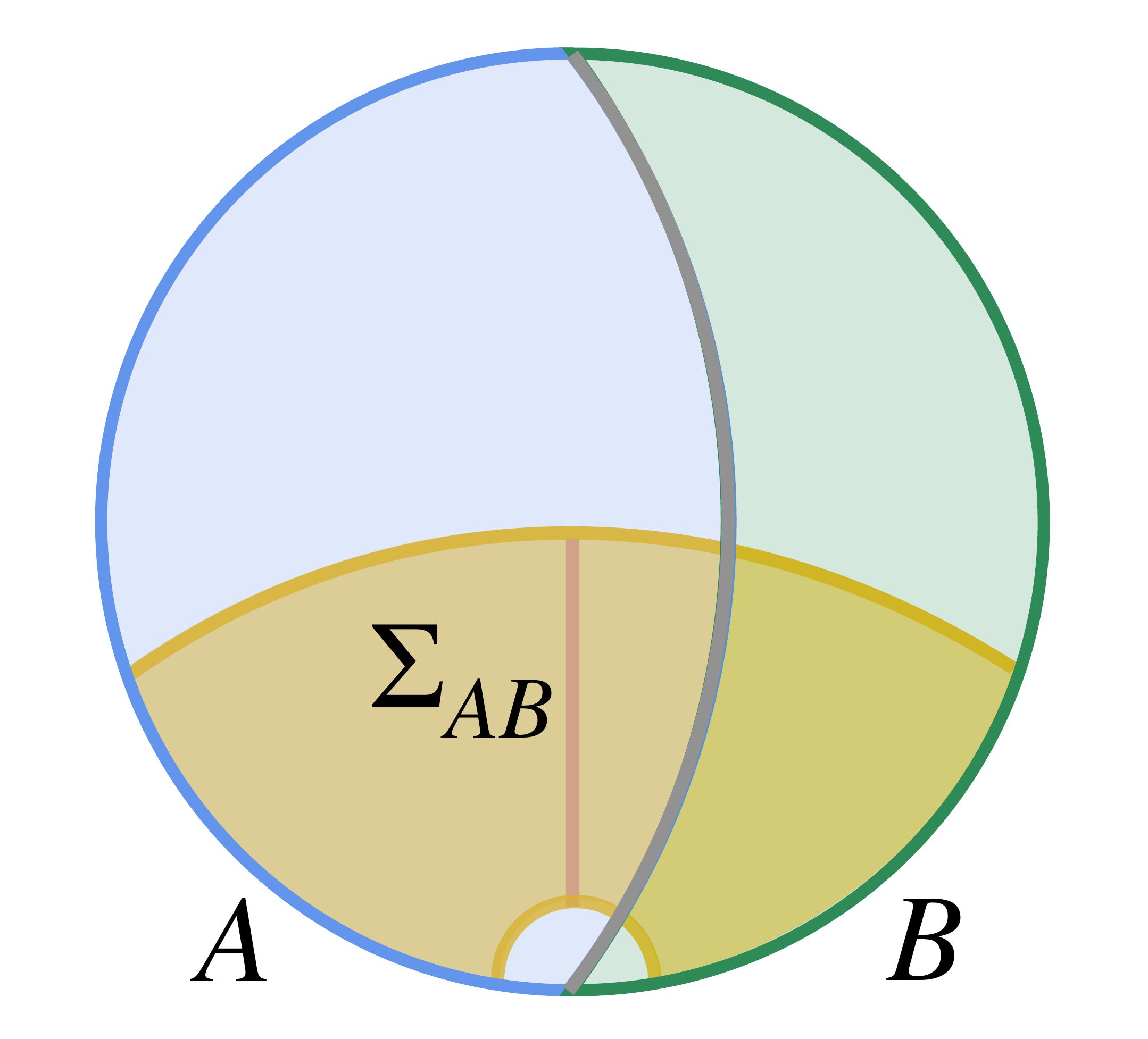}
        \includegraphics[width=7.5cm]{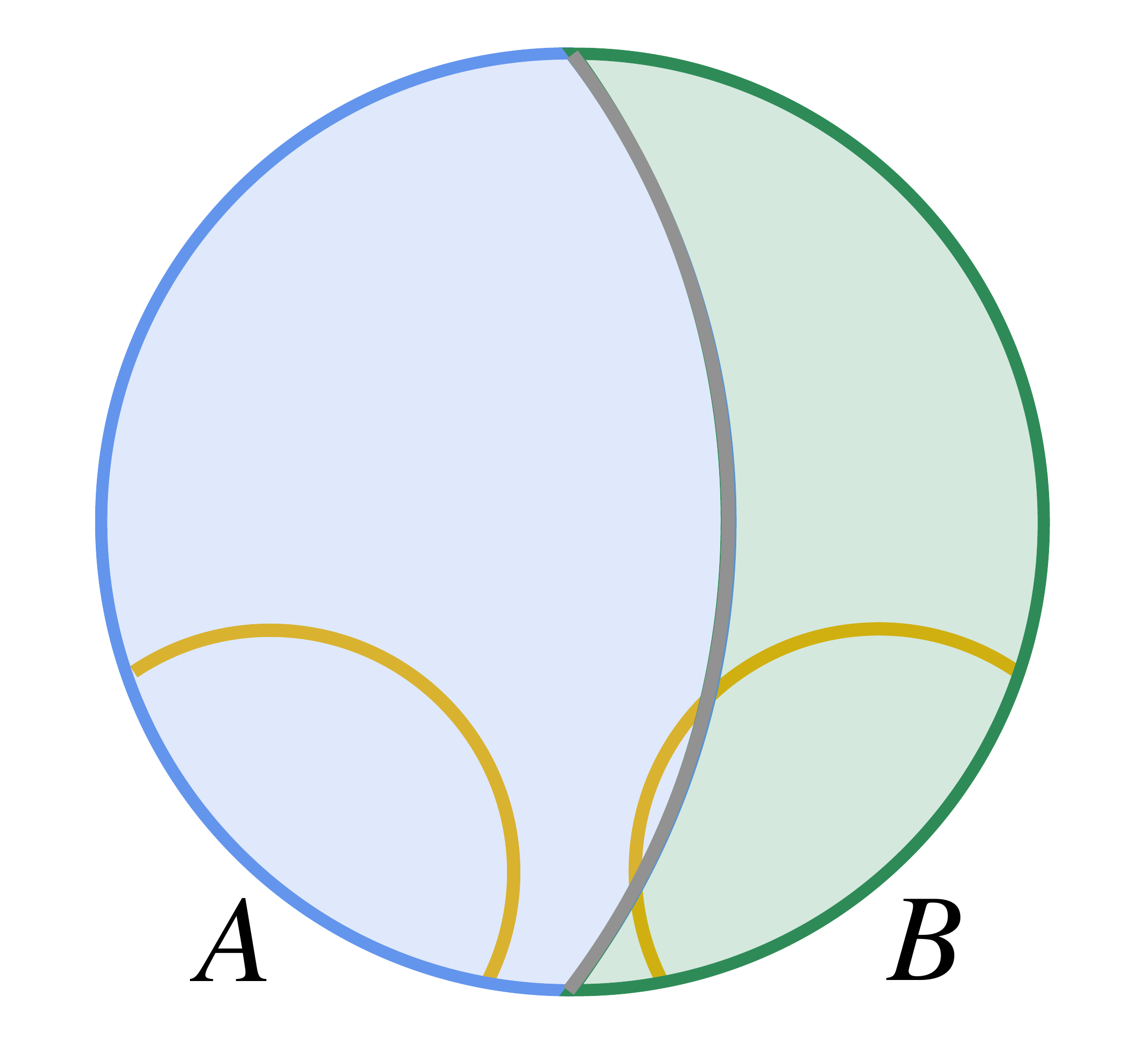}	\caption{(Left) The entanglement wedge of $A\cup B$ is shaded in orange and the minimal $\Sigma_{AB}$ which gives the entanglement wedge cross section is shown in pink. (Right) Both the RT surface of $A$ and that of $B$ are shown in orange. Since the left edge of $B$ is very close to the interface, the RT surface of $B$ goes into $\mathcal{M}^{\rm (I)}$. 
        \label{fig:ICFT_EW}	
 }
\end{figure*}

\paragraph{Reflected entropy and entanglement entropy in symmetric subsystems.} Let us then consider two subsystems $A = \{x|-l\leq x \leq -\epsilon \}$ and $B = \{x| \epsilon \leq x \leq l \}$, where $\epsilon$ is a UV cutoff. These two subsystems are symmetric with respect to the interface at $x=0$. The RT surface of the subsystem $A\cup B$ is composed of two segments: one is the segment given by $\hat{y}^{\rm (I)}_P = \hat{y}^{\rm (II)}_P = \tan\left(\frac{\pi l }{2L}\right)$, and the other is the segment given by $\hat{y}^{\rm (I)}_P = \hat{y}^{\rm (II)}_P = \tan\left(\frac{\pi \epsilon }{2L}\right)$. See Fig.~\ref{fig:ICFT_EW} for a sketch of the entanglement wedge of $A\cup B$. 

Moving $\Sigma_{AB}$ inside of the entanglement wedge, there are two local minima when $\xi_*^{\rm (II)}$ is positive. One sits in $\mathcal{M}^{\rm (I)}$ and is given by $\xi^{\rm (I)} = 0$, and the other one sits in $\mathcal{M}^{\rm (II)}$ and is given by $\xi^{\rm (II)} = 0$. Here, since $\alpha^{\rm (I)} < \alpha^{\rm (II)}$, the global minimum is given by the one sitting at $\xi^{\rm (I)} = 0$. On the other hand, when $\xi_*^{\rm (II)}$ is negative, there is only one local minimum sitting at $\xi^{\rm (I)} = 0$ in $\mathcal{M}^{\rm (I)}$. See Fig.~\ref{fig:ICFT_EW} for a sketch. As a result, in the thin brane model, regardless of the value of $T$, the reflected entropy between $A$ and $B$ in this case is always given by 
\begin{align}
    S^R_{A:B} = 2 E^W_{A:B} = \frac{c^{\rm (I)}}{3} \ln \left(\frac{2L}{\pi\epsilon} \tan\left(\frac{\pi l}{2L}\right) \right). 
\end{align}
Note that the central charge of CFT$^{\rm (II)}$ does not appear in this expression, i.e. the reflected entropy is determined only by the smaller central charge $c^{\rm (I)}$.

\subsection{Finite size effects from holographic perspectives}

\begin{figure*}[h]
	\centering
	\includegraphics[width=8cm]{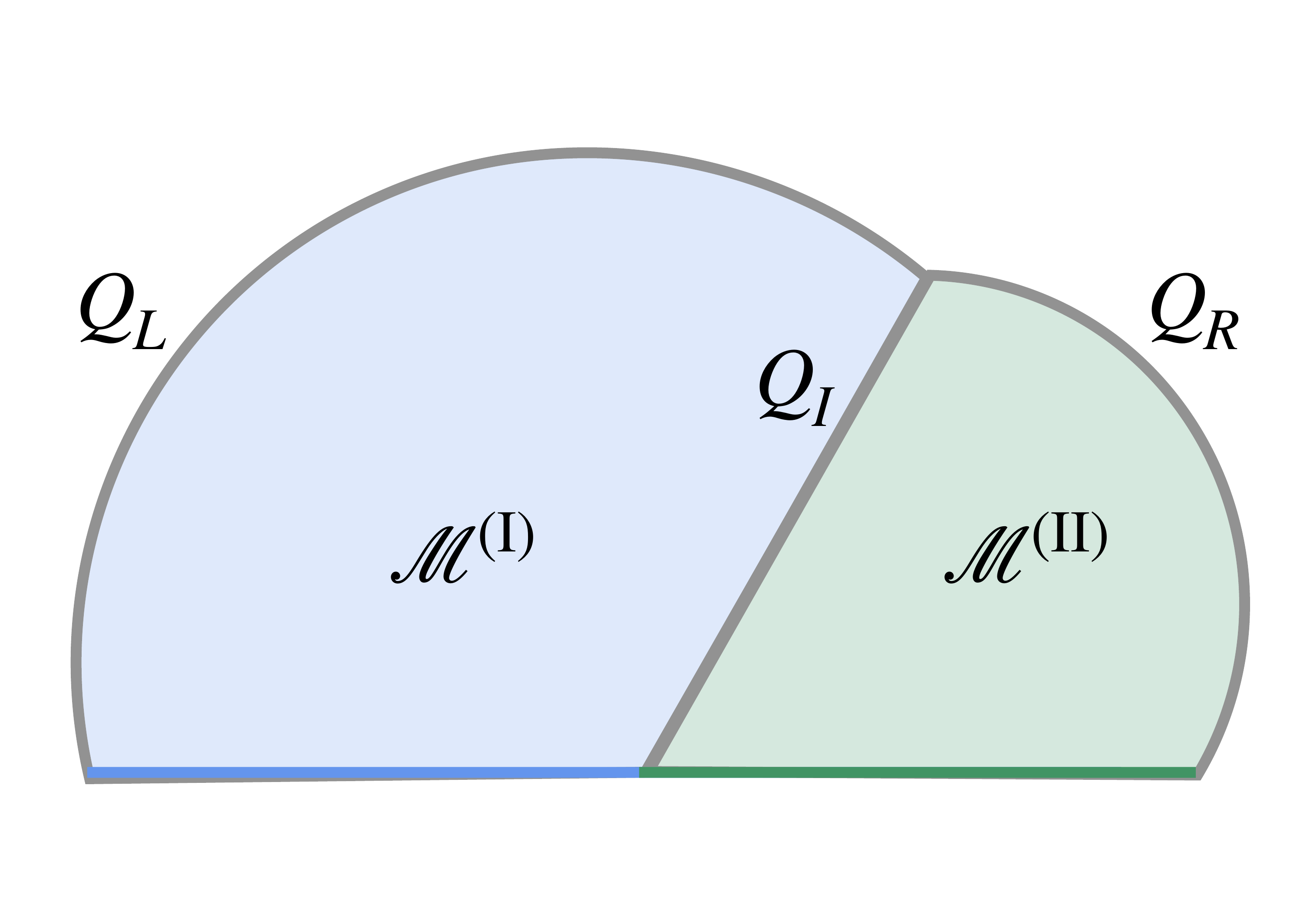}
	\includegraphics[width=8cm]{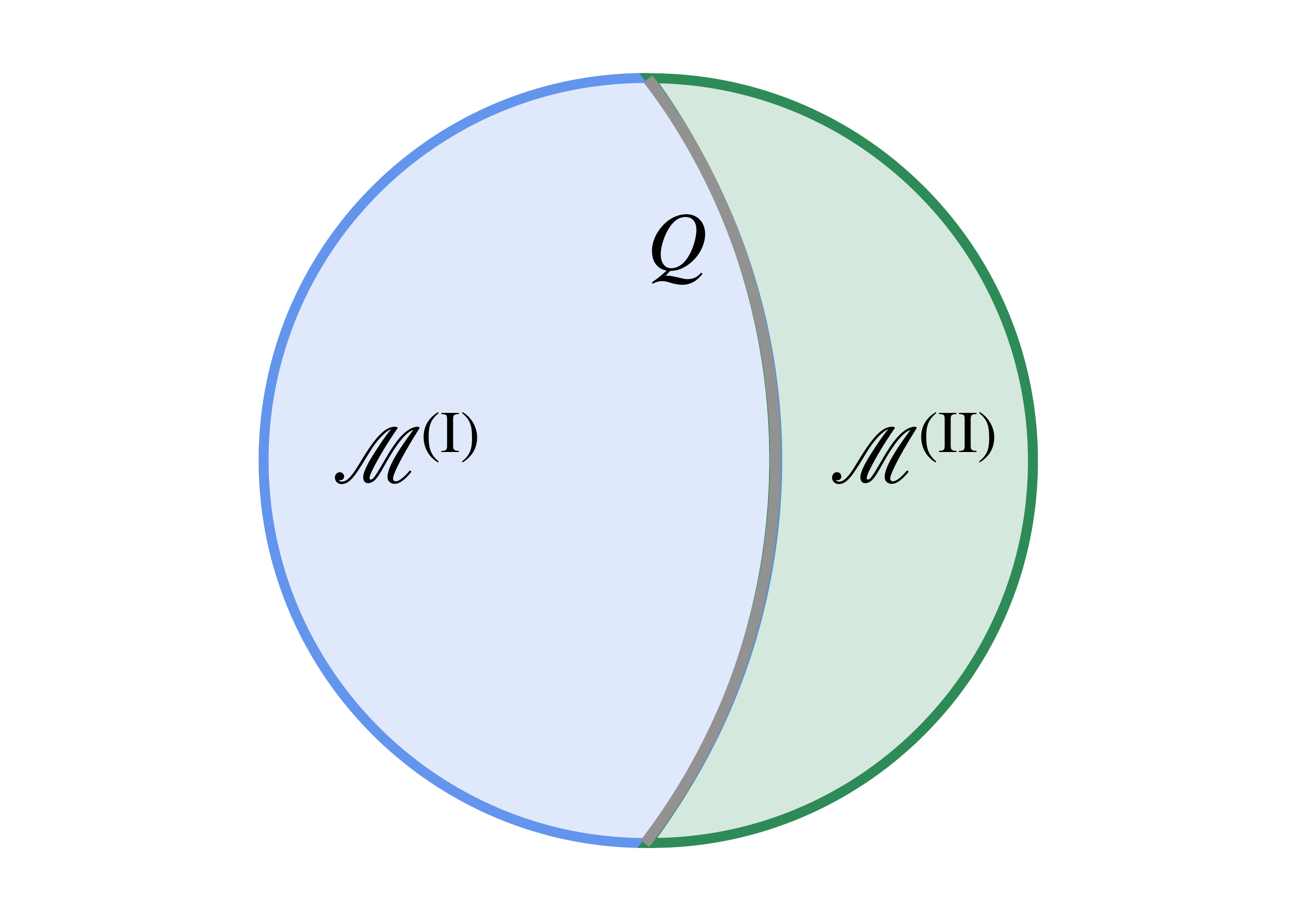}	\caption{(Left) The gravity dual of an ICFT defined on a interval. There are three thin branes extending from the two boundaries and the interface, respectively. The three thin branes nontrivially interact with each other and back reacts on the geometry. (Right) The gravity dual of an ICFT defined on a circle with the periodic boundary condition has only one thin brane in the bulk. 
		\label{fig:interval_circle}	
	}
\end{figure*}

In a homogenous system, the effect of the boundary condition becomes small when the system size goes large. Therefore, it does not matter so much whether we take open boundary conditions or periodic boundary conditions. However, when numerically simulating ground states of ICFTs, we find that taking periodic boundary conditions is much easier than taking open boundary conditions to recover the behavior expected in infinite systems. This indicates that the finite size effect turns out to be more significant when taking the open boundary conditions.

This effect is manifest in the thin brane model of holographic ICFTs. Let us consider an one dimensional ICFT defined on a finite interval, then there exist three thin branes in the gravity dual. As one can see from Fig.~\ref{fig:interval_circle}, two of them extend from the two boundaries of the finite interval and one extends from the interface. Since the thin branes are massive objects, they interact with each other in the bulk spacetime and back react on the background spacetime. As a result, the brane extending from the interface is bended from the original shape and the geometry nearby the interface is greatly affected. On the other hand, as we have already seen in Fig.~\ref{fig:Hol_ICFT}, there is only one thin brane in the bulk if we take the periodic boundary condition and the situation turns out to be much more simple than the open boundary condition case. Therefore, the finite size effect is more significant in the open boundary condition case.

We can also rephrase this holographic explanation in terms of ICFT. When taking open boundary conditions, there are three different line defects in the corresponding ICFT. These three line defects interact with each other through long-range interactions, which greatly affect the physics in the system. On the other hand, if taking periodic boundary conditions, there are only two identical line defects in the corresponding ICFT, corresponding to the two interfaces. Moreover, the apparently two line defects are actually intrinsically a single line defect because we can use a conformal transformation to map the cylinder to a sphere where the two line defects are mapped to the east half and the west half of the equator respectively, and smoothly merge into each other.

\newpage 
\section{CFT derivation of the symmetric EE}

In this appendix, we use a conformal field theory approach to derive the result of symmetric EE in \eqref{eq:symmetic_EE_SM} (\eqref{eq:symmetic_EE} in the main text). 
%(({\blue 4}) in the main text).
This means our analytical results on symmetric EE are not limited to holographic CFTs and apply to general CFTs.

\bigskip
Here we consider the approach as used by Cardy and Tonni in \cite{2016CardyTonni}. As an illustration, we will consider an infinite system first, and then consider a finite system of total length $2L$. 
For the infinite system, we consider a CFT$^{\rm (I)}$ living on $(-\infty,0)$ and a CFT$^{\rm (II)}$ living on $(0,+\infty)$, with a conformal interface inserted along $x=0$.
At zero temperature, the reduced density matrix $\rho_A$ for subsystem $A=[-l,l]$ can be represented as follows:

\begin{equation}
\label{eq:DensityMatrix}
\begin{tikzpicture}[x=0.75pt,y=0.75pt,yscale=-1,xscale=1]
%uncomment if require: \path (0,235); %set diagram left start at 0, and has height of 235

%Shape: Circle [id:dp38559002894856076] 
\draw  [line width=1.0]  (93,110) .. controls (93,107.24) and (95.24,105) .. (98,105) .. controls (100.76,105) and (103,107.24) .. (103,110) .. controls (103,112.76) and (100.76,115) .. (98,115) .. controls (95.24,115) and (93,112.76) .. (93,110) -- cycle ;
%Shape: Circle [id:dp2735217591050322] 
\draw  [line width=1.0]  (184.17,110.5) .. controls (184.17,107.74) and (186.41,105.5) .. (189.17,105.5) .. controls (191.93,105.5) and (194.17,107.74) .. (194.17,110.5) .. controls (194.17,113.26) and (191.93,115.5) .. (189.17,115.5) .. controls (186.41,115.5) and (184.17,113.26) .. (184.17,110.5) -- cycle ;
%Shape: Rectangle [id:dp31251132770802015] 
\draw  [line width=1.0] [color={rgb, 255:red, 255; green, 255; blue, 255 }  ,draw opacity=1 ][fill={rgb, 255:red, 255; green, 255; blue, 255 }  ,fill opacity=1 ] (101,107.33) -- (114.67,107.33) -- (114.67,113) -- (101,113) -- cycle ;
%Shape: Rectangle [id:dp099946135921956] 
\draw  [line width=1.0] [color={rgb, 255:red, 255; green, 255; blue, 255 }  ,draw opacity=1 ][fill={rgb, 255:red, 255; green, 255; blue, 255 }  ,fill opacity=1 ] (172.5,107.5) -- (186.17,107.5) -- (186.17,113.17) -- (172.5,113.17) -- cycle ;
%Straight Lines [id:da15347510519284158] 
\draw   [line width=1.0]  (101,107.33) -- (186.17,107.5) ;
%Straight Lines [id:da925333480451196] 
\draw    [line width=1.0] (101,113) -- (186.17,113.17) ;
%Straight Lines [id:da4452381522489558] 
\draw [line width=1.0] [color={rgb, 255:red, 74; green, 144; blue, 226 }  ,draw opacity=1 ]   (139.67,32) -- (140.33,106.67) ;
%Straight Lines [id:da0587298600352385] 
\draw [line width=1.0] [color={rgb, 255:red, 74; green, 144; blue, 226 }  ,draw opacity=1 ]   (140.33,114) -- (141,188.67) ;
%Straight Lines [id:da11140087753505667] 
\draw    (180,14.33) -- (180.33,34.67) ;
%Straight Lines [id:da37098122615361084] 
\draw    (200,34.33) -- (180.33,34.67) ;
%Straight Lines [id:da11459846032140786] 
\draw    (260.33,109.33) -- (337.33,109.98) ;
\draw [shift={(340.33,110)}, rotate = 180.48] [fill={rgb, 255:red, 0; green, 0; blue, 0 }  ][line width=0.08]  [draw opacity=0] (8.93,-4.29) -- (0,0) -- (8.93,4.29) -- cycle    ;
%Shape: Ellipse [id:dp2891756831946881] 
\draw  [line width=1.0]  (387.33,111) .. controls (387.33,97.56) and (392.18,86.67) .. (398.17,86.67) .. controls (404.15,86.67) and (409,97.56) .. (409,111) .. controls (409,124.44) and (404.15,135.33) .. (398.17,135.33) .. controls (392.18,135.33) and (387.33,124.44) .. (387.33,111) -- cycle ;
%Straight Lines [id:da13733496172603588] 
\draw    [line width=1.0] (398.17,86.67) -- (539,86.67) ;
%Straight Lines [id:da23597682044018908] 
\draw    [line width=1.0] (398.17,135.33) -- (539,135.33) ;
%Shape: Ellipse [id:dp2692021778320167] 
\draw   [line width=1.0] (528.17,111) .. controls (528.17,97.56) and (533.02,86.67) .. (539,86.67) .. controls (544.98,86.67) and (549.83,97.56) .. (549.83,111) .. controls (549.83,124.44) and (544.98,135.33) .. (539,135.33) .. controls (533.02,135.33) and (528.17,124.44) .. (528.17,111) -- cycle ;
%Shape: Ellipse [id:dp17167492383158212] 
\draw  [line width=1.0] [color={rgb, 255:red, 74; green, 144; blue, 226 }  ,draw opacity=1 ][line width=1.2]  (457.75,111) .. controls (457.75,97.56) and (462.6,86.67) .. (468.58,86.67) .. controls (474.57,86.67) and (479.42,97.56) .. (479.42,111) .. controls (479.42,124.44) and (474.57,135.33) .. (468.58,135.33) .. controls (462.6,135.33) and (457.75,124.44) .. (457.75,111) -- cycle ;
%Straight Lines [id:da8510152368952185] 
\draw   (520.67,15.67) -- (521,36) ;
%Straight Lines [id:da9982222657860673] 
\draw    (540.67,35.67) -- (521,36) ;

% Text Node
\draw (82.67,67.07) node [anchor=north west]  {CFT$^{\rm (I)}$};
% Text Node
\draw (142.67,66.07) node [anchor=north west]   {CFT$^{\rm (II)}$};
% Text Node
\draw (179.33,12.4) node [anchor=north west]   {$z$};
% Text Node
\draw (86.67,117.73) node [anchor=north west][inner sep=0.75pt]    {$|a\rangle $};
% Text Node
\draw (182.67,118.4) node [anchor=north west][inner sep=0.75pt]    {$|b\rangle $};
% Text Node
\draw (267.33,85.73) node [anchor=north west][inner sep=0.75pt]    {$w=f( z)$};
% Text Node
\draw (414,103.07) node [anchor=north west][inner sep=0.75pt]    {CFT$^{\rm (I)}$};
% Text Node
\draw (480.67,102.73) node [anchor=north west][inner sep=0.75pt]    {CFT$^{\rm (II)}$};
% Text Node
\draw (520,15.73) node [anchor=north west]    {$w$};
% Text Node
\draw (365.33,103.07) node [anchor=north west][inner sep=0.75pt]    {$|a\rangle $};
% Text Node
\draw (552.67,102.07) node [anchor=north west][inner sep=0.75pt]    {$|b\rangle $};

\end{tikzpicture}
\end{equation}
where the blue solid line corresponds to the conformal interface. 
Here $z=x+i\tau$ with $x$ being the spatial coordinate and $\tau$ being the imaginary time.
The reduced density matrix $\rho_A=\text{Tr}_{\bar A}(\rho)$ is obtained by sewing together the 
degrees of freedom in $\bar A$, and then there is a branch cut along $C=\{i\tau+x| \tau=0, -l\le x\le l \}$.
To introduce regularization, we remove a small disc of radius $\epsilon$ at the entangling point $z=-l+i 0$ and similarly at $z=l+i0$. 
Two conformal boundary conditions $|a\rangle$ and $|b\rangle$ are imposed along these two small discs respectively.
Then the $z$-plane (with two small disks removed) can be mapped to an annulus (where the branch cuts are not shown explicitly) in the $w$-plane after a conformal mapping 
\begin{equation}
w=f(z)=\ln\left(
\frac{z+l}{l-z}
\right).
\end{equation}
The circumference along the periodic $\text{Im}(w)$ direction is $2\pi$, 
and the width of the annulus along the $\text{Re}(w)$ direction is denoted by $W$.
More explicitly, $W=f(l-\epsilon)-f(-l+\epsilon)=2\ln\left(\frac{l}{\epsilon}\right)+\mathcal O(\epsilon)$. 

The $n$-th Renyi entropy of subsystem $A$ is defined by
\begin{equation}
\label{eq:Sn}
S_A^{(n)}=\frac{1}{1-n}\ln\frac{\text{Tr}(\rho_A^n)}{(\text{Tr}\rho_A)^n}=\frac{1}{1-n}\ln\frac{Z_n}{(Z_1)^n},
\end{equation}
Here the partition function $Z_n$ is obtained by gluing $n$ $w$-annulus in \eqref{eq:DensityMatrix} along their branch cuts.
After the gluing, the circumference along the periodic $\text{Im}(w)$ direction becomes $2n\pi$. By considering the Re$(w)$ direction 
in the $w$-annulus as `time', then we have
\begin{equation}
\label{eq:Zn}
Z_n=\langle a |e^{-H^{(n)}_{\text{CFT}^{\rm (I)}}\frac{W}{2}} \hat I e^{-H^{(n)}_{\text{CFT}^{\rm (II)}}\frac{W}{2}}|b\rangle,
\end{equation}
where $\hat I$ denotes the interface operator.
Here $H^{(n)}_{\text{CFT},i}$ are the Hamiltonian of CFT$^{(i)}$ of length $2n \pi$ with periodic boundary conditions.
One can insert complete bases into $Z_n$.
Since $W\gg 1$, only the ground states of CFT$^{\rm (I)}$ and CFT$^{\rm (II)}$ dominate in \eqref{eq:Zn}. Then one can obtain
\begin{equation}
\label{eq:Zn_approx}
Z_n\simeq \langle a |0_{\rm I}\rangle e^{\frac{c^{\rm (I)}}{12n} \frac{W}{2}}\langle 0_{\rm I} | \hat{I}|0_{\rm II}\rangle\langle 0| e^{\frac{c^{\rm (II)}}{12 n} \frac{W}{2}} \langle 0_{\rm II}|b\rangle,
\end{equation}
where $|0_i\rangle$ denotes the ground state of CFT$^{(i)}$ with central charge $c^{\rm (i)}$.
Based on \eqref{eq:Sn} and \eqref{eq:Zn_approx}, one can obtain
\begin{equation}
\begin{split}
S_A^{(n)}\simeq \frac{1+n}{n}\cdot \frac{c^{\rm (I)}+c^{\rm (II)}}{12}\cdot \ln\left(\frac{l}{\epsilon}\right)+\ln(\langle a|0_{\rm I}\rangle)+\ln (\langle 0_{\rm II}|b\rangle)
+\ln (\langle 0_{\rm I}|\hat I|0_{\rm II}\rangle).
\end{split}
\end{equation}
The von-Neumann entropy can be obtained by taking $n=1$. 
The second and third terms correspond to the boundary entropy, and the last term corresponds to the interface entropy.

\bigskip
Now let us come to the case of a finite system of length $2L$ with periodic boundary conditions.
CFT$^{\rm (I)}$ lives on $(-L,0)$ and CFT$^{\rm (II)}$ lives on $(0,L)$.
We glue $x=-L$ and $x=L$ by inserting a conformal interface with the same property as that along $x=0$.
That is, there are two conformal interfaces inserted along $x=0$ and $x=L$ respectively.
Following the previous procedure, we can map the reduced density matrix $\rho_A$ for subsystem $A=[-l, l]$
to an annulus
(where the branch cuts are not explicitly shown):
\begin{equation}
\label{eq:map2}
\begin{tikzpicture}[x=0.75pt,y=0.75pt,yscale=-1,xscale=1]
%uncomment if require: \path (0,235); %set diagram left start at 0, and has height of 235

%Shape: Circle [id:dp38559002894856076] 
\draw  [line width=1.0] (138,110) .. controls (138,107.24) and (140.24,105) .. (143,105) .. controls (145.76,105) and (148,107.24) .. (148,110) .. controls (148,112.76) and (145.76,115) .. (143,115) .. controls (140.24,115) and (138,112.76) .. (138,110) -- cycle ;
%Shape: Circle [id:dp2735217591050322] 
\draw   [line width=1.0](203.17,110.5) .. controls (203.17,107.74) and (205.41,105.5) .. (208.17,105.5) .. controls (210.93,105.5) and (213.17,107.74) .. (213.17,110.5) .. controls (213.17,113.26) and (210.93,115.5) .. (208.17,115.5) .. controls (205.41,115.5) and (203.17,113.26) .. (203.17,110.5) -- cycle ;
%Shape: Rectangle [id:dp31251132770802015] 
\draw [line width=1.0] [color={rgb, 255:red, 255; green, 255; blue, 255 }  ,draw opacity=1 ][fill={rgb, 255:red, 255; green, 255; blue, 255 }  ,fill opacity=1 ] (146,107.33) -- (159.67,107.33) -- (159.67,113) -- (146,113) -- cycle ;
%Shape: Rectangle [id:dp099946135921956] 
\draw [line width=1.0] [color={rgb, 255:red, 255; green, 255; blue, 255 }  ,draw opacity=1 ][fill={rgb, 255:red, 255; green, 255; blue, 255 }  ,fill opacity=1 ] (191.5,107.5) -- (205.17,107.5) -- (205.17,113.17) -- (191.5,113.17) -- cycle ;
%Straight Lines [id:da4452381522489558] 
\draw [line width=1.0][color={rgb, 255:red, 74; green, 144; blue, 226 }  ,draw opacity=1 ]   (171.05,56.5) -- (171.33,106.67) ;
%Straight Lines [id:da0587298600352385] 
\draw [line width=1.0][color={rgb, 255:red, 74; green, 144; blue, 226 }  ,draw opacity=1 ]   (171.33,117.5) -- (171.5,190.75) ;
%Straight Lines [id:da11140087753505667] 
\draw    (237,14.33) -- (237.33,34.67) ;
%Straight Lines [id:da37098122615361084] 
\draw    (257,34.33) -- (237.33,34.67) ;
%Straight Lines [id:da11459846032140786] 
\draw   (260.33,109.33) -- (337.33,109.98) ;
\draw  [line width=1.0][shift={(340.33,110)}, rotate = 180.48] [fill={rgb, 255:red, 0; green, 0; blue, 0 }  ][line width=0.08]  [draw opacity=0] (8.93,-4.29) -- (0,0) -- (8.93,4.29) -- cycle    ;
%Shape: Ellipse [id:dp2891756831946881] 
\draw    [line width=1.0](387.33,111) .. controls (387.33,97.56) and (392.18,86.67) .. (398.17,86.67) .. controls (404.15,86.67) and (409,97.56) .. (409,111) .. controls (409,124.44) and (404.15,135.33) .. (398.17,135.33) .. controls (392.18,135.33) and (387.33,124.44) .. (387.33,111) -- cycle ;
%Straight Lines [id:da13733496172603588] 
\draw    [line width=1.0] (398.17,86.67) -- (539,86.67) ;
%Straight Lines [id:da23597682044018908] 
\draw     [line width=1.0](398.17,135.33) -- (539,135.33) ;
%Shape: Ellipse [id:dp2692021778320167] 
\draw    [line width=1.0](528.17,111) .. controls (528.17,97.56) and (533.02,86.67) .. (539,86.67) .. controls (544.98,86.67) and (549.83,97.56) .. (549.83,111) .. controls (549.83,124.44) and (544.98,135.33) .. (539,135.33) .. controls (533.02,135.33) and (528.17,124.44) .. (528.17,111) -- cycle ;
%Shape: Ellipse [id:dp17167492383158212] 
\draw  [line width=1.0] [color={rgb, 255:red, 74; green, 144; blue, 226 }  ,draw opacity=1 ][line width=0.75]  (457.75,111) .. controls (457.75,97.56) and (462.6,86.67) .. (468.58,86.67) .. controls (474.57,86.67) and (479.42,97.56) .. (479.42,111) .. controls (479.42,124.44) and (474.57,135.33) .. (468.58,135.33) .. controls (462.6,135.33) and (457.75,124.44) .. (457.75,111) -- cycle ;
%Straight Lines [id:da8510152368952185] 
\draw    (520.67,15.67) -- (521,36) ;
%Straight Lines [id:da9982222657860673] 
\draw    (540.67,35.67) -- (521,36) ;
%Shape: Ellipse [id:dp5179818303626411] 
\draw    [line width=1.0](175.05,56.5) .. controls (146.59,56.55) and (123.5,51.75) .. (123.49,45.76) .. controls (123.48,39.78) and (146.55,34.89) .. (175.01,34.83) .. controls (203.48,34.78) and (226.56,39.58) .. (226.57,45.57) .. controls (226.58,51.55) and (203.52,56.44) .. (175.05,56.5) -- cycle ;
%Shape: Ellipse [id:dp7695004341606474] 
\draw    [line width=1.0](175.55,190.5) .. controls (147.09,190.55) and (124,185.75) .. (123.99,179.76) .. controls (123.98,173.78) and (147.05,168.89) .. (175.51,168.83) .. controls (203.98,168.78) and (227.06,173.58) .. (227.07,179.57) .. controls (227.08,185.55) and (204.02,190.44) .. (175.55,190.5) -- cycle ;
%Straight Lines [id:da519522057053101] 
\draw     [line width=1.0](123.49,45.76) -- (123.99,179.76) ;
%Straight Lines [id:da2212037498518955] 
\draw     [line width=1.0](226.57,45.57) -- (227.07,179.57) ;
%Straight Lines [id:da7550280667802713] 
\draw  [line width=1.0][color={rgb, 255:red, 74; green, 144; blue, 226 }  ,draw opacity=1 ] [dash pattern={on 4.5pt off 4.5pt}]  (180.55,35.5) -- (181,168.75) ;
%Straight Lines [id:da95804427759614] 
\draw    [line width=1.0] (146.67,107.17) -- (205.17,107.5) ;
%Straight Lines [id:da9187341169086694] 
\draw    [line width=1.0] (146.67,114.17) -- (205.17,114.5) ;

% Text Node
\draw (125,80.73) node [anchor=north west]   {CFT$^{\rm (I)}$};
% Text Node
\draw (177,81.07) node [anchor=north west]   {CFT$^{\rm (II)}$};
% Text Node
\draw (239.33,14.4) node [anchor=north west]   {$z$};
% Text Node
\draw (131.67,117.73) node [anchor=north west][inner sep=0.75pt]    {$|a\rangle $};
% Text Node
\draw (200.67,117.4) node [anchor=north west][inner sep=0.75pt]    {$|b\rangle $};
% Text Node
\draw (267.33,85.73) node [anchor=north west][inner sep=0.75pt]    {$w=f( z)$};
% Text Node
\draw (410,103.07) node [anchor=north west]   {CFT$^{\rm (I)}$};
% Text Node
\draw (477.67,102.73) node [anchor=north west]   {CFT$^{\rm (II)}$};
% Text Node
\draw (521,15.73) node [anchor=north west]   {$w$};
% Text Node
\draw (365.33,103.07) node [anchor=north west][inner sep=0.75pt]    {$|a\rangle $};
% Text Node
\draw (552.67,102.07) node [anchor=north west][inner sep=0.75pt]    {$|b\rangle $};

\end{tikzpicture}
\end{equation}
where the conformal map is
\begin{equation}
w=f(z)=\ln\left(
\frac{e^{i\frac{2\pi z}{2L}}-e^{-i\frac{2\pi l}{2L}}}{e^{i\frac{2\pi l}{2L}}-e^{i\frac{2\pi z}{2L}}}
\right).
\end{equation}
The width of the $w$-annulus in \eqref{eq:map2} is 
\begin{equation}
W=f(l-\epsilon)-f(-l+\epsilon)=2\ln\left(
\frac{2L}{\pi \epsilon}\,\sin\left(\frac{\pi l}{L}\right)
\right)+\mathcal O(\epsilon).
\end{equation}
Then by repeating the above procedure, one can obtain the $n$-th Renyi entropy 
 \begin{equation}
\begin{split}
S_A^{(n)}\simeq \frac{1+n}{n}\cdot \frac{c^{\rm (I)}+c^{\rm (II)}}{12}\cdot \ln\left(\frac{2L}{\pi \epsilon} \sin\left(\frac{\pi l}{L}\right) \right)
+\ln(\langle a|0_{\rm I}\rangle)+\ln (\langle 0_{\rm II}|b\rangle)
+\ln (\langle 0_{\rm I}|\hat I|0_{\rm II}\rangle).
\end{split}
\end{equation}
Taking $n=1$, one obtains the von-Neumann entropy 
 \begin{equation}
\begin{split}
S_A\simeq  \frac{c^{\rm (I)}+c^{\rm (II)}}{6}\cdot \ln\left(\frac{2L}{\pi \epsilon} \sin\left(\frac{\pi l}{L}\right) \right)
+\ln(\langle a|0_{\rm I}\rangle)+\ln (\langle 0_{\rm II}|b\rangle)
+\ln (\langle 0_{\rm I}|\hat I|0_{\rm II}\rangle),
\end{split}
\end{equation}
where the leading term is the one we give in %({\blue 4}) % 
\eqref{eq:symmetic_EE} 
in the main text.

\bigskip
As a remark, although we are mainly interested in the ground state in this work, it is straightforward to apply the above approach to interface CFTs of an infinite length at finite temperature $\beta^{-1}$. By choosing the subsystem $A=[-l, l]$ with a conformal interface inserted along $x=0$, now there are two length scales $\beta$ and $l$. 
One can find the von Neumann entropy as
 \begin{equation}
\begin{split}
S_A\simeq  \frac{c^{\rm (I)}+c^{\rm (II)}}{6}\cdot \ln\left(\frac{\beta}{\pi \epsilon} \sinh\left(\frac{2\pi l}{\beta}\right) \right)
+\ln(\langle a|0_{\rm I}\rangle)+\ln (\langle 0_{\rm II}|b\rangle)
+\ln (\langle 0_{\rm I}|\hat I|0_{\rm II}\rangle).
\end{split}
\end{equation}
It may be interesting to study the entanglement properties of ICFT at finite temperature with general choices of subsystem $A$ in the future.

\newpage

\section{Numerical calculation of the mutual information and reflected entropy in matrix product states}

In this appendix, we present details of calculating the mutual information $I_{A,B}$ and the reflected entropy $S_{A,B}^R$ for a given many-body wave function $| \psi \rangle$ in terms of matrix product state (MPS). 
Here we will focus on the case of adjacent subsystems $A$ and $B$ with sizes $l_A$ and $l_B$, but an extension to disjoint cases is straightforward. 

\subsection{A standard coarse-grain procedure in the MPS language}

\begin{figure*}\centering
	\includegraphics[width=\columnwidth]{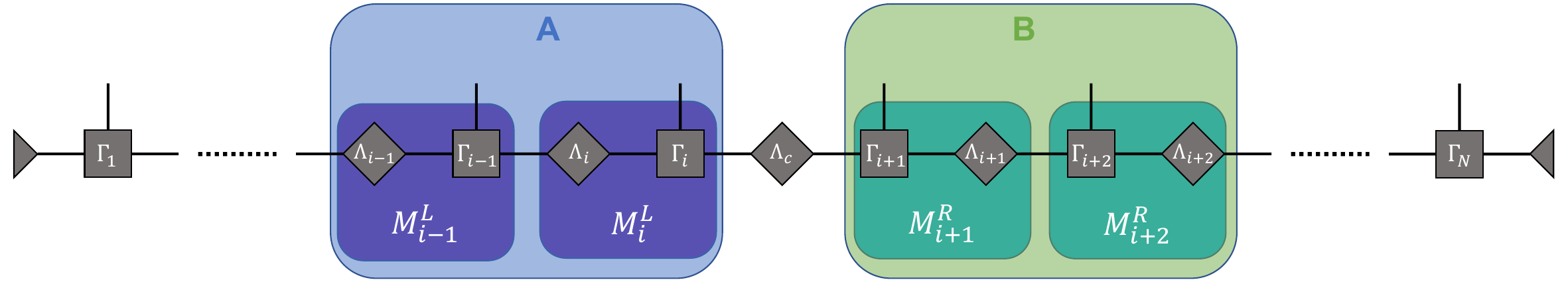}
	\caption{
		\label{fig:MPS_1_app}
		A schematics of an MPS in the canonical form, with two adjacent two subsystem $A$ and $B$ that are separated at the center (the position of $\Lambda_c$) of the MPS. 
		Here the very left/right triangle represents the trivial identity on left/right side, 
		$\Lambda_i$ is the singular value matrix on $i$-th bond, 
		and $\Gamma_i$ is a rank-3 tensor with two virtual (bond) legs and one physical legs. 
		For each physical index (index of the local Hilbert space), $\Gamma_i^{[p_i]}$ is an orthogonal matrix, such that the matrices $M^L = \Lambda \Gamma$ on the left side of the center are in the left canonical form and $M^R = \Gamma \Lambda$ on the right side are in the right canonical form. 
	}
\end{figure*}

As shown in Fig.~\ref{fig:MPS_1_app}, a standard representation of the MPS is defined as the canonical form, for which the target wavefunction has the following expansion
\begin{equation}\label{eq:MPS_canonical_app}
	| \psi \rangle^{[p_1, p_2, \cdots, p_N]} 
	= 
	M_1^{L, [p_1]} \cdots 
	M_{i-1}^{L, [p_{i-1}]} 
	M_i^{L, [p_i]} 
	\Lambda_c 
	M_{i+1}^{R, [p_{i+1}]} 
	M_{i+2}^{R, [p_{i+2}]} 
	\cdots M_{N}^{R, [p_{N}]} .
\end{equation}
Here $M_i^{L/R, [p_i]}$ is the matrix on $i$-th site in the left/right canonical form,
$p_i = 1, 2, \dots, d$ represents the indices of local Hilbert space on $i$-th site, such that each site contains a rank-3 tensor with two virtual (bond) legs $v_{i, L}, v_{i, R}$ and a physical leg $p_i$. 
The form of  \eqref{eq:MPS_canonical_app} represents a Schmidt decomposition that separates left and right parts of the MPS, with a diagonal matrix $\Lambda_c$ of the singular value that located at the ``center'' of the MPS. 
We denote the left and right legs of $\Lambda_c$ by $v_{L, c}, v_{R, c}$. 
The most left and right virtual legs have dimension $1$, so that the product of $N$ matrices for each binary of physical indices $(p_1, p_2, \cdots, p_N)$ is reduced to a number of the wavefunction on the corresponding computational basis. 
The dimensions of virtual legs $\chi_{i, L}$ and $\chi_{i, R}$ of the rank-3 tensor $M_i^{L/R}$ are called bond dimension. 
By definition $\chi_{i, L} = \chi_{i-1, R}$, for convenience we will use the notation of $\chi_{i} = \chi_{i, R}$ and $\chi_{i-1} = \chi_{i, L}$ below. 
An explicit description of a given wavefunction $| \psi \rangle$  requires exponentially growing bond dimensions as $\chi_i = d^{ {\rm{min}} \{ i, N-i \} }$, which is the rank of reduced density matrix with tracing out one side of the $i$-th bond. 
For reducing the computational complexity to an acceptable level, 
in the standard MPS coarse-grain procedure, 
the dimensions of virtual legs are compressed with setting a maximum number of bond dimensions $\chi_{\rm{max}}$. 
For example, in  \eqref{eq:MPS_canonical_app} the global wavefunction $| \psi \rangle$  is represented in a form of Schmidt decomposition. 
By keeping only $\widetilde{\chi}_i = {\rm{min}} \{ {\chi_{\rm{max}}}, d^{ {\rm{min}} \{ i, N-i \} } \}$ largest singular values in $\Lambda_c$ and corresponding modes in $M_{i}$ and $M_{i+1}$, the Schmidt decomposition provides a perfect realization of the aimed compression. 
After applying this approach to each virtual legs with moving the center around all sites, most matrices in the MPS are compressed to have dimensions $\chi_{\rm{max}} \times \chi_{\rm{max}}$, where the finite number $\chi_{\rm{max}}$ controls the accuracy for describing a given wavefunction $| \psi \rangle$ in the MPS form.

\subsection{Coarse-grain of the local wavefunction and reduced density matrix for continuous regions}

\begin{figure*}\centering
	\includegraphics[width=\columnwidth]{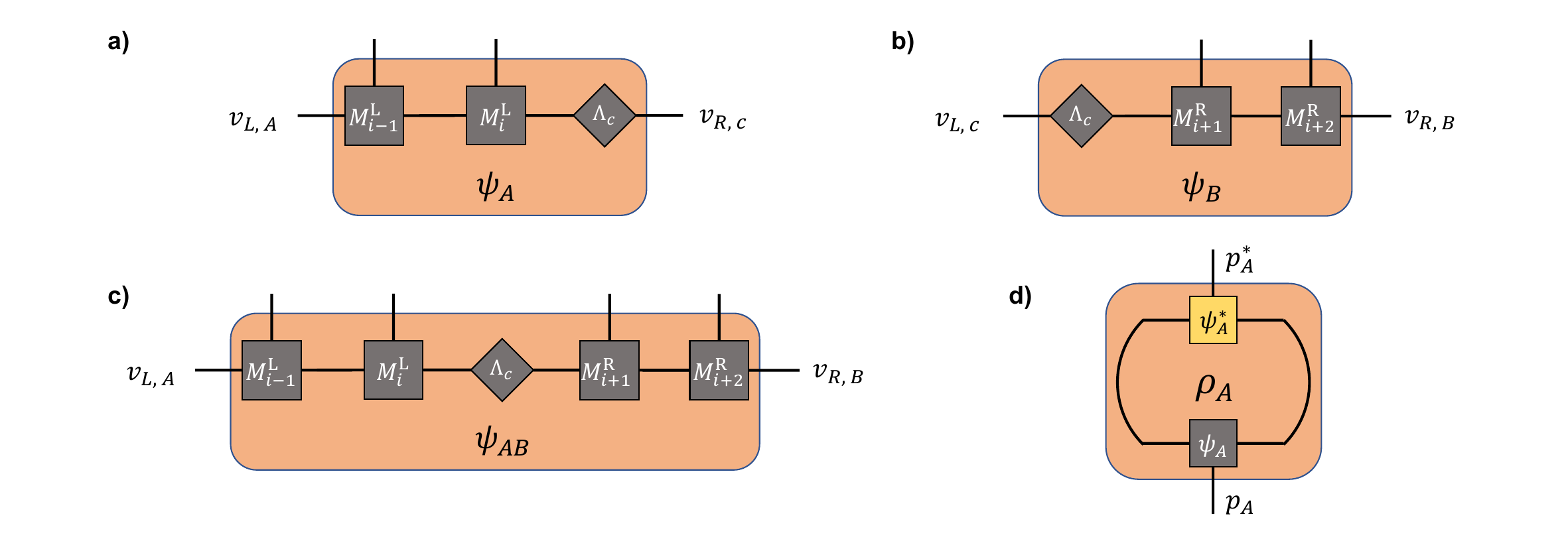}
	\caption{
		\label{fig:MPS_2_app}
		(a-c) The local wavefunctions for two adjacent subsystems $A$ (a), $B$ (b) and their union $AB$ (c).  
		(d) The reduced density matrix $\rho_A$ from the local wavefunction $\psi_A$. 
		Here $A$ and $B$ are assumed to be separated by the center of the MPS, as shown in Fig.~\ref{fig:MPS_1_app}. 
	}
\end{figure*}

\begin{figure*}\centering
	\includegraphics[width=\columnwidth]{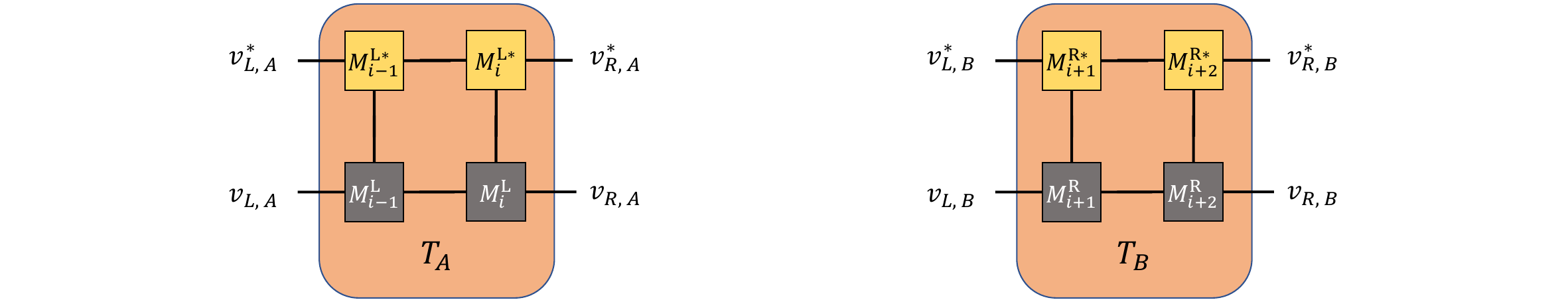}
	\caption{
		\label{fig:MPS_3_app}
		The transfer matrices for subsystem $A$ and $B$. 
		Here the matrices $M$ are assumed to be in the left canonical form in $A$ and in the right canonical form in $B$. 
	}
\end{figure*}

\begin{figure*}\centering
	\includegraphics[width=\columnwidth]{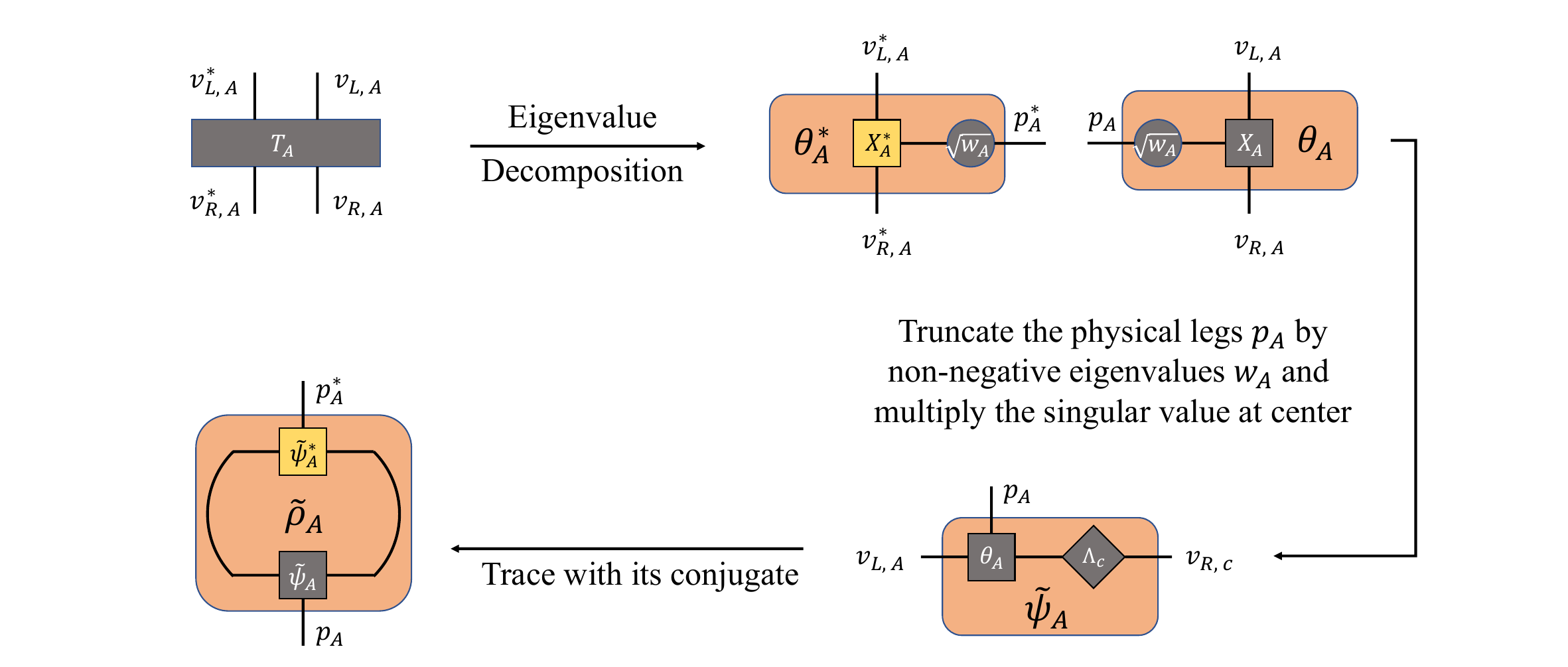}
	\caption{
		\label{fig:MPS_4_app}
		The coarse-grain procedure of local wavefunction $\psi_A$ from the transfer matrix $T_A$. 
		Here we assumed that all matrices in $A$ are in the left canonical form.  
	}
\end{figure*}

Here we are interested in solving the entanglement properties for subsystems that  are located in the middle of the chain. 
Let us start with calculating the reduced density matrix for a continuous subsystem $A$. 
Suppose the MPS is converted to be in the canonical form with a center located at $i$-th site (the most right of $A$) , as shown in Fig.~\ref{fig:MPS_2_app}. 
The local wavefunction of subsystem $A$ is given by a direct product 
\begin{equation}
	| \psi_A \rangle^{ [p_{i-l_A}, p_{i-l_A+1}, \cdots, p_{i}] } 
	= \Tr 
	\psi_A^{ [p_{i-l_A}, p_{i-l_A+1}, \cdots, p_{i}] } 
	= \Tr 
	\left( 
	M^{L, [p_i-l_A]}_{i-l_A} 
	\cdots 
	M^{L, [p_i]}_{i} 
	\Lambda_c 
	\right).
\end{equation}
Here $\psi_A$ is a high-rank tensor with two virtual legs $v_{L, A} = v_{i-l_A}, v_{R, c} = v_{i}$ and $l_A$ physical legs, where $l_A$ is the number of sites in $A$. 
If group all the physical legs into one, then $\psi_A$ is a rank-3 tensor with dimensions $\chi_{L,A} \times d^{l_A} \times \chi_{R,A}$. 
The reduced density matrix can be then obtained by definition $\rho_A = | \psi_A \rangle \langle \psi_A |$, i.e. tracing the left and right virtual legs in $\psi_A$ with its conjugate, as shown in Fig.~\ref{fig:MPS_2_app}(d). 
While the explicit form of $\rho_A$ has exponentially growing dimensions $d^{l_A} \times d^{l_A}$, the rank of $\rho_A$ has been limited to at most $\chi_{\rm{max}}^2$ in the MPS representation. 
To see this, consider the transfer matrix for the continuous subsystem $A$, which is defined as 
\begin{equation}
	T_A  = \sum_{p_{i-l_A+1}, p_{i-l_A+1}, \cdots, p_{i}} 
	\left[ 
	\left( M^{L}_{i-l_A+1} \cdots M^{L}_{i} \right) 
	\left( M^{L}_{i-l_A+1} \cdots M^{L}_{i} \right)^* 
	\right] , 
\end{equation}
where the summation represents a tensor contraction of the physical legs, as show in Fig.~\ref{fig:MPS_3_app}. 
The transfer matrix $T_A$ is a rank-4 tensor with four virtual legs $v_{L, A}, v_{R, A}$ and their conjugate $v_{L, A}^*, v_{R, A}^*$. 
By grouping $(v_{L, A}, v_{R, A})$ and $(v_{L, A}^*, v_{R, A}^*)$, $T_A$ is converted to a matrix with dimensions at most $\chi_{\rm{max}}^2 \times \chi_{\rm{max}}^2$. 
An eigenvalue decomposition of 
\begin{equation}
	T_A = X_A^\dagger w_A X_A = \left( X_A^\dagger \sqrt{w_A} \right)  \left( \sqrt{w_A} X_A \right) = \theta_A^\dagger \theta_A
\end{equation}
leads to an effective description of the local wavefunction as 
\begin{equation}
	\widetilde{\psi}_A = \theta_A \Lambda_c . 
\end{equation}
Here $\widetilde{\psi}_A$ is a rank-3 tensor with dimensions $\chi_{L, A} \times \left( \chi_{L, A} * \chi_{R, A} \right) \times \chi_{R, A}$, where $\left( \chi_{L, A} * \chi_{R, A} \right)$ corresponds to the physical leg of $\widetilde{\psi}_A$ and represents the effective dimension of the local Hilbert space ${\mathcal{H}}_A$. 
Moreover, the eigenvalue decomposition provides a proper way to further compress the dimension of ${\mathcal{H}}_A$. 
As $T_A$ is hermitian, the eigenvalues $w_A$ are real non-negative numbers that allows to sort them and their corresponding modes (eigenvectors). 
Then a truncation can be  made by keeping at most $d_{\rm{max}}$ modes with  largest amplitude, resulting a compressed Hilbert space with dimension $\widetilde{d}_A = {\rm{min}} \{ d_{\rm{max}}, \chi_{L, A} * \chi_{R, A} \}$,  such that  the approximated reduced density matrix  has dimensions $\widetilde{d}_A \times \widetilde{d}_A$ that bounded by a finite number of $d_{\rm{max}}$. 
The above coarse-grain procedure of local wavefunction is summarized in Fig.~\ref{fig:MPS_4_app}. 
This allows an efficient evaluation of the entanglement properties of cutting a subsystem $A$ in the middle of the chain.

\begin{figure*}\centering
	\includegraphics[width=\columnwidth]{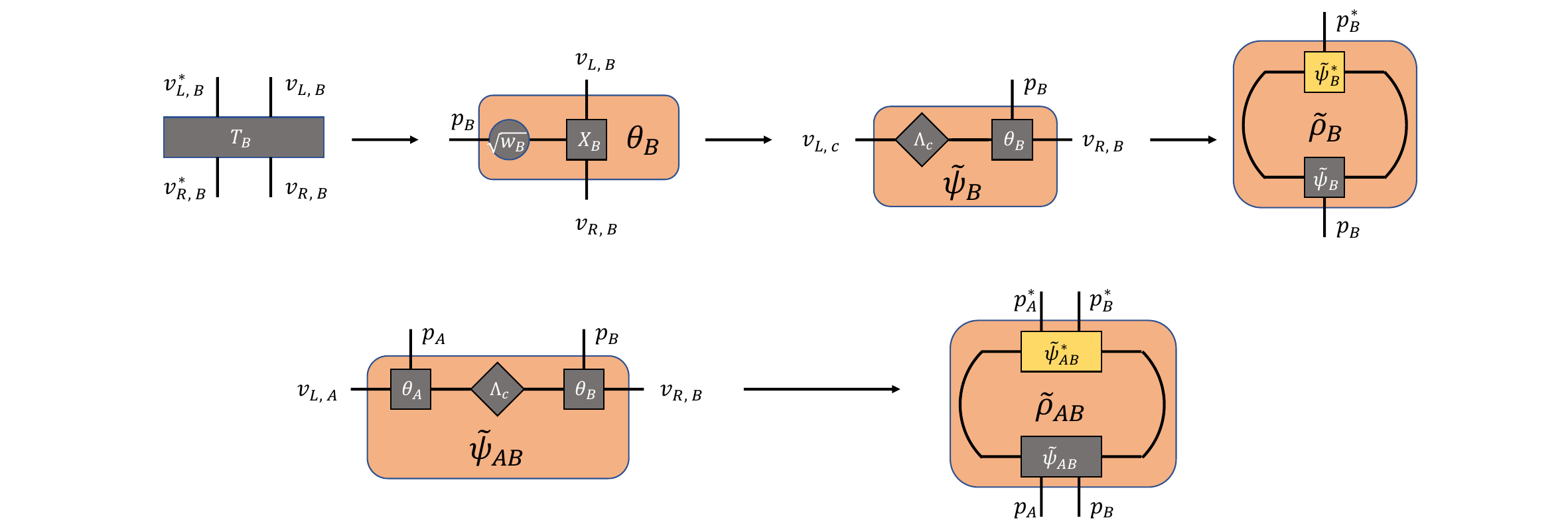}
	\caption{
		\label{fig:MPS_5_app}
		(a) The calculation of approximated reduced density matrix $\widetilde{\rho}_B$ from the transfer matrix $T_B$. 
		(b) A construction of the approximated local wavefunction $\widetilde{\psi}_{AB}$ and reduced density matrix $\widetilde{\rho}_{AB}$ from $\widetilde{\psi}_A$ and $\widetilde{\psi}_B$. 
		Here $A$ and $B$ are assumed to be separated by the center of the MPS, as shown in Fig.~\ref{fig:MPS_1_app}. 
	}
\end{figure*}

\subsection{Calculation of the mutual information}

Now let us move to calculate mutual information $I_{A,B} = S_A + S_B - S_{AB}$ for adjacent two subsystems $A$  and $B$, which requires knowing reduced density matrices $\rho_A$, $\rho_B$ and $\rho_{AB}$. 
Here for convenience, we set the center of the MPS to separate $A$ and $B$, as shown in Fig.~\ref{fig:MPS_1_app}. 
In this case, the calculation of local wavefunction $\widetilde{\psi}_A$ is just the same as we have discussed above. 
For the right subsystem $B$, $\widetilde{\psi}_B$ can be obtained in a similar way as 
\begin{equation}
	\begin{aligned}
		T_B  = \sum_{p_{i+1}, p_{i+2}, \cdots, p_{i+l_B}} 
		\left[ 
		\left( M^{R}_{i+1} \cdots M^{R}_{i+l_B} \right) 
		\left( M^{R}_{i+1} \cdots M^{R}_{i+l_B} \right)^* 
		\right] 
		% \\ & = X_B^\dagger w_B X_B 
		\stackrel{eigenvalue}{\underset{decomposition}{\Longrightarrow}} 
		\left( X_B^\dagger \sqrt{w_B} \right)  \left( \sqrt{w_B} X_B \right) = \theta_B^\dagger \theta_B , 
	\end{aligned}
\end{equation}
and 
\begin{equation}
	\widetilde{\psi_B} = \Lambda_c \theta_B , 
\end{equation}
with dimensions $\chi_{L, B} \times \widetilde{d}_B \times \chi_{R, B}$ after the truncation of $w_B$. 
Meanwhile, we have the joining subsystem $A \cup B$, which can be evaluated as 
\begin{equation}
	\widetilde{\psi}_{AB} 
	= \theta_{A} \Lambda_c \theta_{B} ,
\end{equation}
with dimensions $\chi_{L,A} \times \left( \widetilde{d}_A * \widetilde{d}_B \right) \times \chi_{R,B}$ (here we consider the multiple physical legs are grouped into one $(p_A, p_B) \to (p_{AB})$). 
It should be noticed that, since $A$ and $B$ are adjacent, $A \cup B$ is still a continuous region, for which the effective Hilbert space dimension grows polynomially with the system size for a one-dimensional quantum critical chain. 
In this sense, we can do further compression for $\widetilde{\psi}_{AB}$ as 
\begin{equation}
	\sum_{p_A, p_B} 
	\left[ 
	\left( \theta_A \Lambda_c \theta_B \right) 
	\left( \theta_A \Lambda_c \theta_B \right)^* 
	\right] 
	\stackrel{eigenvalue}{\underset{decomposition}{\Longrightarrow}} 
	\left( X_{AB}^\dagger \sqrt{w_{AB}} \right)  \left( \sqrt{w_{AB}} X_{AB} \right) 
	= \widetilde{\psi}_{AB}^\dagger \widetilde{\psi}_{AB} , 
\end{equation}
where the physical leg of $\widetilde{\psi}_{AB}$ can be truncated to have an effective dimension $\widetilde{d}_{AB} = {\rm{min}} \{ d_{\rm{max}}, \widetilde{d}_A * \widetilde{d}_B \}$. 
Once we get all the targeted local wavefunctions, the approximated reduced density matrices $\widetilde{\rho}_{A/B/AB}$ can be obtained by definition, i.e. tracing the virtual legs of $\widetilde{\psi}_{A/B/AB}$ and its conjugate. 
Their entanglement entropy is 
\begin{equation}
	S_{A/B/AB} = - \sum_{i} \xi_{A/B/AB, i} \ln \xi_{A/B/AB, i} ,
\end{equation}
where $\xi_{A/AB/AB, i}$ are eigenvalues of $\widetilde{\rho}_{A/B/AB}$. 
These lead to the final result of mutual information $I_{A,B}$. 

\subsection{Calculation of the reflected entropy}

\begin{figure*}\centering
	\includegraphics[width=\columnwidth]{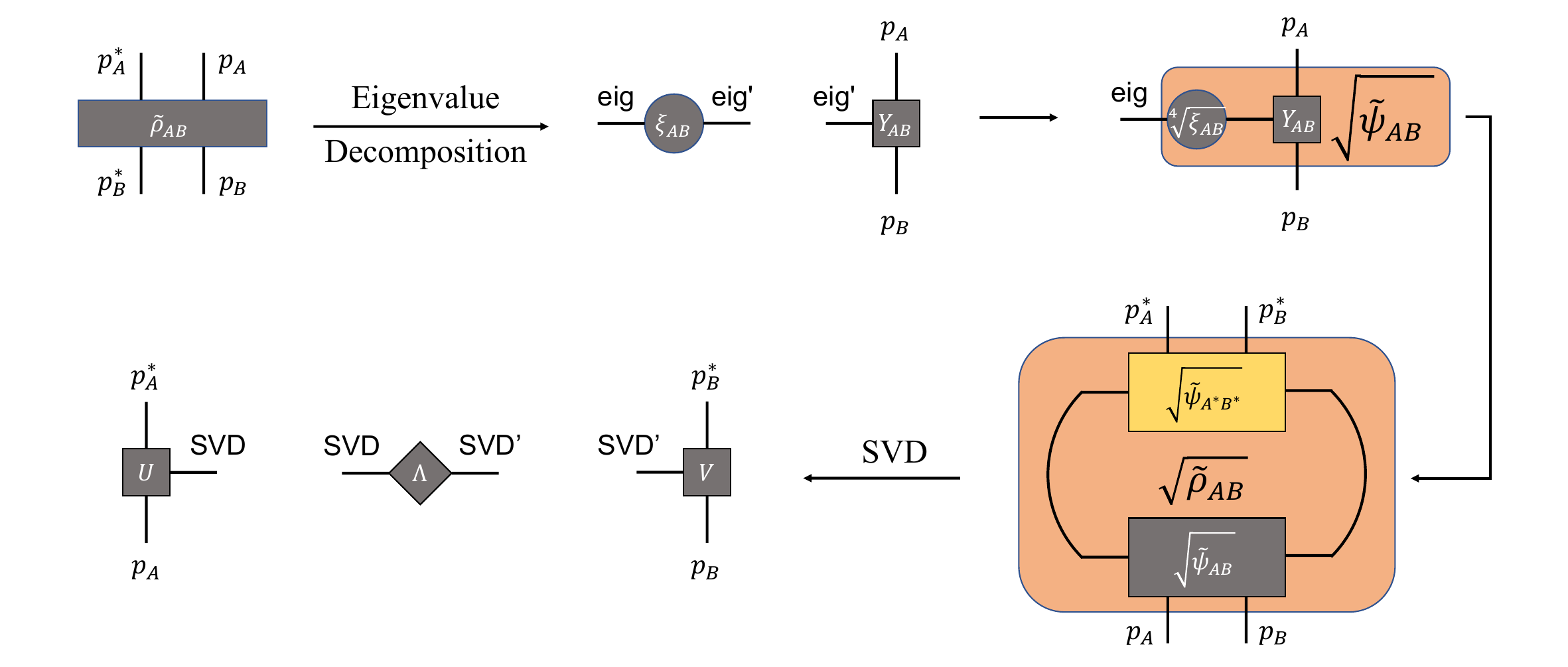}
	\caption{
		\label{fig:MPS_6_app}
		The scheme for calculating the reflected entropy from a given reduced density matrix $\widetilde{\rho}_{AB}$ for two subsystems $A$ and $B$. 
		The reflected entropy can be obtained as $S^R_{A,B} = - \sum_i \Lambda_i^2 \ln \Lambda_i^2 $. 
	}
\end{figure*}

The reflected entropy is defined as the entanglement entropy of separating subsystems $A$ and $B$ for the canonical purification $| \sqrt{\rho_{AB}} \rangle$ of a mixed state $\rho_{AB}$ in a doubled Hilbert space $(\mathcal{H}_{A} \otimes \mathcal{H}_{A}^*) \otimes (\mathcal{H}_{B} \otimes \mathcal{H}_{B}^*)$. 
In this part, we will briefly introduce our method of calculating the reflected entropy from a given MPS, as summarized in Fig.~\ref{fig:MPS_6_app}. 

Suppose we have already obtained the approximated reduced density matrix $\widetilde{\rho}_{AB}$ with dimensions $\widetilde{d}_A \times \widetilde{d}_A \times \widetilde{d}_B \times \widetilde{d}_B$ for the four legs $p_A, p_A^*, p_B, p_B*$. 
After grouping legs $(p_A, p_B)$ and $(p_A^*, p_B^*)$ in $\widetilde{\rho}_{AB}$, applying an eigenvalue decomposition gives 
\begin{equation}
	\widetilde{\rho}_{AB} 
	= Y_{AB}^\dagger \xi_{AB} Y_{AB} . 
\end{equation}
The component of canonical purification $| \sqrt{\rho_{AB}} \rangle$  is the original Hilbert space $(\mathcal{H}_{A} \otimes \mathcal{H}_{B}$ is 
\begin{equation}
	\sqrt{\widetilde{\psi}_{AB}} = \sqrt[4]{\xi_{AB}} Y_{AB} . 
\end{equation}
A product with its conjugate $\sqrt{\widetilde{\psi}_{A^* B^*}}$ leads to the canonical purification 
\begin{equation}
	\sqrt{ \widetilde{\rho}_{AB} } 
	= % \sum_{\text{virtual legs}} 
	\sqrt{\widetilde{\psi}_{A^* B^*}} 
	\sqrt{\widetilde{\psi}_{AB}} ,
\end{equation}
which is a vector in the doubled Hilbert space $(\mathcal{H}_{A} \otimes \mathcal{H}_{A}^*) \otimes (\mathcal{H}_{B} \otimes \mathcal{H}_{B}^*)$. 
By definition, separating $(\mathcal{H}_{A} \otimes \mathcal{H}_{A}^*)$ and $(\mathcal{H}_{B} \otimes \mathcal{H}_{B}^*)$ through a singular value decomposition (SVD) gives 
\begin{equation}
	\left( \sqrt{ \widetilde{\rho}_{AB} } \right)_{AA^* BB^*} 
	= U_{AA*} \Lambda V_{BB^*} .
\end{equation}
The reflected entropy is then obtained as
\begin{equation}
	S^R_{A,B} = - \sum_i \Lambda_i^2 \ln \Lambda_i^2 .
\end{equation}

\subsection{Extension to disjoint subsystems}

\begin{figure*}\centering
	\includegraphics[width=\columnwidth]{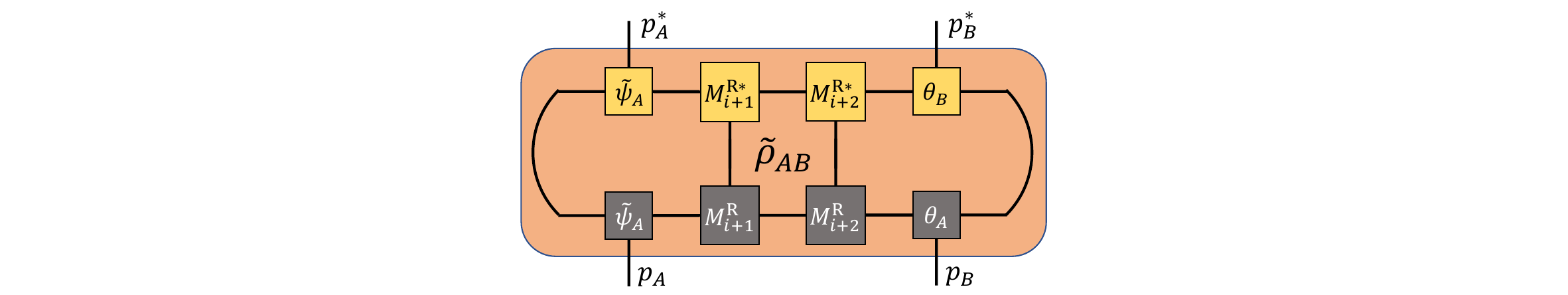}
	\caption{
		\label{fig:MPS_7_app}
		The scheme for calculating the approximated reduced density matrix $\rho_{AB}$ for disjoint two subsystem $A$ and $B$. 
		Here we assumed that the center of the MPS is located at the very right of subsystem $A$. 
	}
\end{figure*}

Up to now, we have discussed the approach to mutual information and reflected entropy for adjacent two subsystems $A$ and $B$ with a given MPS. 
Here we point out  that these discussions can be directly extended to the case of disjoint subsystems. 
The only difference is: when calculating the reduced density matrix for the union $\rho_{AB}$, there are additional matrices that need to be contracted, as shown in Fig.~\ref{fig:MPS_7_app}. 
Any other steps remain the same as the adjacent case.

\clearpage

\section{More numerical results of entanglement properties about the interface}

In this Supplementary Material, we present more numerical results of entanglement properties, including entanglement entropy (EE), mutual information (MI), and reflected entropy (RE), of an interface gluing two distinct CFTs.
Two lattice models are considered in the present work. 
The first one is a spin-$\frac{1}{2}$ model of gluing a transverse Ising chain (realizes the Ising CFT with central charge $c^{\rm (I)} = \frac{1}{2}$) and an O'Brien-Fendley chain (realizes the tricritical Ising CFT with central charge $c^{\rm (II)} = \frac{7}{10}$). 
The second one is a spinless free fermionic model of gluing a Kitaev chain with central charge $c^{\rm (I)} = \frac{1}{2}$ and a tight-binding chain with central charge $c^{\rm (II)} = 1$. 

We explore possible universal information about the interface from various entanglement measures, based on the holographic expectation of a selection rule of the effective central charge as a prefactor of the logarithmic RE~\cite{yuya2022_reflected}. 
Before we present our detailed model, we would like to comment on the challenges we meet: 1) In general, the coupling between glued two distinct gapless theories would introduce a gap, making it difficult to prevent a massive RG flow to a trivial IR. 
2) The (sufficient) condition of approaching the selection rule $c_{\rm{eff}} = {\rm min} \{ c^{\rm (I)}, c^{\rm (II)} \}$, which should be the upper bound of $c_{\rm{eff}}$, is not aware in the context of conformal field theory. 
Under this ground, we consider a CFT$^{\rm (II)}$ with central charge $c^{\rm (II)}$, for which an additional interaction triggers a flow to another fixed point CFT$^{\rm (I)}$ with a finite value of the central charge $c^{\rm (I)}$, and the additional interaction is always irrelevant to CFT$^{\rm (I)}$. (In this paper, we always let $c^{\rm (I)} < c^{\rm (II)}$.)
Apparently, there is a suitable playground: the $A$-series minimal models, for which an RG flow between near-by $A$-models can be triggered by adding $\Phi_{1,3}$ operator~\cite{huse1984_rg_A_series}. 
Well-known examples of $A$-series minimal models include:  Ising CFT $A_{4,3}$, tricritical Ising CFT $A_{5,4}$, and so on. 
Here we adopt the O'Brien-Fendley model 
\begin{equation}
    H_{\rm{OF}} = \sum_{n} 
    H_{\rm{TFI}}(n) + g H_{\rm{int}}(n) 
    = \sum_{n} 
    (\sigma_{n}^{x} \sigma_{n+1}^{x} - \sigma_{n}^{z}) + 
    g (\sigma_{n-1}^{x} \sigma_{n}^{x} \sigma_{n+1}^{z} + \sigma_{n-1}^{z} \sigma_{n}^{x} \sigma_{n+1}^{x}) 
\end{equation}
for realizing Ising ($g < g_c$) and tricritical Ising ($g = g_c$) CFTs as two near-by $A$-models, and the RG flow between them is triggered by modifying the coupling constant $g$ to be less than the value of tricritical Ising fixed point $g = g_c$. 
As expected by CFT, a first-order phase transition to a gapped phase can be triggered by letting $g > g_c$, i.e. changing the sign of the additional $\Phi_{1,3}$ operator. 
The value of $g_c \approx 0.428$ was numerically determined via a two-point correlation function of scaling operators, however, in the case of gluing two half-chain, we find that this value is shifted to be $g_c \approx 0.41$. 
This is fully a lattice effect and does not influence the emergent universal entanglement signatures that are reported in this paper. 

For realizing an interface between Ising and tricritical Ising CFTs, we let the coupling constant $g$ in the O'Brien-Fendley (OF) model be inhomogeneous, i.e.
\begin{equation}\label{eq:Ham_inter_OF_app}
H_{\rm{OF, inter}} = 
\sum_{n \in [-L, L-1]} H_{\rm TFI}(n) 
% \left( \sigma^{x}_{n} \sigma^{x}_{n+1} - \sigma^{z}_{n} \right) 
+ g_L \sum_{n \in [-L, -1]} H_{\rm int} (n) 
+ g_R \sum_{n \in [0, L-1]} H_{\rm int} (n) 
% + g_{\rm inter} \left[ H_{\rm int} (0) + H_{\rm int} (L-1) \right]
\end{equation}
Here the site number ranges from $-L$ to $L-1$ with a total system size of $2L$, and a periodic boundary condition is assumed. 
The inhomogeneous coupling strength ($g_L < g_c$, $g_R = g_c$; if not specified, we will always set $g_L = 0$ for realizing the Ising fixed point) creates an interface bond between $n = -1$ and $n = 0$ sites, there is also another symmetric interface bond between $n = -L$ and $n=L-1$ sites due to the periodic setting, we will referee to these bonds simply as \textit{the interface} in the following discussions.

We then try to extend the construction onto another independent model: a spinless free fermionic model of gluing a real fermion CFT with central charge $c^{\rm (I)} = \frac{1}{2}$) and a complex fermion CFT with central charge $c^{\rm (II)} = 1$ 
\begin{equation}\label{eq:Ham_inter_fermion_app}
\begin{aligned}
    H_{\rm Fermion, inter} 
    & = \sum_{n \in [-L, -1]} H_{\rm RF} (n)
    + \sum_{n \in [0, L-1]} H_{\rm CF} (n) \\ 
    & = \sum_{n \in [-L, -1]} \left( 
    - f^\dagger_n f_{n+1} + \Delta f_n f_{n+1} + h.c. 
    - \mu f^\dagger_n f_n 
    \right) 
    + \sum_{n \in [0, L-1]} \left( 
    - f^\dagger_n f_{n+1} + h.c. 
    \right) \\ 
    & = - \sum_{n \in [-L, L-1]} \left( 
    f^\dagger_n f_{n+1} + h.c. \right) 
    + \Delta \sum_{n \in [-L, -1]} \left( 
    f_n f_{n+1} + h.c. \right) 
    - \mu \sum_{n \in [-L, -1]} f^\dagger_n f_n , 
\end{aligned}
\end{equation}
where we set $\Delta = 1, \mu = 2$ to let the left half-chain be gapless. 
As the model is Gaussian, we can solve its entanglement entropy and mutual information exactly for a large system size with the help of the \textit{correlation matrix} techniques~\cite{Peschel_2003calculation, Peschel_2009reduced}. 
This allows us to perform a large-scale numerical calculation and explore faithful entanglement scaling behaviors that are free of finite-size effects. 
However, we also point out that the reflected entropy of this model can not be straightforwardly obtained from the correlation matrix. 
For this part, we again apply the MPS techniques discussed in the previous Supplementary Material.

Besides, it is generally hard to check whether or not a conformal interface is realized in lattice models. 
For the free fermionic chain with a bond defect, the left- and right-moving wavefunctions can be easily solved, which provide an entrance to the transmission and reflection coefficients for each mode that is labeled by the wavenumber~\cite{Eisler_2012_defect}. 
The conformal/scale invariance can be then understood as the independence of transmission and reflection coefficients on the wavenumber. 
In contrast, when gluing two distinct theories, obtaining such an analytical solution is not an easy task even for free theories (e.g. the junction of a Kitaev chain and a tight-binding model) due to the intrinsic inhomogeneity. 
However, there is several strong evidence of realizing an ICFT. 
First, the EE in the system always exhibits logarithmic scaling and does not saturate to a finite value as increasing the system size. 
Second, in the non-conformal interface case, there is energy loss across the interface, such that one expects the upper bound of effective central charge $c_{\rm{eff}} = {\rm min}\{ c^{\rm(I)}, c^{\rm(II)} \}$ cannot be approached. 
In our numerical simulations, a robust selection rule of $c_{\rm{eff}} = {\rm min}\{ c^{\rm(I)}, c^{\rm(II)} \}$ is observed and appears to be robust under various settings. 
% For example, one can modify the coupling constant (irrelevant to the CFT with a smaller central charge) on the interface with leaving the bulk coupling terms unchanged, to see whether or not the entanglement scaling behaviors are strongly changed.
% 
Below we will provide numerical results for supporting the discussion in the main text and the above.

\subsection{Entanglement properties in the O'Brien-Fendley chain}

In this part, we present some numerical results for entanglement properties in the inhomogeneous O'Brien-Fendley model of  \eqref{eq:Ham_inter_OF_app}, under various settings of the entanglement-cut configuration and the coupling constant in the model. 
For obtaining the ground state MPS, we adopt a bond dimension of $\chi_{\rm max} = 100$. 
As discussed in the previous Supplementary Material, for calculating the entanglement properties for subsystems located in the middle of an MPS, we use the coarse-grain procedure of compressing the reduced density matrix with another cut-off dimension $d_{\rm max} = 100$.

\subsubsection{Bipartite entanglement entropy: subsystem-size dependence}

\begin{figure*}[b]\centering
	\includegraphics[width=\columnwidth]{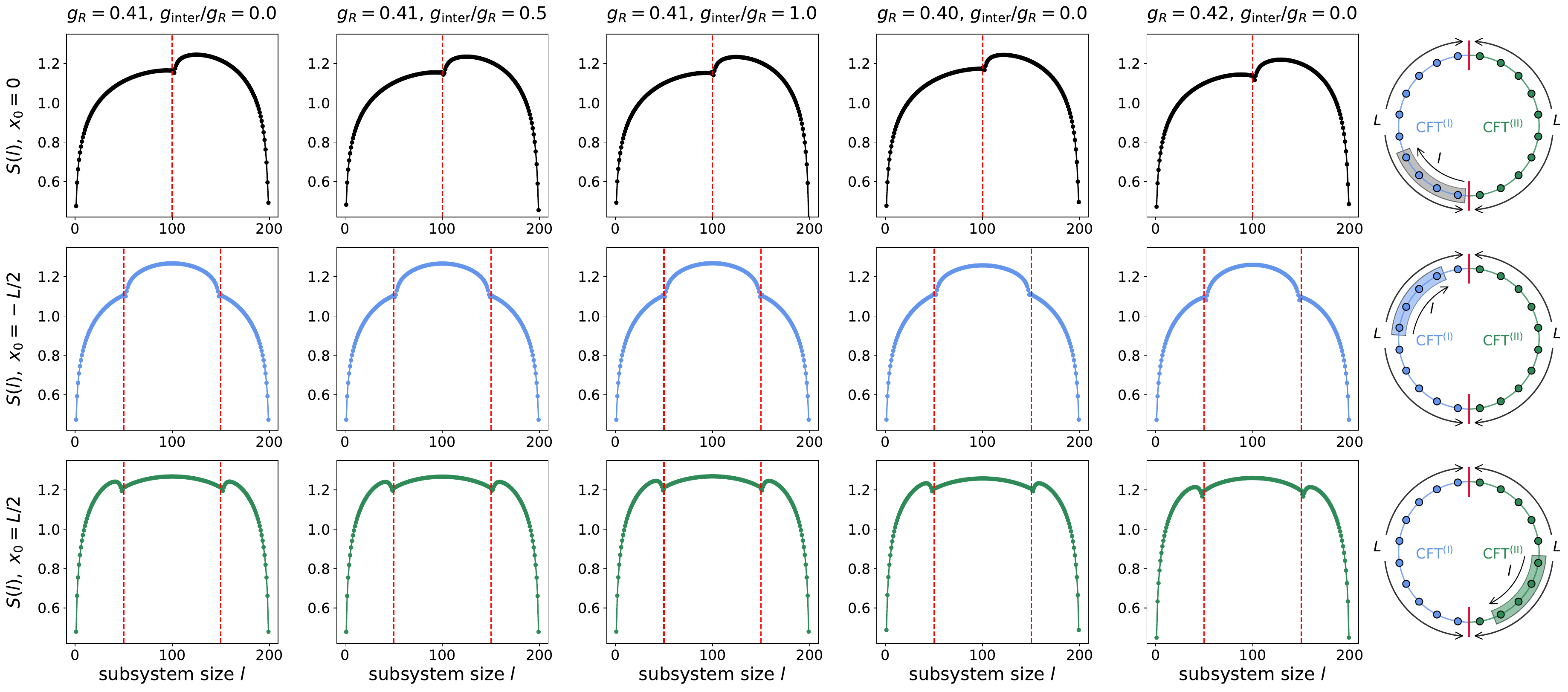}
	\caption{
		\label{fig:EE_sub_dep_OF_app}
        The subsystem-size dependence of EE $S(l)$ in the inhomogenous OF model, for a subsystem with one end located 
        at the interface (upper panel)
        in the middle of CFT$^{\rm (I)}$ (middle panels), 
        and in the middle of CFT$^{\rm (II)}$ (lower panels). 
        The red lines represent the position of the interface between two critical half-chains. 
        Here we choose $L = 100$, such that the length of the total system is $2L = 200$. 
	}
\end{figure*}

Let us begin with the dependence of EE on the subsystem size $l$ with a fixed total system size $L$. 
In Fig.~\ref{fig:EE_sub_dep_OF_app}, we consider: 1. the EE for a subsystem with one end located at the interface; 2. the EE for a subsystem with one end located in the middle of one of the critical half-chain. 
As varying the subsystem size $l$, we observe a dramatic change in the EE across the interface. 
Moreover, in the case of one end located in CFT$^{\rm (II)}$ with a larger central charge, we observe an additional phase when the subsystem size $l$ approaches the position of the interface, exhibiting a non-trivial reduction of EE with increasing the subsystem size. 
This observation is quantitatively consistent with the holographic calculations.

Next, we move to the scaling behavior of EE on the subsystem size $l$. 
One of the most typical entanglement-cut configurations of EE in interface theories is applying two entanglement cuts that are located in the bulk of two half-chains and symmetrically around the interface, dubbed as the \textit{symmetric EE}. 
As the EE is mainly contributed by quantum correlation around the entanglement cut locally, one expects
\begin{equation}
    S_{AB} = \frac{k_{\rm symm}}{6} \ln 2l + b
    = \frac{c^{\rm (I)} + c^{\rm (II)}}{6} \ln 2l + b, 
\end{equation}
where $A$ and $B$ are two adjacent subsystems that touch at the interface, with $l_A = l_B = l$.

Another interesting case is applying one entanglement cut at the interface with varying the subsystem size. 
Similarly, in \cite{Karch2021_interfaceEE}, Karch, Luo, and Sun suggested the following scaling behaviors of EE from a holographic calculation
\begin{equation}\label{eq:KLS_EE_conjecture}
    S_A = \frac{\kappa^{\rm (I)}}{6} \ln l_A + b' = \frac{c^{\rm (I)} + f^{\rm (I)}}{6} \ln l_A + b', \qquad
    S_B = \frac{\kappa^{\rm (II)}}{6} \ln l_B + b'' =  \frac{c^{\rm (II)} + f^{\rm (II)}}{6} \ln l_B + b'', 
\end{equation}
where $A$ and $B$ are two subsystems that adjacent at the interface, $f$ is the interface contribution to the EE. 
Consequently, they proposed the following universal relation
\begin{equation}
    \kappa^{\rm (I)} - \kappa^{\rm (II)} 
    = c^{\rm (I)} - c^{\rm (II)}
\end{equation}
for generic holographic interfaces~\cite{karch_2022_universal_EE_ICFT}.

Nevertheless, as we have discussed in the previous section, even for the simple thin-brane model, the scaling of $S_B$ could be tricky and in general does not have a simple analytical form. 
It is therefore important to test Karch-Luo-Sun's proposal on lattice models.

\begin{figure*}\centering
	\includegraphics[width=\columnwidth]{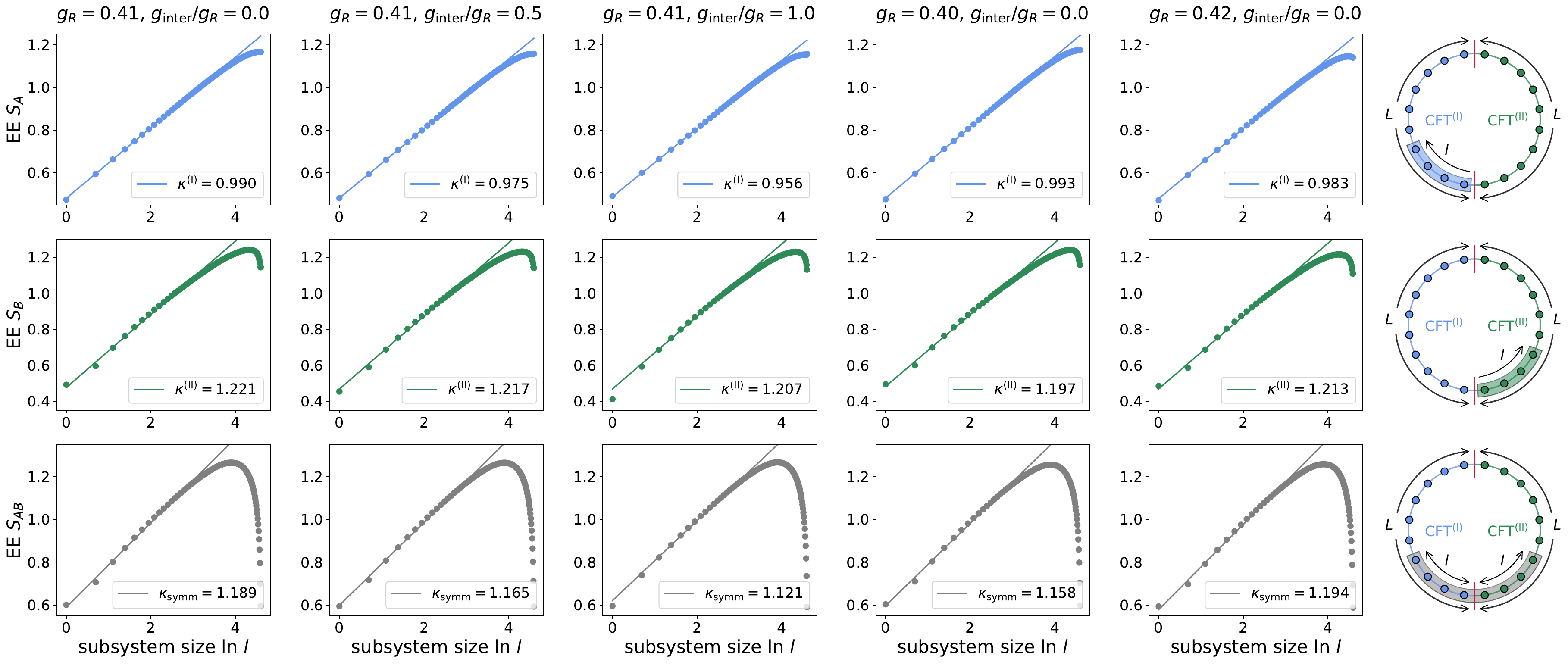}
	\caption{
		\label{fig:EE_sub_scaling_OF_app}
        The subsystem-size dependence of EE in the inhomogenous OF model, under various entanglement-cut configurations. (upper panels) $S_A$ with an entanglement cut at the interface and another entanglement cut in the CFT$^{\rm (I)}$. (middle panels) $S_B$ with an entanglement cut at the interface and another entanglement cut in the CFT$^{\rm (II)}$. (lower panels) The symmetric EE $S_{AB}$. 
        Here we choose $L = 100$, such that the length of the total system is $2L = 200$. 
	}
\end{figure*}

In Fig.~\ref{fig:EE_sub_scaling_OF_app}, we test these expected scaling behaviors by performing numerical calculations on a lattice model. 
For $S_A$ with one end at the interface and another in the bulk of the left half-chain with a smaller central charge, it exhibits a perfect logarithmic dependence on the subsystem size $l$. 
A scaling of $S_A = \frac{\kappa^{\rm (I)}}{6} \ln l$ gives $\kappa^{\rm (I)} \approx 1.0$, indicating that the interface contribution $f \approx 0.5$. 
In contrast, the $S_B$ (with one end at the interface and another in the bulk of the right half-chain with a larger central charge) and the symmetric EE $S_{AB}$ do not have a clear logarithmic dependence of the subsystem size $l$. 
A scaling for small subsystem sizes $l \le L/10$ results in a prefactor larger than the expected value $f + c^{\rm (II)} \approx 1.2$. 

There are two possible reasons for the inconsistency. 
First, we are lacking of a finite-size scaling form of EE. 
In pure CFTs, we have $S = \frac{c}{3} \ln \frac{L}{\pi} \sin (\pi l / L)$ that works well for a finite system with total system size $L$. 
This relation can not be easily extended to interface theories, making it hard to extract possible universal information about the interface. 
Even for the simple thin-brane model considered in the previous section, $S_B$ does not give a simple analytical dependence on $l$. 
Second, the interaction between CFT$^{\rm (I)}$ and CFT$^{\rm (II)}$ may lead to new critical exponents near the interface and modify the quantum correlation across the interface. 
In the later discussion on the free fermionic model of gluing real fermion CFT and complex fermion CFT, we will show that a large system size without interaction gives a chance to capture universal features from bipartite EE.

\begin{figure*}[h]\centering
	\includegraphics[width=\columnwidth]{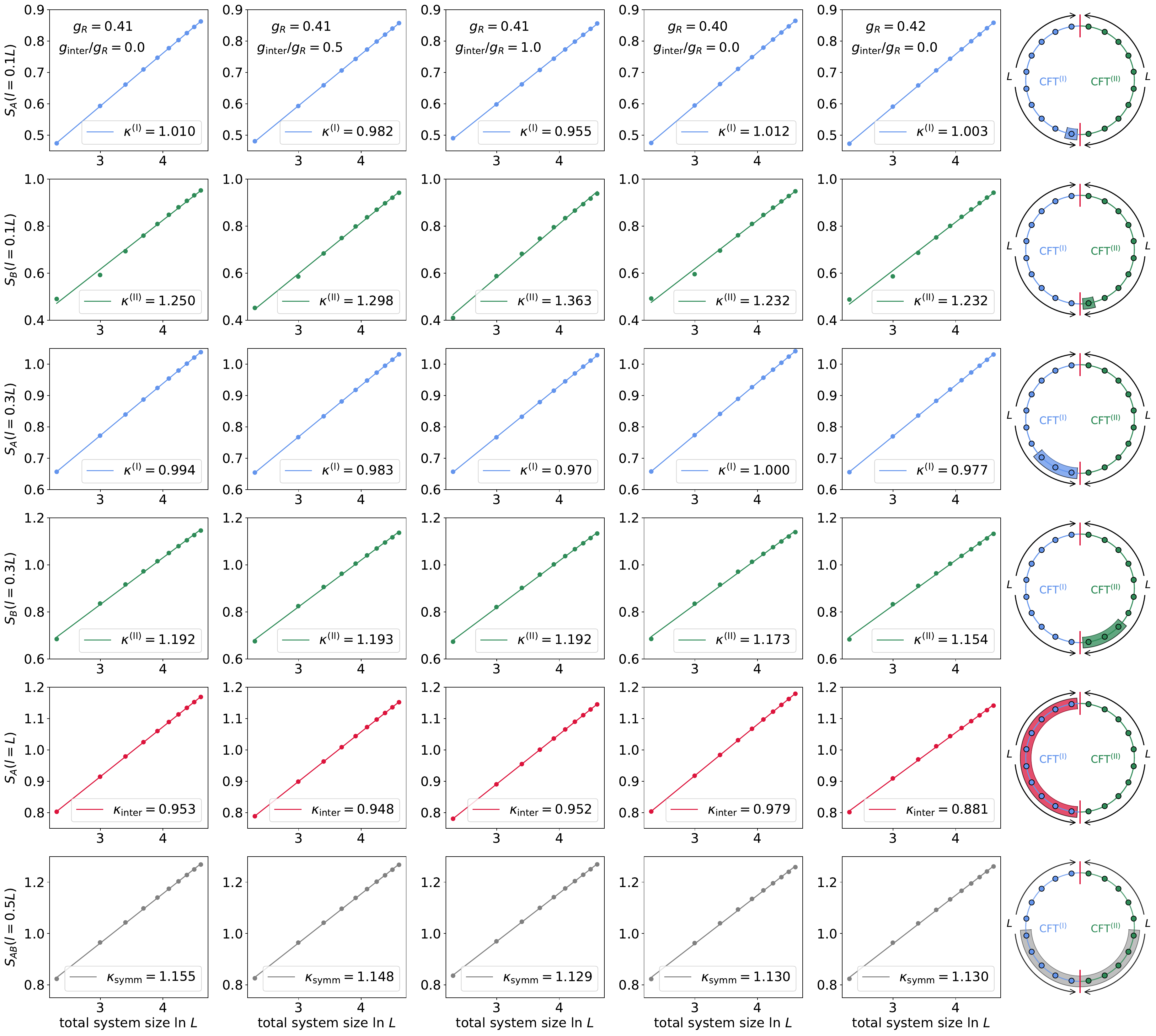}
	\caption{
		\label{fig:EE_total_scaling_OF_app}
        The total-system-size dependence of EE $S_A$, $S_B$, and $S_{AB}$ in the inhomogeneous OF model, under various ratios of $l/L$. 
        Different columns represent data of various settings of the coupling constants $g_R$ and $g_{\rm inter}$. 
        Here we choose $L \in [10, 100]$, such that the length of the total system is $2L \in [20, 200]$. 
	}
\end{figure*}

\subsubsection{Bipartite entanglement entropy: total-system-size dependence}

We have shown that the dependence of EE on the subsystem size $l$ in a near-interface region would be subtle, here we turn to consider the total-system-size dependence with a fixed ratio of $l/L$. 
As shown in Fig.~\ref{fig:EE_total_scaling_OF_app}, under various settings, the EE exhibits a robust logarithmic dependence on the total system size, giving appreciated prefactors of $\kappa^{\rm (I)} \approx 1.0$, $\kappa^{\rm (II)} \approx 1.2$, and $\kappa_{\rm symm} \approx 1.0$. 
It should be noticed that the scaling behavior is somewhat sensitive to the choice of $l/L$. 
For a small ratio of $l/L = 0.1$, the logarithmic dependence of $S_B$ is not so perfect, and the prefactor also has a shift from the expected value of $\kappa^{\rm (II)} \approx 1.2$. 
We also investigate the total-system-size dependence of the symmetric EE $S_{AB}$. 
Although the logarithmic scaling is good, its prefactor is not precise as the expected value of $c^{\rm (I)} + c^{\rm (II)} = 1.2$.

\begin{figure*}\centering
	\includegraphics[width=\columnwidth]{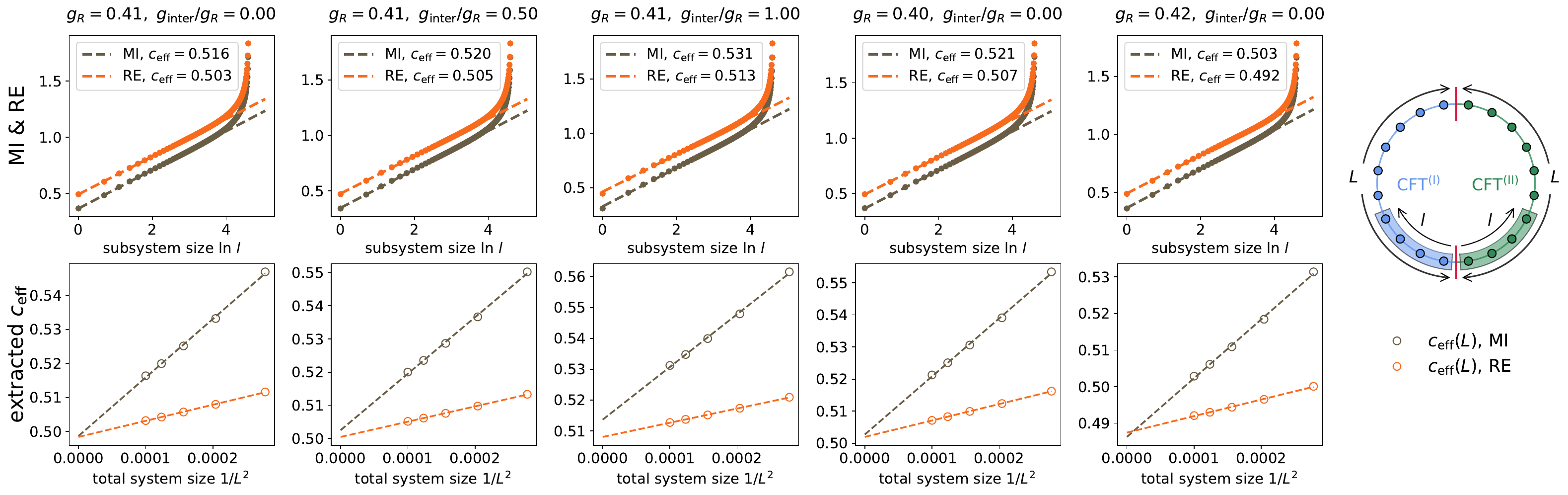}
	\caption{
		\label{fig:MI_RE_OF_app}
        The mutual information (MI) and reflected entropy (RE) in the inhomogenous OF model under various settings of the bulk coupling constant $g_R$ and interface coupling constant $g_{\rm inter}$. 
        (upper panel) The MI and RE as a function of the subsystem size $\ln l$, with fixed total system size $L = 100$. 
        (lower panel) A finite-size scaling of the extracted effective central charge $c_{\rm eff}(L)$ from $\propto \ln l, l \ll L$.  
        The very right panel shows a scheme for the entanglement-cut configuration. 
        Two adjacent subsystems $A$ and $B$ are symmetrically around the interface. 
        The data presented in the main text was $g_R = 0.41, g_{\rm inter} / g_R = 0$. 
	}
\end{figure*}

\begin{figure*}\centering
	\includegraphics[width=\columnwidth]{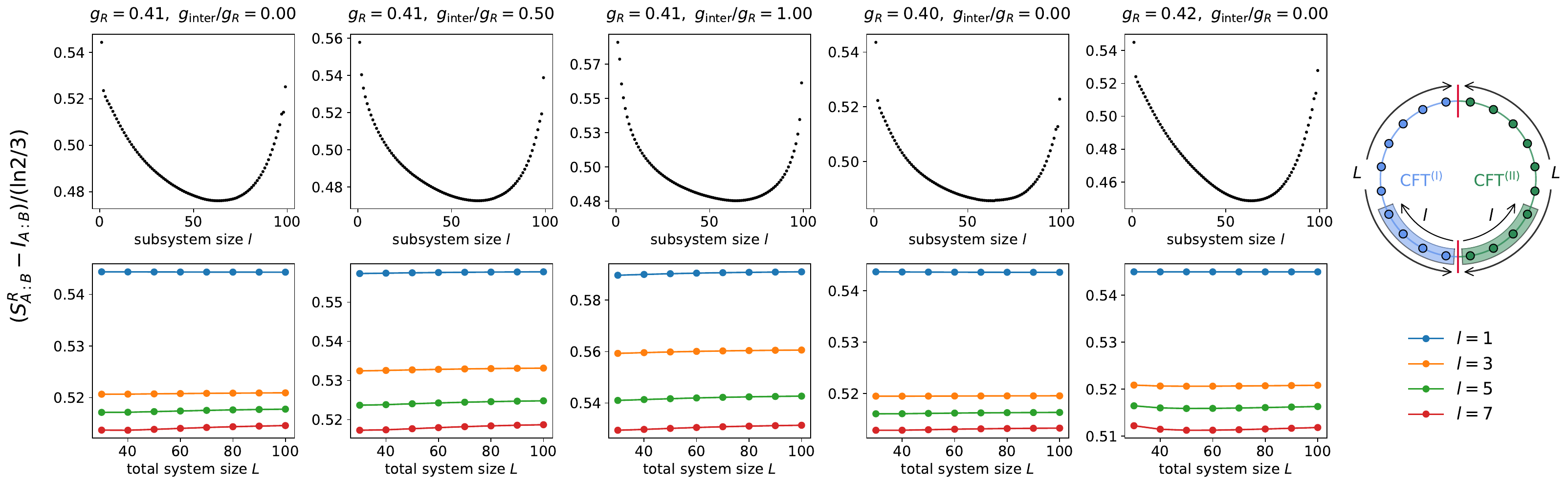}
	\caption{
		\label{fig:Markov_Gap_OF_app}
        The Markov gap (difference between RE and MI $S^R_{A:B} - I_{A:B}$) in the inhomogenous OF model under various settings of the bulk coupling constant $g_R$ and interface coupling constant $g_{\rm inter}$. 
        (upper panel) The Markov gap as a function of the subsystem size $\ln l$, with fixed total system size $L = 100$. 
        (lower panel) A finite-size scaling of the extracted effective central charge $c_{\rm eff}(L)$ from various $l$.  
        The very right panel shows a scheme for the entanglement-cut configuration. 
        Two adjacent subsystems $A$ and $B$ are symmetrically around the interface. 
	}
\end{figure*}

\subsubsection{Mutual information and reflected entropy}

In Fig.~\ref{fig:MI_RE_OF_app}, we present numerical results of the MI and RE of the inhomogenous OF model under various settings. 
All considered parameters of the bulk coupling constant $g_R$ and interface coupling constant $g_{\rm inter}$ exhibit similar scaling behaviors and similar values of extracted effective central charge $c_{\rm eff} \sim {\rm min} \{ c^{\rm (I)}, c^{\rm (II)} \}$.

In pure CFTs, the Markov gap (difference between the reflected entropy and mutual information) is found to be system-size-independent as $S^R_{A:B} - I_{A:B} = \frac{c}{3} \ln 2$. 
Here, an explicit calculation can not be obtained in the context of ICFT, even in holography the Markov gap is complicated. 
Nevertheless, here we try to find signatures of $S^R_{A:B} - I_{A:B} = \frac{c_{\rm eff}}{3} \ln 2$ from the numerical simulations. 
As shown in Fig.~\ref{fig:Markov_Gap_OF_app}, we investigate the Markov gap $S^R_{A:B} - I_{A:B}$ as a function of the total and sub-system size. 
For all considered system sizes, it is closed to the expected value of $(S^R_{A:B} - I_{A:B}) / \frac{\ln 2}{3} \approx 0.5$, but with a complicated dependence on the subsystem size, so that fails to be universal. 
Another notable observation is that, with a given subsystem size $l$, the numerically obtained Markov gap is nearly unchanged with increasing the total system size $L$. 

\subsection{Entanglement properties in the fermionic interface model}

In this part, we present numerical results of entanglement properties in the free fermionic model of gluing a real fermion CFT with central charge $c^{\rm (I)} = \frac{1}{2}$ and a complex fermion CFT with central charge $c^{\rm (II)} = 1$, as shown in  \eqref{eq:Ham_inter_fermion_app}. 
By performing a large-scale numerical simulation based on the correlation matrix technique, we confirm that the EE does not saturate to a finite value as increasing total system size. 
This provides strong evidence that the interface model does not flow to a massive theory. 
Besides, in this non-interacting model, we find that the logarithmic EE (both on subsystem size and total system size) gives a prefactor of $c_{\rm eff}$ closed to the appreciated value of $\text{min}\{ c^{\rm (I)}, c^{\rm (II)} \}$. 
This model could be a good starting point for understanding the emergence of a selection rule of $c_{\rm eff} = \text{min}\{ c^{\rm (I)}, c^{\rm (II)} \}$, which is left to future investigations. 
Below we present the numerical results from an exact solution (for EE and MI) and an approximated solution (for RE). 
Just like the inhomogenous OF model, here we test various settings of the interface coupling constant $\Delta_{\rm inter}$, which is the coupling strength of the pairing term on the interface.

\begin{figure*}[b]\centering
	\includegraphics[width=\columnwidth]{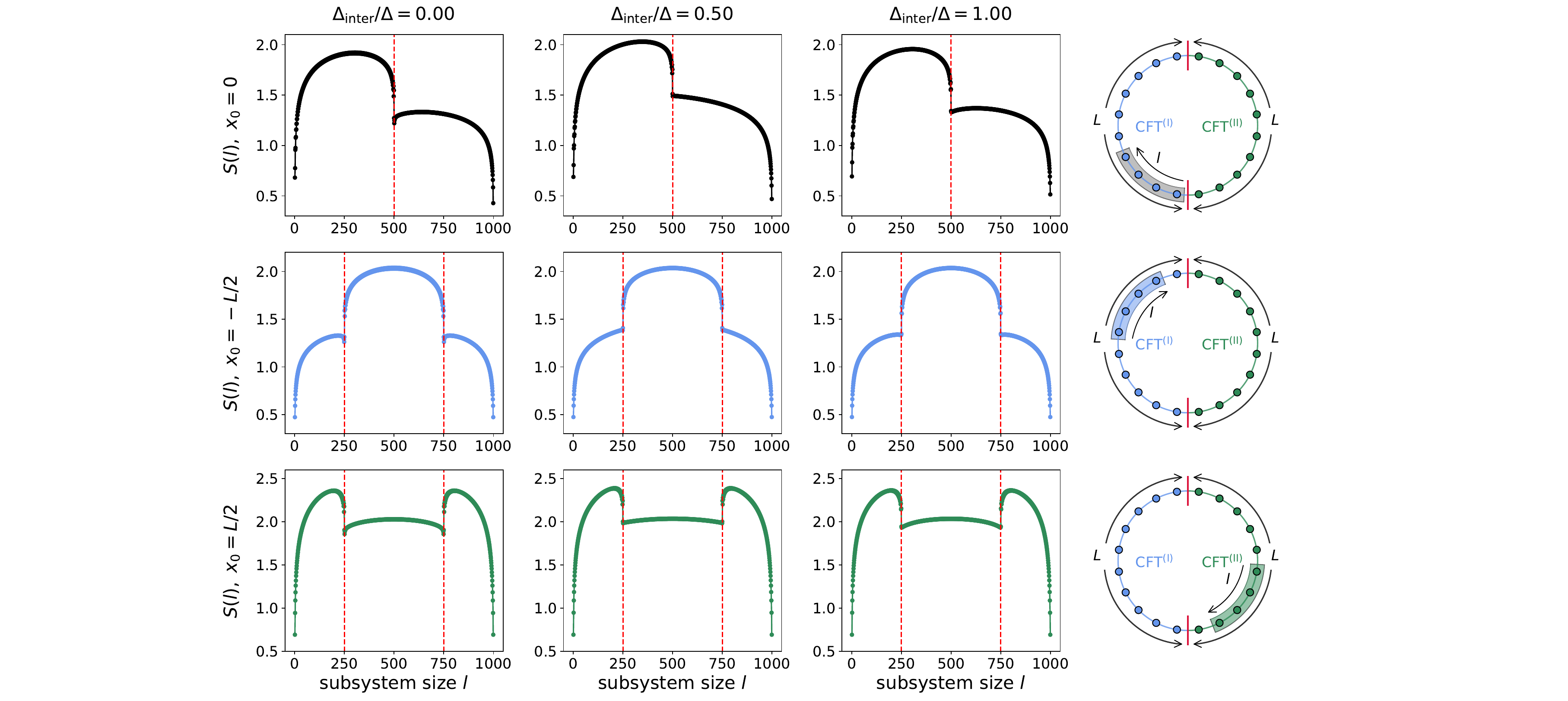}
	\caption{
		\label{fig:EE_sub_dep_Fermion_app}
        The subsystem-size dependence of EE $S(l)$ in the fermionic interface model, for a subsystem with one end located 
        at the interface (upper panel)
        in the middle of CFT$^{\rm (I)}$ (middle panels), 
        and in the middle of CFT$^{\rm (II)}$ (lower panels). 
        The red lines represent the position of the interface between two critical half-chains. 
        Here we choose $L = 500$, such that the length of the total system is $2L = 1000$. 
        The results are calculated exactly by using the correlation matrix technique. 
	}
\end{figure*}

\begin{figure*}\centering
	\includegraphics[width=\columnwidth]{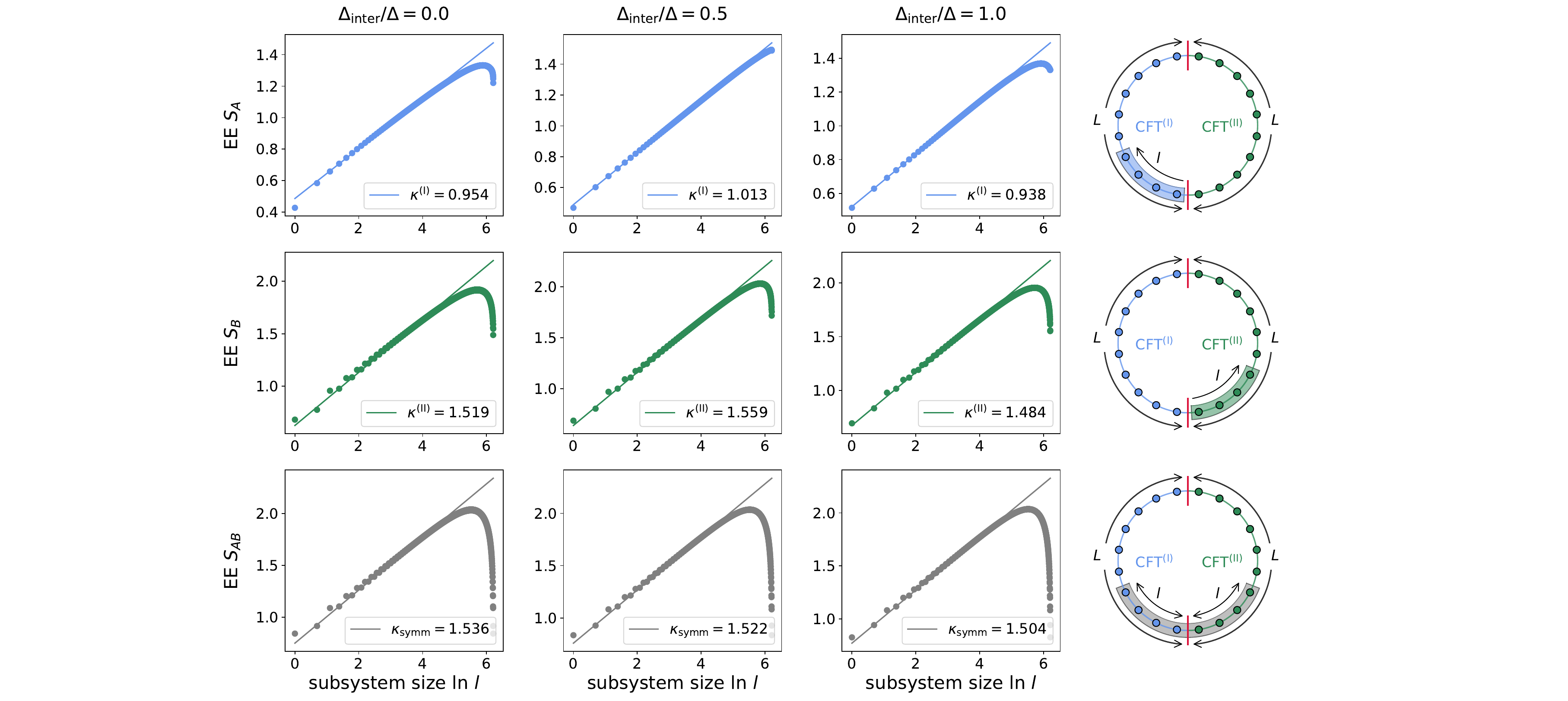}
	\caption{
		\label{fig:EE_sub_scaling_Fermion_app}
        The subsystem-size dependence of EE in the fermionic interface model, under various entanglement-cut configurations. (upper panels) $S_A$ with an entanglement cut at the interface and another entanglement cut in the CFT$^{\rm (I)}$. (middle panels) $S_B$ with an entanglement cut at the interface and another entanglement cut in the CFT$^{\rm (II)}$. (lower panels) The symmetric EE $S_{AB}$. 
        Here we choose $L = 500$, such that the length of the total system is $2L = 1000$. 
        The results are calculated exactly by using the correlation matrix technique. 
	}
\end{figure*}

\subsubsection{Bipartite entanglement entropy: subsystem-size dependence}

In Fig.~\ref{fig:EE_sub_dep_Fermion_app}, we consider 1. the EE for a subsystem with one end located at the interface; 2. the EE for a subsystem with one end located in the middle of one of the critical half-chain. 
The results indicate a change in the bulk central charge across the interface, akin to the inhomogeneous OF model in Fig.~\ref{fig:EE_sub_dep_OF_app}. 
Notably, here the EE has a sudden jump at the interface, which is similar to the holographic interfaces.

For extracting possible universal entanglement signatures about the interface, in Fig.~\ref{fig:EE_sub_scaling_Fermion_app} we present a scaling of $S_A$, $S_B$, and $S_{AB}$ and find a good agreement with the expected forms of $S_A \sim \frac{1}{6} (c^{\rm (I)} + f) \ln l$, $S_B \sim \frac{1}{6} (c^{\rm (II)} + f) \ln l$, and $S_{AB} \sim \frac{1}{6} (c^{\rm (I)} + c^{\rm (II)}) \ln l$, with $f \approx \text{min}\{ c^{\rm (I)}, c^{\rm (II)} \} = 0.5$ approximately. 
% This is consistent with the latter observation of a universal effective central charge $c_{\rm eff} = \text{min} \{ c^{\rm (I)}, c^{\rm (II)} \}$ from mutual information. 
% 
Such a universal-like scaling behavior of bipartite EE is not observed in the inhomogenous OF model. 
In that case, even a logarithmic dependence on the subsystem size is not precise, and we do not observe a signature of improvement with increasing the total system size up to $2L = 300$. 

We would like to point out that, even for the free fermion model considered here, the expected form of  \eqref{eq:KLS_EE_conjecture} does not precisely hold. 
It is possible that the scaling of $S_A$ with $A \in {\rm CFT^{(I)}}$ should be interpreted as $S_A \sim \frac{1}{3} (c^{\rm (I)}) \ln l$ instead of $S_A \sim \frac{1}{6} (c^{\rm (I)} + f) \ln l$, just as suggested by our holographic calculation of the thin-brane model. 
In this sense, it is hard to say our lattice observations suggest the universality of Karch-Luo-Sun's proposal, for which a demonstration needs further theoretical investigations. 
It is also important to test other interacting interface theories. 
We leave these to further investigations.

\begin{figure*}[h]\centering
	\includegraphics[width=\columnwidth]{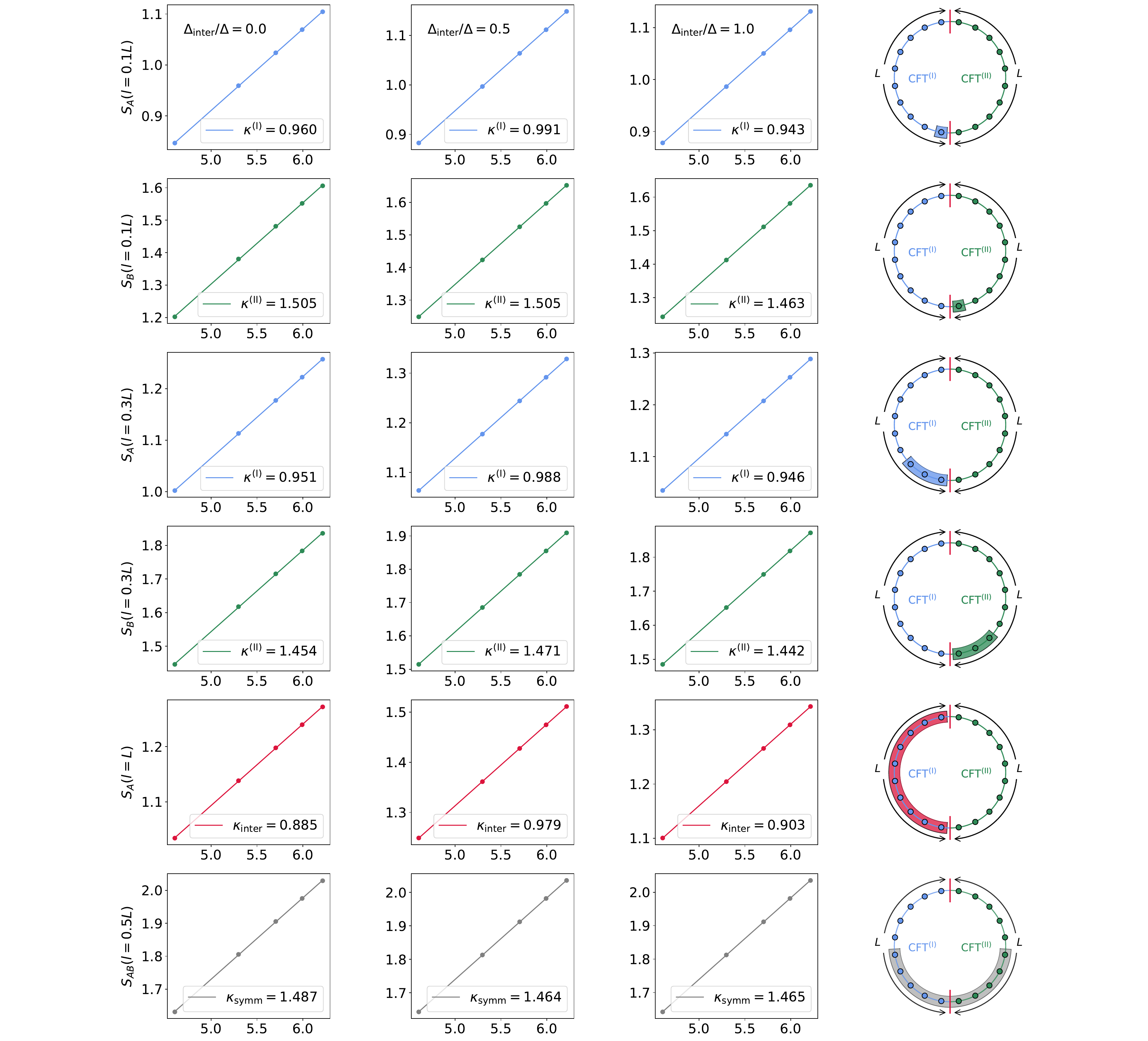}
	\caption{
		\label{fig:EE_total_scaling_Fermion_app}
        The total-system-size dependence of EE $S_A$, $S_B$, and $S_{AB}$ in the inhomogeneous OF model, under various ratios of $l/L$. 
        Different columns represent data of various settings of the coupling constants $g_R$ and $g_{\rm inter}$. 
        Here we choose $L \in [100, 500]$, such that the length of the total system is $2L \in [200, 1000]$. 
        The results are calculated exactly by using the correlation matrix technique. 
	}
\end{figure*}

\subsubsection{Bipartite entanglement entropy: total-system-size dependence} 

In Fig.~\ref{fig:EE_total_scaling_Fermion_app}, we present numerical results of bipartite EE with fixed ratios of $l/L$, which exhibits a logarithmic dependence on the total system size $L$.
It is clear to see that the prefactor of $\kappa_{\rm inter}$ is sensitive to the coupling constant $\Delta_{\rm inter}$ on the interface bond. 
In contrast, the prefactor of $\kappa^{\rm (I)}$, $\kappa^{\rm (II)}$, and $\kappa_{\rm symm}$ is robust under various settings. 
This means that the expectation of $\kappa^{\rm (I)} = c^{\rm (I)} + f$, $\kappa^{\rm (II)} = c^{\rm (II)} + f$, and $\kappa_{\rm inter} = 2f$ is not satisfied, indicating the breaking of a simple picture that the interface contributes locally to the EE.

\begin{figure*}\centering
	\includegraphics[width=\columnwidth]{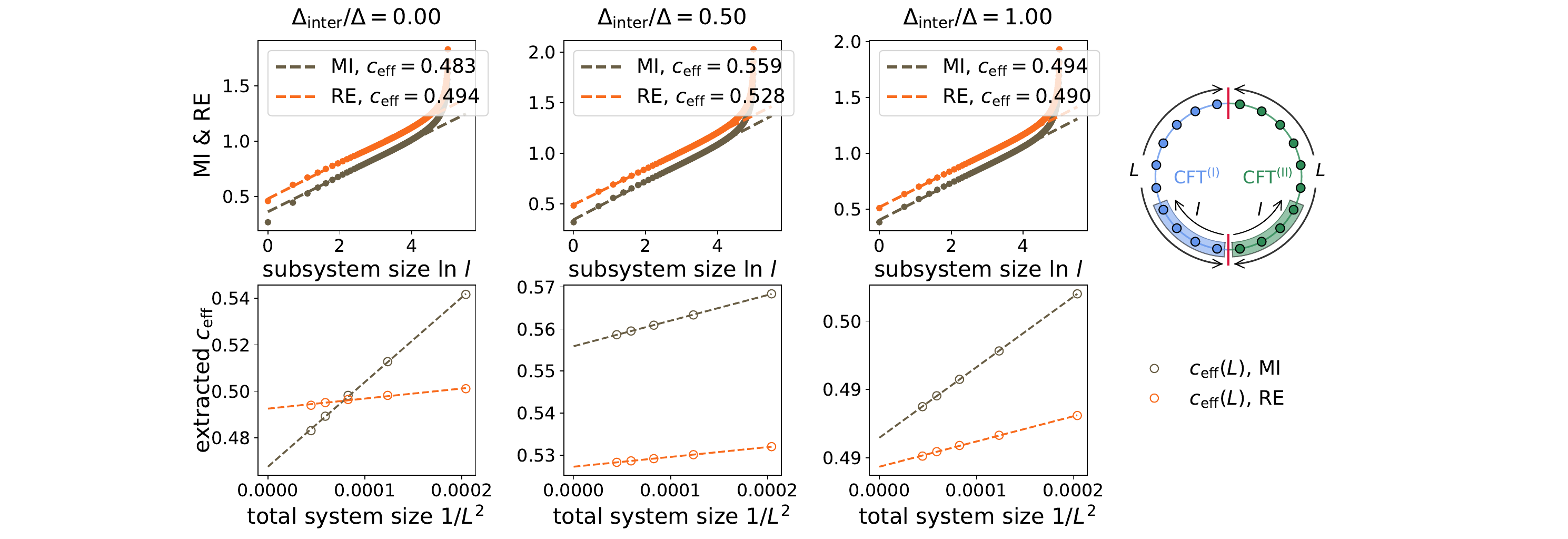}
	\caption{
		\label{fig:MI_RE_Fermion_app}
        The mutual information (MI) and reflected entropy (RE) in the fermionic interface model under various settings of the interface coupling constant $\Delta_{\rm inter}$. 
        (upper panel) The MI and RE as a function of the subsystem size $\ln l$, with fixed total system size $L = 150$. 
        (lower panel) A finite-size scaling of the extracted effective central charge $c_{\rm eff}(L)$ from $\propto \ln l, l \ll L$.  
        The very right panel shows a scheme for the entanglement-cut configuration. 
        Two adjacent subsystems $A$ and $B$ are symmetrically around the interface. 
        The data presented in the main text was $\Delta_{\rm inter} = \Delta$. 
	}
\end{figure*}

\begin{figure*}\centering
	\includegraphics[width=\columnwidth]{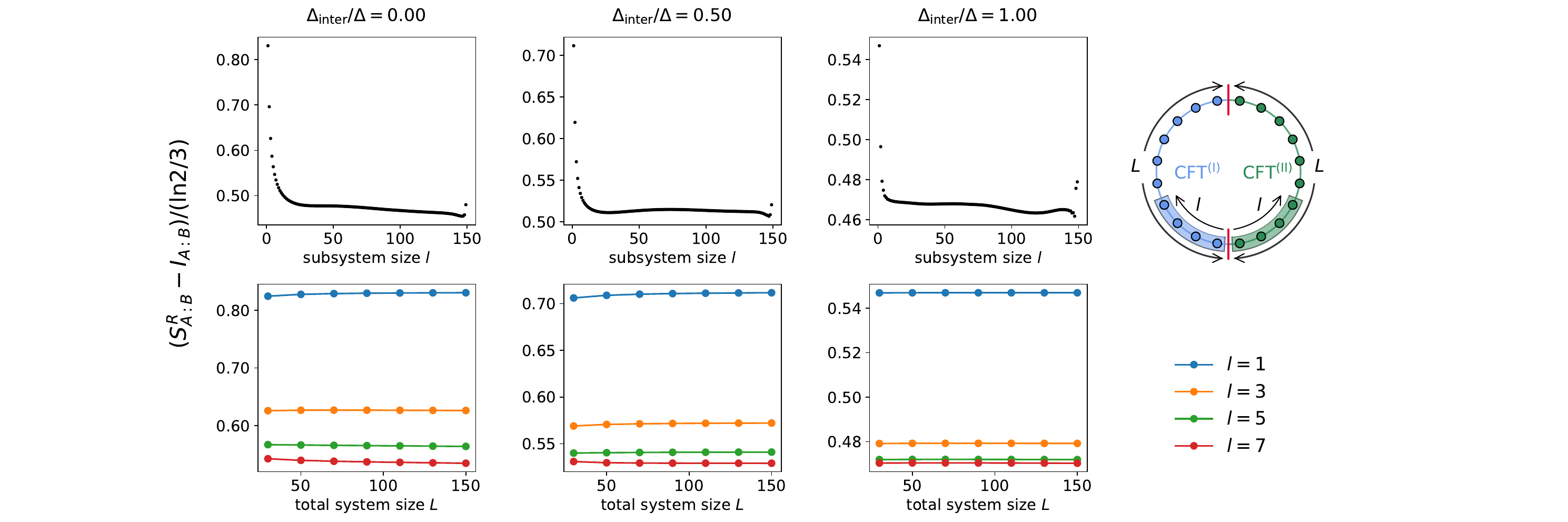}
	\caption{
		\label{fig:Markov_Gap_Fermion_app}
        The Markov gap (difference between RE and MI $S^R_{A:B} - I_{A:B}$) in the fermion interface model under various settings of interface coupling constant $\Delta_{\rm inter}$. 
        (upper panel) The Markov gap as a function of the subsystem size $\ln l$, with fixed total system size $L = 150$. 
        (lower panel) A finite-size scaling of the extracted effective central charge $c_{\rm eff}(L)$ from various $l$.  
        The very right panel shows a scheme for the entanglement-cut configuration. 
        Two adjacent subsystems $A$ and $B$ are symmetrically around the interface. 
	}
\end{figure*}

\subsubsection{Mutual information and reflected entropy}

In Fig.~\ref{fig:MI_RE_Fermion_app}, we present the numerical results of MI and RE in the fermionic interface model by using the MPS coarse-grain procedure discussed in the previous section. 
For the case of $\Delta_{\rm inter} = \Delta = 0, 1$, the logarithmic scaling of both MI and RE gives a prefactor of $c_{\rm eff}$ close to $\text{min} \{ c^{\rm (I)}, c^{\rm (II)} \} = 0.5$. 
For the case of $\Delta_{\rm inter} / \Delta = 0.5$, the MI gives a larger prefactor of $c_{\rm eff} \approx 0.559$ and the RE gives $c_{\rm eff} \approx 0.528$, indicating that the RE would be more robust under various settings of microscopic details. 
We also investigate the dependence of the Markov gap (difference between RE and MI). 
Akin to the inhomogeneous OF model, for a given subsystem size $l$, the value of the Markov gap is nearly unchanged with varying the total system size. 
For a fixed total system size $2L=300$, $(S^R_{A:B} - I_{A:B})/\frac{\ln 2}{3}$ has a larger value For small subsystem size $l$ and becomes stable for most subsystems size with a value close to $0.5$. 
The complicated subsystem-size dependence makes is hard to use the Markov gap to extract universal information about the interface.

\begin{figure*}\centering
	\includegraphics[width=\columnwidth]{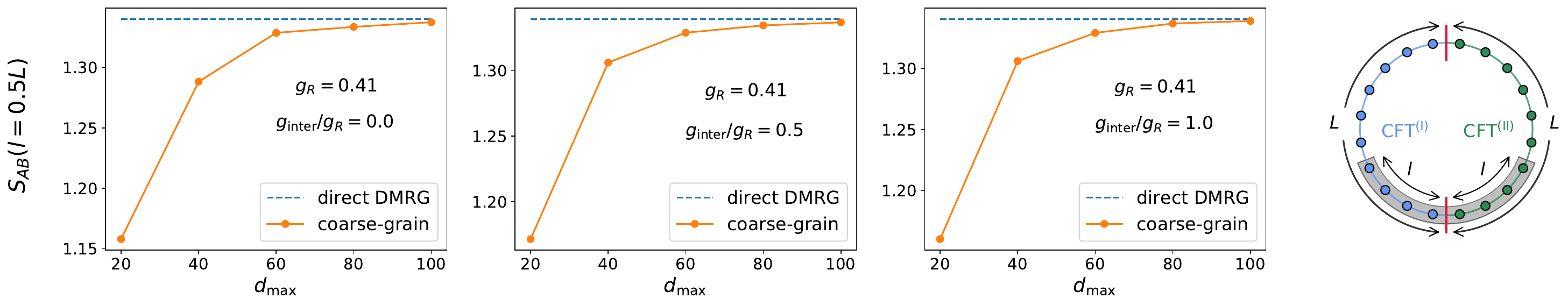}
	\caption{
		\label{fig:EE_OF_converge_app} 
        The convergence of symmetric EE $S_{AB}(l=0.5L)$ in the inhomogeneous OF model, obtained from the coarse-grain procedure for compressing the reduced density matrix of a subsystem located in the middle of the chain. 
        The blue dash lines represent the result from a direct DMRG solution via shifting the starting point of the periodic Hamiltonian to the middle of the left half-chain (the position of $x = l/2$), as a benchmark. 
        The orange dots are numerical results from the coarse-grain procedure for compressing the reduced density matrix of $\rho_{AB}$, plotted as a function of the bond dimension $d_{\rm max}$. 
        The very right panel shows a scheme for the entanglement-cut configuration. 
        Two adjacent subsystems $A$ and $B$ are symmetrically around the interface. 
        Here we choose $L = 150$, and the length of the total system size is $2L = 300$.  
	}
\end{figure*}

\begin{figure*}\centering
	\includegraphics[width=\columnwidth]{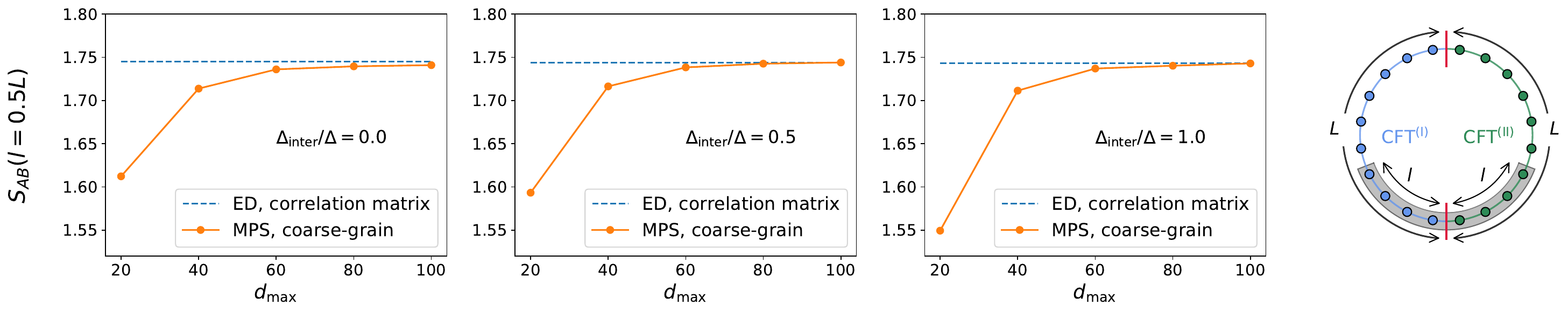}
	\caption{
		\label{fig:benchmark_symmEE_Fermion_deff_app}
        The convergence of symmetric EE $S_{AB}$ in the fermionic interface model, calculated from the coarse-grain procedure for compressing the reduced density matrix of a subsystem located in the middle of an MPS. 
        The blue dash lines represent the exact result from the correlation matrix technique as a benchmark. 
        The orange dots are numerical results from the coarse-grain procedure for compressing the reduced density matrix of $\rho_{AB}$, plotted as a function of the bond dimension $d_{\rm max}$. 
        The very right panel shows a scheme for the entanglement-cut configuration. 
        Two adjacent subsystems $A$ and $B$ are symmetrically around the interface. 
        Here we choose $L = 150$, such that the length of the total system is $2L = 300$. 
	}
\end{figure*}

\begin{figure*}\centering
	\includegraphics[width=\columnwidth]{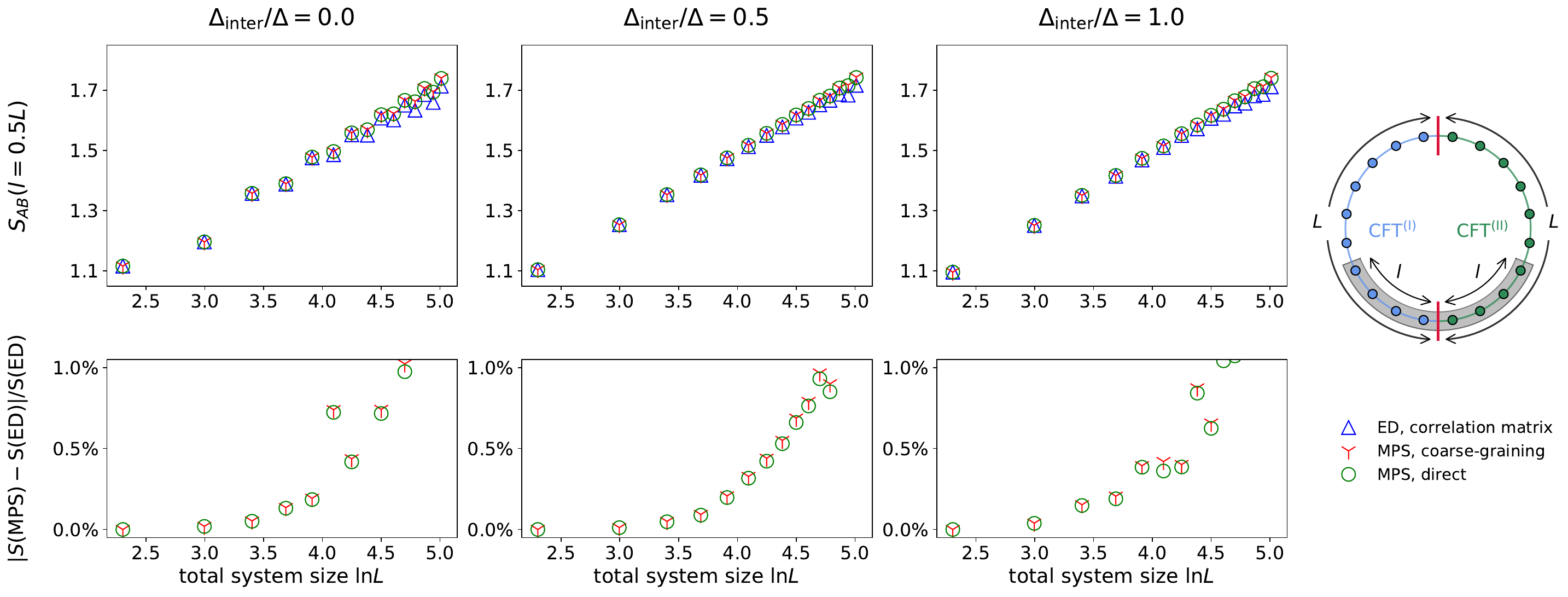}
	\caption{
		\label{fig:benchmark_symmEE_Fermion_app}
        A comparison of symmetric EE $S_{AB}$ in the fermionic interface model obtained from various numerical methods, under various settings of the interface coupling constant $\Delta_{\rm inter}$. 
        The very right panel shows a scheme for the entanglement-cut configuration. 
        Two adjacent subsystems $A$ and $B$ are symmetrically around the interface. 
        Here we choose $L \in [10, 150]$, such that the length of the total system is $2L \in [20, 300]$. 
	}
\end{figure*}

\subsection{Benchmark of the numerical approach}

In this part, we compare the numerical result of the symmetric EE $S_{AB}$ that is obtained from a direct DMRG calculation and the coarse-grain procedure discussed in the previous Supplementary Material. 
For direct DMRG solution, we mean that: as the model is in a periodic setting, we can spatially shift the starting point of our simulated Hamiltonian, such that we are only required to calculate the ordinary bipartite EE. 
Here we consider the case of $l_A = l_B = l = L/2$ for a benchmark. 
As shown in Fig.~\ref{fig:EE_OF_converge_app} and~\ref{fig:benchmark_symmEE_Fermion_deff_app}, the results from the coarse-grain procedure quickly converges as increasing the cut-off dimension $d_{\rm max}$ of the local Hilbert space $\mathcal{H}_{AB}$, showing a good convergence of our compression algorithm for $\rho_{AB}$. 
Moreover, in Fig.~\ref{fig:benchmark_symmEE_Fermion_app} we make a direct comparison between the exact result, a direct DMRG solution, and the coarse-grain procedure for the free fermionic interface model. 
Surprisingly, we find retreating the MPS via the coarse-grain procedure gives more accurate results than a direct DMRG solution.

%
% ****** End of file apssamp.tex ******